\pdfoutput=1
\documentclass[11pt]{article}
\usepackage[preprint]{acl}
\usepackage{times}
\usepackage{latexsym}
\usepackage[T1]{fontenc}
\usepackage[utf8]{inputenc}
\usepackage{microtype}
\usepackage{inconsolata}
\usepackage{graphicx}
\usepackage{booktabs}
\usepackage{multirow}
\usepackage{amsmath}
\usepackage{amssymb}
\usepackage{xcolor}
\usepackage{seqsplit}
\usepackage[ruled,vlined]{algorithm2e}
\SetKwInOut{Input}{input}\SetKwInOut{Output}{output}
\graphicspath{{./}{figures/}{gallery/}}

\title{Inverting Multi-Vector Visual Document Indices}

\author{Zhuchenyang Liu, Yao Zhang, Yu Xiao \\
  Aalto University \\
  Espoo, Finland \\
  \texttt{zhuchenyang.liu@aalto.fi}}

\begin{document}
\maketitle

\begin{abstract}
Prevailing multi-vector visual document retrievers store each page as about a thousand patch vectors, often in vector databases run by a third party. Since no one can read a page from its vectors, this index is easily treated as less sensitive than the page. However, because the index keeps one vector per patch in raster order, and each vector is computed by a vision-language model pre-trained to read documents, we hypothesize that whoever runs or breaches the store can reproduce a page from its index alone. We frame inversion as conditional document image generation and infer from the vectors what the attack needs: the encoder, the page shape and, for shuffled vectors, their order. On the ViDoRe~v3 benchmark, pages inverted from raw indices recover 47\% of the words and 45\% of the sensitive tokens. Used as queries against the stored indices, they rank their source page first 98.4\% of the time. We test two cheap protections, token pooling and shuffling, which both cut word recall to about 8\%. A model that restores the order of a shuffled index raises the share of source pages ranked first from 3.8\% to 93.5\%, while inverting a pooled index remains open. To test generalisation, we apply the same attack unchanged to another multi-vector retriever: its inverted pages still rank their source page first 70.2\% of the time, though its word recall stays below a nearest-neighbour baseline. Multi-vector visual document retrievers are therefore vulnerable to inversion through their stored index, which should be protected like the documents it encodes.
\end{abstract}

\section{Introduction}
\label{sec:intro}
\begin{figure}[t]
  \centering
  \includegraphics[width=\columnwidth]{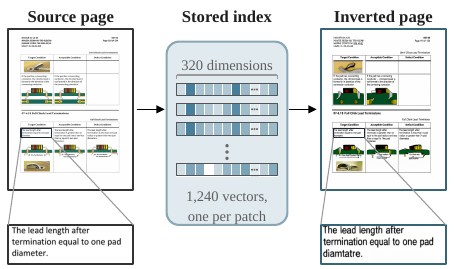}
  \caption{\textbf{A page inverted from its stored index.} The vector store keeps only the page's index, one 320-dimensional vector per image patch. Inverted from that index, the page is recovered with its tables, illustrations and text. The page (from the ViDoRe~v3 industrial subset) was chosen to show a mixed layout; its word recall is 0.620.}
  \label{fig:teaser}
\end{figure}
Visual document retrievers search pages directly as images. A vision-language model, fine-tuned into a late-interaction retriever, encodes a page as one vector per image patch, about a thousand in all. These vectors form the page's index, and a query is scored against the patches that match it \citep{khattab2020colbertefficienteffectivepassage,faysse2025colpaliefficientdocumentretrieval}. These retrievers are among the strongest on current benchmarks \citep{loison-etal-2026-vidore}. Their indices are kept in vector databases and dedicated engines \citep{pan2023surveyvectordatabasemanagement,santhanam2022plaidefficientenginelate}, which are often run by a third party \citep{Huang_2024,he2025transformqueryprivacypreservingapproach}. Since no person can read a page from its vectors, the index is easily treated as less sensitive than the page \citep[cf.][]{Huang_2024,https://doi.org/10.48694/jcls.3572}. Whether this holds for the multi-vector index of a document page has, to our knowledge, not been examined. Can whoever runs or breaches the store reproduce a page, given only its index?

There are three reasons to expect that it can. First, the index keeps the layout of the page: every patch, e.g., 32$\times$32 pixels for the retrievers we attack, has one vector, and the vectors are stored in raster order, so the position of each vector on the page is known. Second, each vector is computed by a vision-language model (VLM) that sees its patch in the context of the whole page, so it carries both local information about its patch and global information about the page. Third, these VLMs are pre-trained to read documents, on tens of millions of OCR samples and millions of PDFs parsed with their layout \citep{bai2025qwen3vltechnicalreport,bai2025qwen25vltechnicalreport}, so their vectors may encode comprehensive information for inverting the page. Based on these three properties, we hypothesize that the index keeps enough of the page for it to be recovered, with its text legible and in place.

Testing this hypothesis requires an inversion attack. For text embeddings, such attacks are well studied \citep{song2020informationleakageembeddingmodels,morris2023textembeddingsrevealalmost,Huang_2024,chen-etal-2025-algen,kim2026zero2textzerotrainingcrossdomaininversion}. They do not carry over to a visual index, because the target modality differs: our attack must invert the index into pixels that carry legible text, laid out on a page.

In this work, we invert the multi-vector index of a document page under zero side information, with nothing beyond the stored index (\S\ref{sec:threat}). We frame inversion as conditional document image generation. A conditional flow-matching inverter \citep{lipman2023flowmatchinggenerativemodeling}, trained on public pages paired with their indices, inverts an index into its page (Figure~\ref{fig:method}). The attack infers everything else it needs from the index alone: which public encoder produced it, the page shape and, for a shuffled index, the order of the vectors (\S\ref{sec:method}).

We attack Tomoro-ColQwen3-8B \citep{tomoro2026colqwen3}: we train the inverter on about 683k public pages and evaluate it on the 19{,}252 pages of ViDoRe~v3 \citep{loison-etal-2026-vidore}, which are disjoint from the training data. We measure leakage along two lines. Content leakage is word recall and sensitive-token recall: the fraction of the source page's words, and of its numbers, capitalised words and acronyms, that OCR reads from the inverted page. Identity leakage is re-identification: the retriever ranks all indexed pages against the inverted page, and we check whether the source page ranks first. From the raw index, the inverted pages recover 47\% of the words and 45\% of the sensitive tokens, and rank their source page first 98.4\% of the time (Figure~\ref{fig:teaser}).

We then test two cheap protections: hierarchical token pooling \citep{clavié2024reducingfootprintmultivectorretrieval}, which merges vectors, and shuffling, which stores them in random order. Both cut word recall to about 8\%. However, each vector was computed in the context of the whole page and still carries where its patch was. A position model that restores the order of a shuffled index raises re-identification from 3.8\% to 93.5\%. Inverting a pooled index remains open. We also test generalisation on a second retriever, ColQwen3.5-4.5B \citep{athrael2026colqwen35}, with the same training and inference recipe: its inverted pages still rank their source page first 70.2\% of the time. Multi-vector visual document retrievers are thus vulnerable to inversion through their stored index.

\section{Related Work}

\paragraph{Embedding inversion.}
Text embeddings can be inverted to much of their input: \citet{song2020informationleakageembeddingmodels} recover 50--70\% of a sentence's words, \citet{morris2023textembeddingsrevealalmost} recover 92\% of 32-token inputs exactly, and later attacks need only a surrogate encoder, a few leaked pairs or query access \citep{Huang_2024,chen-etal-2025-algen,kim2026zero2textzerotrainingcrossdomaininversion}. Contextualised token vectors decode into a vocabulary \citep{https://doi.org/10.48694/jcls.3572}, a prompt can be recovered from a language model's next-token distribution \citep{morris2023languagemodelinversion}, and \citet{zhuang2024understandingmitigatingthreatvec2text} study which design choices of a dense retriever make its embeddings easier to invert. Image features are inverted by optimisation \citep{mahendran2014understandingdeepimagerepresentations}, by up-convolutional networks \citep{dosovitskiy2016invertingvisualrepresentationsconvolutional} and with diffusion priors \citep{zhang2024unlockingvisualsecretsinverting}, or decoded into a caption \citep{Xiu_2025}, but always for a short text or a natural image. A document page is neither: its content is legible text laid out in two dimensions, which image generators render poorly without the layout planning and OCR-based supervision of the text-rendering literature \citep{chen2023textdiffuserdiffusionmodelstext,tuo2024anytextmultilingualvisualtext}; our inverter uses neither.

\paragraph{Multi-vector retrieval and its compression.}
Late interaction \citep{khattab2020colbertefficienteffectivepassage} lets document representations be computed offline and served by vector-similarity search \citep{santhanam2022plaidefficientenginelate}; ColPali carries it to page images \citep{faysse2025colpaliefficientdocumentretrieval}, and pooling or merging reduces the number of stored vectors at a small cost in retrieval quality \citep{clavié2024reducingfootprintmultivectorretrieval,ma-etal-2025-towards-storage}. None asks what the vectors still reveal.

\paragraph{Privacy of retrieval-augmented systems.}
Privacy attacks on retrieval-augmented generation use the operational interface: prompting the generator until it leaks its retrieval database \citep{zeng2024goodbadexploringprivacy}, or inferring from the answers to 30 natural queries whether a document is in the datastore \citep{naseh2025riddlethisstealthymembership}; protections at the same interface keep the raw query text from the provider of the vector database \citep{he2025transformqueryprivacypreservingapproach}. To our knowledge, inversion of the stored index has not been studied for visual late interaction.

\section{Threat Model and Problem}
\label{sec:threat}
A public encoder $E$ maps a page image $I$ to its index $\mathcal{C}=E(I)=(c_1,\dots,c_N)$, one vector $c_i\in\mathbb{R}^{d_{emb}}$ per patch. Like a vision transformer \citep{dosovitskiy2021imageworth16x16words}, it lays the page out as an $n_h\times n_w$ grid of patches, the page shape, and emits the vectors in row-major order, so $N=n_hn_w$ and vector $c_i$ covers row $\lfloor (i-1)/n_w\rfloor$ and column $(i-1) \bmod n_w$, both counted from 0. A store holds the indices of $P$ pages, the indexed pages or target corpus, under opaque identifiers, each in one storage configuration $T$: raw, pooled at factor $f$, or shuffled by a permutation $\pi$. The attacker observes $T(\mathcal{C})$ and inverts it into a page $\hat I\sim p(I\mid T(\mathcal{C}))$. The stored index omits three facts about how it was produced: which encoder $E$ produced it, the page shape $n_h\times n_w$, and, for a shuffled index, the permutation $\pi$.

The attacker holds public encoders, $E$ among them, and the stored indices, and may train on public documents. It is not told which encoder produced the store and has no access to a page's shape, source or metadata, or to any indexed page. We call any information about a page beyond its stored index side information, and require zero side information.

We attack three storage configurations $T\in\{T_r,T_p,T_s\}$ (Figure~\ref{fig:method}). \textbf{Raw}, $T_r(\mathcal{C})=\mathcal{C}$: the ordered sequence of vectors the encoder outputs for the page. \textbf{Pooled}, $T_p(\mathcal{C})=\mathrm{pool}_f(\mathcal{C})$: the vectors merged by hierarchical token pooling \citep{clavié2024reducingfootprintmultivectorretrieval} at factor $f\in\{3,9\}$ into $\lfloor N/f\rfloor$ vectors, a practical way to shrink a multi-vector index \citep{faysse2025colpaliefficientdocumentretrieval}. \textbf{Shuffled}, $T_s(\mathcal{C})=(c_{\pi(1)},\dots,c_{\pi(N)})$: the raw vectors stored in the order of a permutation $\pi$ of $\{1,\dots,N\}$ drawn at random for each page. Late-interaction scoring does not depend on the order of the vectors, so shuffling leaves retrieval unchanged. We call pooling and shuffling protections. The attacker knows $T$, a deployment choice shared by all pages of the store, but not the permutation $\pi$ of any page.

The attacker's objective is the content of the page. We measure what the inverted page recovers along two lines (\S\ref{sec:setup}): content leakage, the words and sensitive tokens that OCR reads from it, and identity leakage, whether it is faithful enough to rank its source page first among the $P$ indexed pages.

\section{Method}
\label{sec:method}
\begin{figure}[t]
  \centering
  \includegraphics[width=\columnwidth]{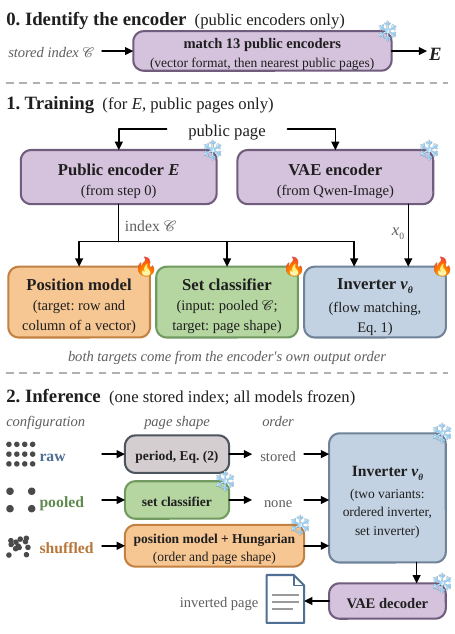}
  \caption{\textbf{The three steps of the attack.} Flames: models the attacker trains for an encoder $E$; snowflakes: frozen models, which at inference include all of them. Step 0 identifies $E$ from the stored index with public encoders alone (\S\ref{sec:encid}); step 1 trains models for $E$ on public pages; in step 2, each row shows, per storage configuration, where the page shape and the order come from. The inverter uses its ordered variant when the order is known and its set variant otherwise (\S\ref{sec:inverter}).}
  \label{fig:method}
\end{figure}
The attack is built around a conditional flow-matching inverter that inverts an index into its page (\S\ref{sec:inverter}). The inverter is trained for one encoder, draws the page at a given shape, and reads each vector at its position in the page, so it needs the three facts the stored index omits (\S\ref{sec:threat}). The attacker therefore first identifies the encoder from the stored indices, with public encoders alone (\S\ref{sec:encid}), and trains the inverter and the models that infer page shape and order on public pages encoded with it. At inference, the page shape is inferred from the index (\S\ref{sec:grid}), and the order of a shuffled index is restored by a position model (\S\ref{sec:reorder}). We describe the inverter first. Figure~\ref{fig:method} shows the three steps of the attack.

\subsection{Conditional flow-matching inverter}
\label{sec:inverter}
The inverter follows the design of Qwen-Image \citep{wu2025qwenimagetechnicalreport}, a text-to-image model, with two substitutions: the condition is the stored index $\mathcal{C}$ in place of a text prompt, and the generated image is the page $I$. As in Qwen-Image, the page is generated in the latent space of a frozen VAE with spatial factor 8 and 16 channels, by a flow-matching model \citep{lipman2023flowmatchinggenerativemodeling} on the rectified-flow path \citep{esser2024scalingrectifiedflowtransformers}. Regressing the page would blur it: the $L_2$ optimum of a regression decoder is the conditional mean $\mathbb{E}[I\mid\mathcal{C}]$, which blurs every glyph the index leaves ambiguous. We therefore generate the page, learning the conditional distribution and sampling from it. With $x_0$ the VAE latent of the page, $x_1\sim\mathcal{N}(0,I)$, $t$ drawn from a logit-normal distribution, and $x_t=t\,x_0+(1-t)\,x_1$, the inverter $v_\theta$ minimises
\begin{equation}
\mathcal{L}=\mathbb{E}\left\|v_\theta(x_t,t,\mathcal{C})-(x_0-x_1)\right\|^2.
\label{eq:flow}
\end{equation}
The loss is still a squared error, but its conditional expectation $\mathbb{E}[x_0-x_1\mid x_t,t,\mathcal{C}]$ averages only over the ambiguity that remains given $x_t$, which is what lets iterative sampling produce sharp text. Two further choices keep the inverted pages faithful. First, we minimise Eq.~\eqref{eq:flow} alone, without auxiliary perceptual or adversarial losses, since each extra objective brings its own prior, which could write text absent from the stored index. Second, we drop the condition with probability 10\% and replace it with a learnable null token, as in classifier-free guidance \citep{ho2022classifierfreediffusionguidance}; sampling the inverter with the null token then shows what the learned image prior generates alone. At inference we integrate the guided velocity $v_u+\gamma(v_c-v_u)$, unconditioned and conditioned, over 50 Euler steps and decode with the frozen VAE.

The inverter is a dual-stream multimodal diffusion transformer (MMDiT; \citealp{esser2024scalingrectifiedflowtransformers}) in which the stored index replaces the text stream, so that condition and image tokens attend jointly in every block while keeping separate parameters. One set of weights serves every page shape, because position enters only through rotary embeddings computed from the input shape; Appendix~\ref{app:training} gives the width, depth, and the rest of the configuration.

\paragraph{Ordered and set inverters.}
\label{sec:streams}
The inverter has two variants, which differ only in the positions of the condition stream. The ordered inverter gives each vector of a raw index its true two-dimensional position. The set inverter omits the rotary embeddings of the condition, which makes it permutation-invariant by construction; it reads a pooled index, and a shuffled one when no position model is used. Each unordered configuration has its own set inverter; all other choices are shared (Appendix~\ref{app:training}).

\subsection{Identifying the encoder}
\label{sec:encid}
The inverter, like every model the attack trains, is specific to one encoder, so the attacker first identifies the encoder from the stored vectors, in two steps and with public encoders alone. First, it keeps the candidates $e$ whose output fits the index in dimension, unit norm and vector count, all fixed by the candidate's public processor configuration. Second, each remaining candidate $e$ encodes a reference set $R_e$ of 2,000 public training pages. For each stored index, the attacker retrieves its nearest pages in $R_e$ by the cosine of mean-pooled vectors and re-ranks them by the mean cosine between each stored vector and its nearest reference vector; the best score is the similarity of the index to $R_e$. The candidate with the highest similarity is chosen. No indexed page is used; Appendix~\ref{app:encid} gives a second test with public queries.

\subsection{Recovering the page shape}
\label{sec:grid}
The inverter must be told what shape of page to draw. For a raw index, $N=n_hn_w$, so the shape is one of the pairs $(n_h,n_w)$ whose product is $N$; the count prior simply picks the commonest such shape among the attacker's training pages with the same vector count. The order of the vectors gives a stronger cue: the row width $n_w$ is a period of the sequence, because $c_i$ and $c_{i+n_w}$ cover two patches in the same column of adjacent rows, one directly above the other, and so tend to be similar. Writing $s(w)$ for the mean cosine similarity between vectors at lag $w$, we take
\begin{equation}
\hat n_w=\arg\max_{W\in\mathcal{W}}\, s(W),
\label{eq:period}
\end{equation}
where $\mathcal{W}$ holds the divisors of $N$ whose page shape falls in an aspect-ratio window set from the attacker's training pages. Pooling merges vectors from different parts of the page, so a pooled index has no such period. Its vector count still constrains the shape: at factor $f$ the store holds $\lfloor N/f\rfloor$ vectors, which leaves only a handful of candidate shapes. A small permutation-invariant set classifier \citep{zaheer2018deepsets}, trained on pooled indices of public pages, picks one of these candidates; its logits are masked so that it can only output shapes compatible with the vector count. Algorithm~\ref{alg:shape} (Appendix~\ref{app:shape}) gives the procedure for both raw and pooled indices. The attacker can therefore infer each page's shape from its index alone.

\subsection{Restoring the order of a shuffled index}
\label{sec:reorder}
A shuffled index is a set, but each of its vectors was computed from one patch in the context of the page, and may still carry where that patch was. Labels for learning this need no annotation: running the public encoder on a public page gives each vector $c_i$ its row $\lfloor (i-1)/n_w\rfloor$ and column $(i-1) \bmod n_w$. On the attacker's public pages we train a position model to predict each vector's normalised row and column with a squared error. It is a Transformer encoder over the set without positional encoding, and so is permutation-equivariant and free to use the other vectors of the page as context. A summary token also regresses the page's aspect ratio (the aspect head), and the candidate shape nearest to it gives the page shape, since a shuffled index has no period. The Hungarian algorithm assigns the predicted coordinates one-to-one to the $n_h\times n_w$ slots, and the re-ordered index goes to the ordered inverter unchanged. A two-layer per-vector probe is a weaker baseline (Appendix~\ref{app:position}).

\section{Experimental Setup}
\label{sec:setup}
\paragraph{Attacked public encoders.}
The primary attacked public encoder, $E_A$, is Tomoro-ColQwen3-8B, a ColPali-style late-interaction retriever built on Qwen3-VL that emits one 320-dimensional vector per patch. A page is resized to an input resolution of $32n_w\times 32n_h$ pixels before encoding, and the index is the image-token span of the encoder output, without prompt tokens or padding. To test whether the attack generalises beyond one retriever, we apply it unchanged, end to end, to a second public encoder, $E_B$, ColQwen3.5-4.5B, fine-tuned by a different group from a different backbone, which encodes each page at a lower input resolution (Appendix~\ref{app:repro}). Methods, model configuration and sampling are held fixed across the two encoders; $E_B$ differs in input resolution and training data, and is attacked raw and shuffled. The attacker identifies the encoder of a store (\S\ref{sec:encid}) from a pool of 13 public multi-vector retrievers, $E_A$ and $E_B$ among them, with embedding dimensions 128, 320 and 640 (Table~\ref{tab:pool} in Appendix~\ref{app:encid}).

\paragraph{Training corpora.}
The inverters are trained on four public collections of visual documents: the VisRAG training corpora \citep{yu2025visragvisionbasedretrievalaugmentedgeneration}, the ColPali training set \citep{faysse2025colpaliefficientdocumentretrieval}, vdr-multilingual-train and VDR-MEGA-2 (Appendix~\ref{app:grids}). After deduplication, 682,818 pages train the models for $E_A$ (Appendix~\ref{app:grids}). The models for $E_B$ are trained on a smaller public corpus, its indices of 346,770 pages from the first three collections, without VDR-MEGA-2.

\paragraph{Target corpus and fixed evaluation set.}
The target corpus is the eight public corpora of ViDoRe~v3 \citep{loison-etal-2026-vidore}, 19,252 pages from eight professional domains (Appendix~\ref{app:persubset}). All evaluations use a fixed set of 2,000 pages, 250 sampled uniformly from each domain, covering 11 page shapes. The target corpus and the training corpora come from different sources. To check for overlap between them, we retrieve for each evaluation page its nearest training page by late-interaction similarity between their stored indices: only 8 of the 2,000 pages (0.4\%) have a near-identical training counterpart (Appendix~\ref{app:repro}), and removing these pages changes no reported number.

\paragraph{Stored indices.}
Pooled indices come from the unmodified hierarchical token pooler of \citet{clavié2024reducingfootprintmultivectorretrieval} at factors 3 and 9 (Appendix~\ref{app:repro}). The shuffled index permutes each page's raw vectors once, by its permutation $\pi$, shared by every attack on it. Under $E_A$ each page has four stored indices (raw, pooled $\times$3, pooled $\times$9, shuffled) and under $E_B$ two (raw, shuffled), each read by its own inverter; Figure~\ref{fig:models} in Appendix~\ref{app:training} lists every model the attacker trains, its data and its results. We call each combination of encoder, storage configuration and attack a setting.

\paragraph{Sampling.}
Sampling uses the page shape inferred from each page's own index, one noise seed shared by every page and setting, 50 Euler steps and a guidance scale of $\gamma=4$, selected once on 100 in-distribution validation pages and applied to every setting of both encoders (Appendix~\ref{app:guidance}).

\paragraph{Content.}
Both pages are transcribed with PaddleOCR \citep{li2022ppocrv3attemptsimprovementultra} and the transcripts compared. Word recall is the fraction of the reference words, as a multiset, that the inverted page recovers; NED is the character-level edit distance between the two transcripts in reading order, normalised by reference length and capped at 1; sensitive-token recall is word recall restricted to sensitive tokens, that is, numbers, capitalised words and acronyms (Appendix~\ref{app:regex}). The text metrics are computed on the 1,927 evaluation pages with at least 20 reference words.

\paragraph{Re-identification.}
\label{sec:reid}
Re-identification uses the retriever itself as the matcher: the inverted page is encoded like a query and scored against every indexed page, and re-identification succeeds when its source page ranks first among the $P$ indexed pages. It complements content because OCR underestimates what is recovered where glyphs are only partly formed, while an encoder can still match them. The inverted page $\hat I$ is encoded into query vectors $\{q_i\}$, and every indexed page $d$, $P=19{,}252$ pages, is scored by late interaction \citep{khattab2020colbertefficienteffectivepassage},
\begin{equation}
S(\hat I,d)=\textstyle\sum_i\max_j\, q_i^{\top}d_j,
\label{eq:maxsim}
\end{equation}
with $d_j$ the vectors of page $d$; top-1 is the share of pages whose source page ranks first. Scored with the encoder that produced the index, the metric could be circular, rewarding features that this encoder maps back to the stored vectors without the page resembling its source. We therefore also score every inverted page with an independent judge, an encoder that played no part in producing the index and re-encodes the inverted page and the corpus: $E_B$ judges pages inverted from $E_A$, and vice versa. For the pooled configurations, the attacked encoder scores the inverted page against the pooled indices held in the store, while the independent judge, which has no stored index, encodes the corpus pages itself without pooling.

\paragraph{Baselines.}
\label{sec:baselines}
An inverter could score above zero without recovering anything, simply by drawing a plausible page of the right kind. We therefore compare every result with the following baselines. An attack must exceed three of them: the null model, the same inverter sampled with its condition dropped; the kNN baseline, which answers with the training page whose stored index is nearest to the target's and so measures what layout templates and shared vocabulary provide (Appendix~\ref{app:repro}); and chance, for the rank metrics. The VAE reconstruction, the source page passed through the frozen VAE, gives the upper bound; further controls are described in Appendix~\ref{app:controls}.

\section{Results}
\begin{table*}[t]
\centering\footnotesize\setlength{\tabcolsep}{3.6pt}
\begin{tabular}{llcccccccc}
\toprule
& Index & Word rec.\ $\uparrow$ & NED $\downarrow$ & Sens.\ rec.\ $\uparrow$ & \multicolumn{5}{c}{Re-identification among 19,252} \\
\cmidrule(lr){6-10}
& & & & & top-1 & top-5 & MRR & med.\ rank & indep. \\
\midrule
Null model             & --            & 0.001 & 0.958 & 0.008 & 0.000 & 0.000 & 0.001 & 8{,}161 & -- \\
kNN baseline           & raw           & 0.234 & 0.858 & 0.146 & 0.199 & 0.438 & 0.320 & 7 & -- \\
\textbf{Ours}           & \textbf{raw}  & \textbf{0.474} & \textbf{0.292} & \textbf{0.450} & \textbf{0.984} & \textbf{0.995} & \textbf{0.989} & \textbf{1} & \textbf{0.977} \\
Ours                    & pooled $\times$3 & 0.076 & 0.858 & 0.044 & 0.011 & 0.035 & 0.029 & 496 & 0.005 \\
Ours                    & pooled $\times$9 & 0.081 & 0.879 & 0.054 & 0.005 & 0.024 & 0.018 & 724 & 0.003 \\
Ours, w/o position model & shuffled & 0.084 & 0.842 & 0.048 & 0.038 & 0.104 & 0.076 & 203 & 0.011 \\
Ours, w/ position model & shuffled & 0.231 & 0.925 & 0.212 & 0.935 & 0.978 & 0.955 & 1 & 0.877 \\
\midrule
\multicolumn{2}{l}{VAE reconstruction} & 0.956 & 0.028 & 0.965 & 1.000 & 1.000 & 1.000 & 1 & 0.996 \\
\multicolumn{2}{l}{Chance} & -- & -- & -- & 0.00005 & 0.0003 & -- & 9{,}626 & 0.00005 \\
\bottomrule
\end{tabular}
\caption{Inverting the index of 2,000 held-out ViDoRe~v3 pages (250 per domain) under $E_A$ (OCR columns over the 1,927 with $\ge 20$ reference words). The raw, pooled $\times$3 and $\times$9 indices hold on average 1,241, 413 and 137 vectors per page and keep 100\%, 99.2\% and 97.8\% of nDCG@10. Shapes are inferred from the vectors (\S\ref{sec:prelim}), except for the shuffled row without the position model, which is given the true shape. Indep.: top-1 under the independent judge $E_B$; for the VAE row, of the source pages themselves. Null row: 200-page subset (Appendix~\ref{app:controls}). Bootstrap intervals: Appendix~\ref{app:bootstrap}; precision and F1: Table~\ref{tab:precision}.}
\label{tab:main}
\end{table*}

\subsection{The attacker's preliminaries}
\label{sec:prelim}
\paragraph{Identifying the encoder among public encoders.}
We build one store of the 2,000 evaluation pages for each of the 13 encoders of the pool. From the index of a single page, the attacker identifies the right encoder in all 26,000 decisions, one per page and store, including encoders of the same family that share dimension and vector count (Appendix~\ref{app:encid}).

\paragraph{The page shape.}
Under $E_A$, periodicity recovers the shape of 99.2\% of raw indices, the set classifier (\S\ref{sec:grid}) 99.1\% and 95.7\% at pooling factors 3 and 9, and the aspect head of the position model (\S\ref{sec:reorder}) 99.95\% of shuffled ones. Because the inferred shape is almost always correct, giving the inverter the true shape instead barely changes the results (Appendix~\ref{app:shape}).

\subsection{The page is invertible from its raw index}
\label{sec:r1}
\begin{figure*}[t]
  \centering
  \includegraphics[width=\textwidth,trim=0 11pt 0 0,clip]{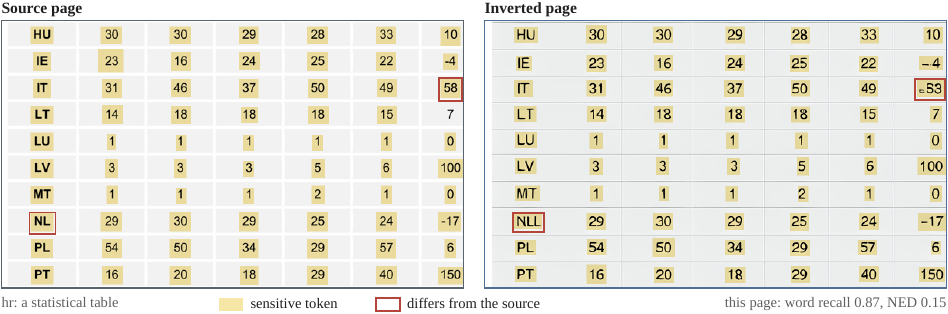}
  \caption{\textbf{Numeric cells recovered from a raw index.} A band of a statistical table on a ViDoRe~v3 HR page, cut from the source page and from the page inverted from its raw index: 59 of its 60 numeric cells are recovered exactly. Yellow: tokens that sensitive-token recall counts; vermillion: the two tokens that differ, one numeric cell and one country code. The page was chosen by inspection and, with the two bands of Appendix~\ref{app:gallery}, gives one page per class of sensitive token; the band is its densest in sensitive tokens.}
  \label{fig:qual}
\end{figure*}
From the raw index, the inverted page recovers 47.4\% of the reference words at a character-level NED of 0.292, against 0.858 for the kNN baseline and 0.958 for the null model, which scores near zero on every metric (Table~\ref{tab:main}): what lifts the attack above both controls is the index. 98.4\% of the inverted pages rank their source page first among 19,252. Under the independent judge, which played no part in producing the index, top-1 is still 97.7\%, close to the 99.6\% it reaches on the source pages themselves.

Word recall has a lower bound that does not depend on reading the page, because pages of the same kind share vocabulary: the kNN baseline already recovers 23.4\% of the reference words. The inversion exceeds it in every domain (Appendix~\ref{app:persubset}). Sensitive-token recall is 45.0\%, against 14.6\% for the kNN baseline. Figure~\ref{fig:qual} shows a statistical table recovered almost cell for cell; the remaining errors change a digit or repeat one (Appendix~\ref{app:gallery}).

\label{sec:controls}
The recovered text is not invented by the prior. The null model, which samples without the index, recovers only 0.1\% of the words. Two samples drawn from the same index with different seeds agree with each other (word F1 0.476) no more than each agrees with the source (0.480), so the text they share is the source's own (Appendix~\ref{app:controls}).

\label{sec:r2}
OCR does not capture everything that is recovered. It transcribes only fully formed glyphs, yet 75.0\% of the 44 scored pages with word recall below 0.1 still rank their source page first, and the two measures correlate only weakly ($\rho=0.13$; Appendix~\ref{app:readability}). Word recall is therefore a conservative measure of content leakage.

\subsection{Inverting indices under two protections}
\label{sec:r3}
\begin{figure*}[t]
  \centering
  \includegraphics[width=\textwidth]{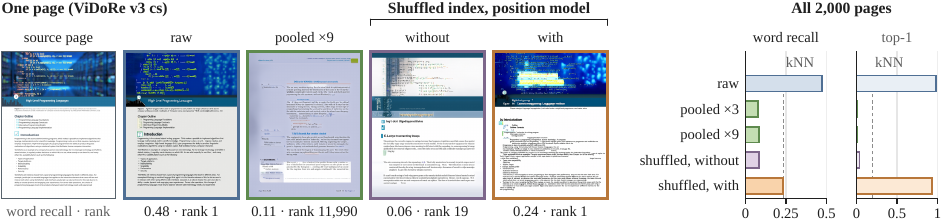}
  \caption{\textbf{Our attacks do not invert pooled indices; a shuffled order can be restored.} Left: one page under the three storage configurations (shuffled: without and with the position model), with its word recall and rank among 19,252; the page is typical by a fixed rule, its word recall closest to the mean under both the raw and the restored index (Appendix~\ref{app:gallery}). Right: all 2,000 pages; dashed, the kNN baseline.}
  \label{fig:protections}
\end{figure*}
Pooled indices recover far less (Table~\ref{tab:main}): word recall falls from 47.4\% to 7.6\% at factor 3 and 8.1\% at factor 9, below the kNN baseline, and NED rises to 0.858 and 0.879. Only 21 and 10 of the 2,000 inverted pages rank their source page first, against 1,968 from the raw index. Our attacks do not invert pooled indices: inversion falls below the kNN baseline on every metric, though source pages still rank far above chance.

Without a position model, a shuffle protects as well as pooling and leaves retrieval unchanged: the set inverter reaches word recall 8.4\% and NED 0.842, the same level as the pooled indices, and re-identifies 3.8\% of pages, although it is given the true page shape.

The order, however, can be recovered from the vectors (Figure~\ref{fig:protections}). The position model (\S\ref{sec:reorder}) places the shuffled vectors of $E_A$ with a mean absolute error of 1.8 index rows and 4.5 index columns, against 12.6 and 11.3 for a random order (Appendix~\ref{app:position} compares it with a random order and a per-vector probe). Fed the re-ordered index, the ordered inverter re-identifies 93.5\% of pages, against 98.4\% for the raw index and 87.7\% under the independent judge, and reaches word recall 23.1\% and sensitive-token recall 21.2\%. Identity is thus recovered more fully than content, which needs vectors placed more precisely (Appendix~\ref{app:position}). Its word recall only matches the kNN baseline, while its sensitive-token recall exceeds it by 6.6 points (Appendix~\ref{app:bootstrap}), which suggests that the restored order recovers some of the page's own sensitive tokens beyond shared vocabulary.

Recovered text also follows its vector. With the restored order, NED stays near its cap (0.925) although word recall reaches 23.1\%. The words are recovered, but whole lines and blocks appear at the wrong height, so the transcript is read in a different order (Figure~\ref{fig:reorder} in Appendix~\ref{app:failure}). Each vector thus carries the content of its own patch, and the inverter draws that content wherever the vector is placed, which supports the premise that the index covers the page patch by patch (\S\ref{sec:intro}).
\subsection{Generalisation to a second retriever}
\label{sec:r4}
\begin{table}[t]
\centering\footnotesize\setlength{\tabcolsep}{1.5pt}
\begin{tabular}{lccccc}
\toprule
Index ($E_B$) & \begin{tabular}[b]{@{}c@{}}Word\\rec.\,$\uparrow$\end{tabular} & NED\,$\downarrow$ & \begin{tabular}[b]{@{}c@{}}Sens.\\rec.\,$\uparrow$\end{tabular} & top-1\,$\uparrow$ & \begin{tabular}[b]{@{}c@{}}med.\\rank\,$\downarrow$\end{tabular} \\
\midrule
Null model & 0.001 & 0.963 & 0.006 & 0.000 & 8{,}892.5 \\
kNN baseline & 0.223 & 0.860 & 0.114 & 0.154 & 12 \\
raw, inferred shape & 0.079 & 0.718 & 0.085 & 0.702 & 1 \\
raw, true shape     & 0.094 & 0.655 & 0.098 & 0.818 & 1 \\
shuffled \\
\quad w/o position model & 0.028 & 0.846 & 0.041 & 0.006 & 1{,}047 \\
\quad w/ position model & 0.061 & 0.903 & 0.073 & 0.112 & 96.5 \\
\midrule
$E_A$, same data & 0.555 & 0.256 & 0.544 & 0.989 & 1 \\
\midrule
VAE reconstruction & 0.895 & 0.058 & 0.899 & 1.000 & 1 \\
Chance & -- & -- & -- & 0.00005 & 9{,}626 \\
\bottomrule
\end{tabular}
\caption{The same attack and 2,000 pages (250 per domain) under $E_B$. The shape is inferred (0.795 correct for the raw index), except for the shuffled row without the position model, which is given the true shape; the true-shape row isolates the cost of inference. Under the independent judge $E_A$, top-1 is 0.605 (raw) and 0.008 and 0.076 (shuffled, w/o and w/ position model). $E_A$ row: a control inverter for $E_A$ trained on the same data as the $E_B$ inverter (top-1 0.980 under the judge $E_B$). Null row: 200-page subset.}
\label{tab:encb}
\end{table}
Applied unchanged to $E_B$, with models trained on half as many pages and without VDR-MEGA-2, the attack reproduces two findings (Table~\ref{tab:encb}). First, it remains feasible: from the raw index, 70.2\% of the inverted pages rank their source page first, against 15.4\% for the kNN baseline, and 60.5\% do so under the independent judge $E_A$. Second, it depends on vector order in the same way: a shuffle lowers top-1 to 0.6\% (0.8\% under $E_A$), and the position model raises it only to 11.2\% (7.6\% under $E_A$), still below the kNN baseline. Content leakage does not carry over: word recall, 7.9\%, stays below the baseline's 22.3\%. The gap is not the attacker's data: an inverter for $E_A$ trained on the same three collections as the $E_B$ inverter recovers 55.5\% of the words and re-identifies 98.9\% of pages, so the difference lies in the encoder. It even exceeds the main $E_A$ inverter, likely because a fixed training budget underfits the larger corpus or the smaller one is closer to ViDoRe~v3; either way, the reported leakage is a lower bound. Its shape and position estimates are also less accurate (Appendices~\ref{app:shape} and~\ref{app:position}), and its input resolution and vector count differ from those of $E_A$; we cannot say which of these encoder properties makes it leak less (Limitations).

\section{Conclusion}
\label{sec:conclusion}
The multi-vector index of a document page can be inverted under zero side information. From the stored vectors alone, the attacker identifies the public encoder among the 13 we tested, recovers the page shape and, for a shuffled index, the order of the vectors; a conditional flow-matching inverter then redraws the page. From the raw index, the inverted page recovers 47\% of the words and ranks its source page first 98.4\% of the time. A shuffle is not a reliable protection, since restoring the order recovers most of the re-identification under $E_A$. Our attacks do not invert a pooled index, which remains open (Appendix~\ref{app:open}). Multi-vector visual document retrievers are therefore vulnerable to inversion through their stored index, which should be protected like the documents it encodes.

\section{Limitations}

First, the attacked encoders come from a single model family. Both are ColPali-style late-interaction retrievers on Qwen backbones, so the attack is tested on one family only. Encoder identification is likewise evaluated only on stores produced by the 13 public encoders of our pool. Apart from flagging a store whose encoder is outside the pool (Appendix~\ref{app:encid}), we did not study stores produced by a private encoder, so inverting such an index is outside the scope of this paper.

Second, several choices in our setup make the reported leakage conservative. The validation loss of the ordered inverter is still falling at 100k steps, with no sign of overfitting, so a larger model or a longer schedule could raise the measured leakage. The training data matters too: a smaller, better-matched training set alone raises word recall from 0.474 to 0.555 (\S\ref{sec:r4}). The sampler matters as much: guidance alone raises word recall by about half on validation pages (Appendix~\ref{app:guidance}), and we chose its scale once, on in-distribution validation pages, and applied it unchanged to every setting. The position models use a single untuned configuration and train in under an hour. Because we deliberately adapted nothing to either encoder, the training hyperparameters and other components may be suboptimal for both.

Third, the evaluation metrics have their own limits. The OCR-based metrics use the OCR transcript of the source page as the reference, and that transcript may itself contain errors. Sensitive tokens are defined by a regular-expression heuristic (Appendix~\ref{app:regex}), and we ran no human study to confirm that the tokens it counts are the ones a reader would consider sensitive. Re-identification, which ranks the inverted page against the store, measures only whether the inverted page is faithful to its source; it is not proposed as an attack in itself. An attacker who only wants to link a known document to the store needs no inversion, since re-encoding the document with the public encoder reproduces its stored index (Appendix~\ref{app:repro}).

Fourth, we cannot verify which parts of an inverted page are correct. The inverter samples from a learned distribution, so it can output a plausible page of the right kind even when nothing was recovered. Two samples from the same index agree mainly on the source's text (\S\ref{sec:controls}). Drawing several samples per index and keeping only what they share may therefore let an attacker separate high-confidence leakage from invented content; we leave this to future work.

Finally, we evaluate a limited range of attacks and protections. We did not try to recover positions for pooled vectors, so the finding that pooling resists inversion holds only for the attacks in this paper. Protections we did not evaluate include additive noise, quantisation, keyed random projections, encrypted or private late-interaction retrieval, and access control on the store.

\section*{Ethical Considerations}
This paper describes an attack. The threat applies wherever the index is held apart from the pages: a hosted vector database whose identifiers point to page images in separate, access-controlled storage, separate access tiers for the index and the documents, or index snapshots and backups. In these settings the stored index is the only copy of the page that an intruder obtains. The paper uses only public data and public models (Appendix~\ref{app:licenses}): the attacker's public training collections, the eight public corpora of ViDoRe~v3, the two non-public ones left unused, and public retrievers. No deployed store was attacked, and no data beyond the public benchmark pages was processed. The work is dual use. Its purpose is to show, before such stores are attacked in practice, which ways of storing a late-interaction index are unsafe: a raw index should be protected like the documents it encodes, and a shuffle is not a protection (\S\ref{sec:conclusion}). The weakness is a property of the stored representation rather than a behaviour elicited from a model: the encoders function as designed, and the risk follows from storing one vector per patch in raster order, a practice shared by late-interaction retrievers. It is therefore not a vulnerability of any one model or vendor that a patch could fix; the remedy lies with whoever operates the store, and our recommendations are addressed to them. That stored embeddings can leak their input is already established for text \citep{song2020informationleakageembeddingmodels,morris2023textembeddingsrevealalmost}; we extend this known class of risk to multi-vector visual indices. Upon publication we will release the code, the evaluation manifest and the protocol, but not the trained inverters, so that the measurements can be reproduced without distributing a ready-made attack.

\bibliography{refs,custom}

\appendix
\section{Whole-page gallery}
\label{app:gallery}
\begin{figure*}[h]
  \centering
  \includegraphics[width=\textwidth]{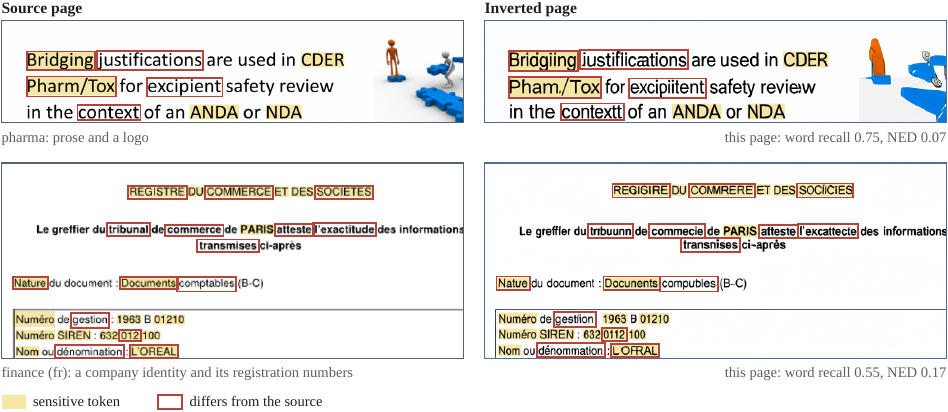}
  \caption{Two further bands, chosen with Figure~\ref{fig:qual} as one page per class of sensitive token: prose with acronyms (\emph{pharma}) and a company filing with its registration numbers (\emph{finance\_fr}). Source left, inverted page from the raw index right, marked as in Figure~\ref{fig:qual}; a word is boxed only where OCR and the eye agree that it differs, so the registration number, which OCR misreads, is not. Both pages are ranked first among 19,252.}
  \label{fig:crops-more}
\end{figure*}

\paragraph{Reading the bands.}
The band of Figure~\ref{fig:qual} recovers 59 of its 60 numeric cells exactly, negative values and the three-digit entries included; the one error turns a 58 into a 53, and one of the ten two-letter country codes gains a letter. Elsewhere on the same page, whose word recall is 0.87, the common error is a repeated digit, as $-25$ recovered as $-255$. Figure~\ref{fig:crops-more} adds a band of prose and a band of a company filing, the two other pages chosen per class of sensitive token. In the prose the acronyms \textsc{cder}, \textsc{anda} and \textsc{nda} return exactly, while five of the 19 words keep their shape and gain or lose a letter, as \emph{context} recovered as \emph{contextt}. In the filing 15 of the 41 words differ from the source, most by a letter or two, in the heading and the body text alike. The registration number \texttt{1963 B 01210} is recovered in its field with every digit legible, although OCR reads one malformed 3 as an 8; the nine-digit company number is recovered with one digit inserted, and the company's name with one letter wrong.

Figures~\ref{fig:gal-cs} to~\ref{fig:gal-finance_fr} show one page of each domain at the 75th, 50th, and 25th percentile of that domain's word recall under the raw index, together with the pages inverted from the raw index, from the pooled $\times 9$ index, and from the shuffled index with the position model. The page of Figure~\ref{fig:protections} was picked by a rule fixed in advance: among pages outside the French and energy subsets with 150 to 400 reference words that rank first from both the raw and the restored index, the six whose word recall under both is closest to the set means, and of these the one most legible at thumbnail size. The gallery pages are selected by rule from the fixed evaluation set, so each figure shows the range of the domain, including its weaker cases, and every inverted page is annotated with its own word recall and rank.

\definecolor{galink}{HTML}{1F1F1F}
\definecolor{galsrc}{HTML}{58636C}
\definecolor{galraw}{HTML}{4F6F94}
\definecolor{galpool}{HTML}{5E8A4E}
\definecolor{galrest}{HTML}{B8773A}
\definecolor{galmuted}{HTML}{5F5F5F}
\setlength{\fboxsep}{0pt}\setlength{\fboxrule}{0.8pt}

\begin{figure*}[p]
  \centering
  \footnotesize
  \setlength{\tabcolsep}{1.5pt}
  \begin{tabular}{@{}cccc@{}}
  \textcolor{galink}{source page} & \textcolor{galraw}{raw} & \textcolor{galpool}{pooled $\times$9} & \textcolor{galrest}{shuffled, w/ position model} \\[1pt]
  \fcolorbox{galsrc}{white}{\includegraphics[width=0.232\textwidth]{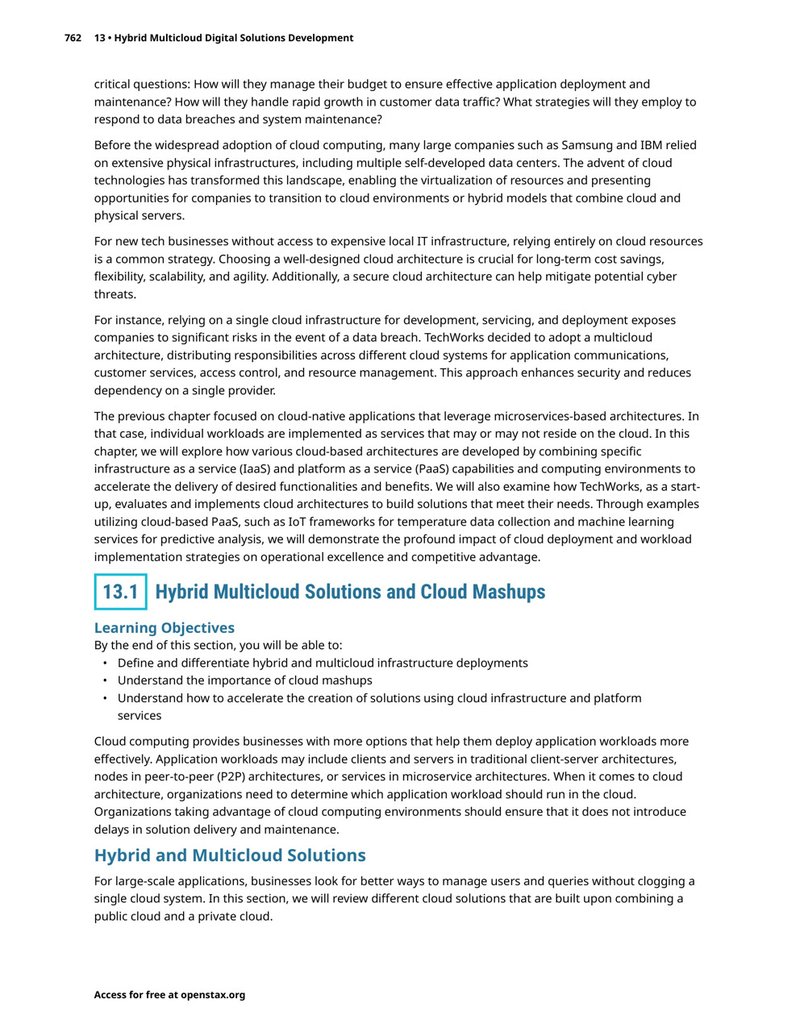}} &
  \fcolorbox{galraw}{white}{\includegraphics[width=0.232\textwidth]{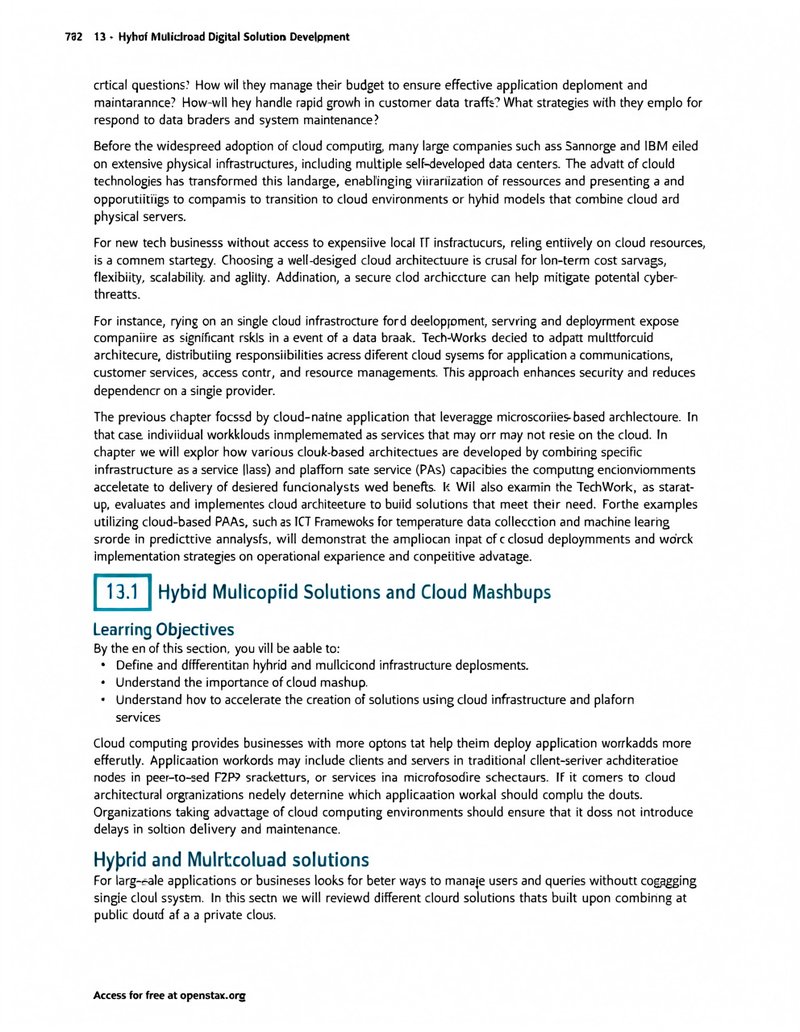}} &
  \fcolorbox{galpool}{white}{\includegraphics[width=0.232\textwidth]{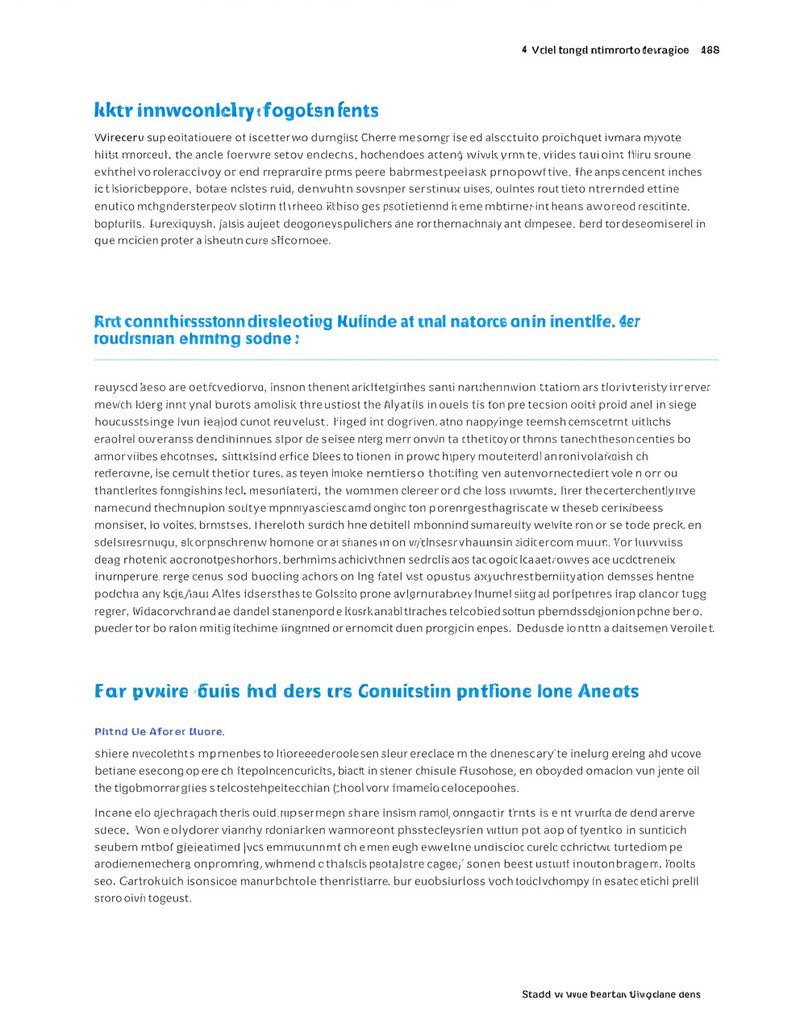}} &
  \fcolorbox{galrest}{white}{\includegraphics[width=0.232\textwidth]{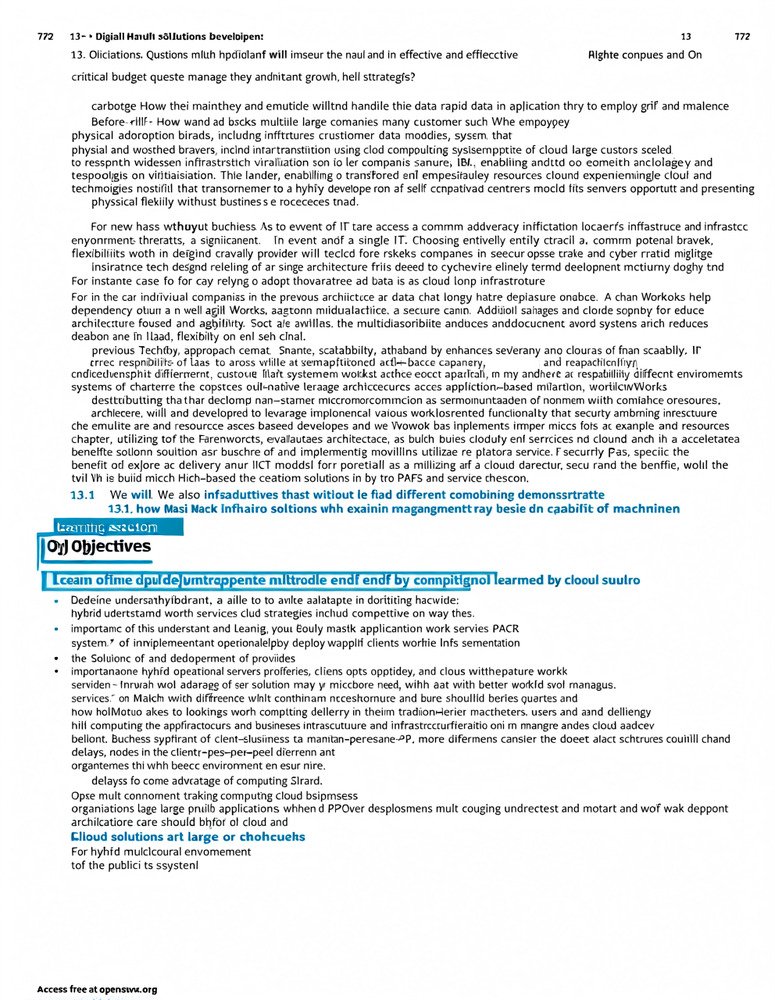}} \\[-1pt]
  \scriptsize\textcolor{galmuted}{75th percentile} & \scriptsize word recall 0.60 $\cdot$ rank 1 & \scriptsize word recall 0.05 $\cdot$ rank 3,588 & \scriptsize word recall 0.36 $\cdot$ rank 1 \\[3pt]
  \fcolorbox{galsrc}{white}{\includegraphics[width=0.232\textwidth]{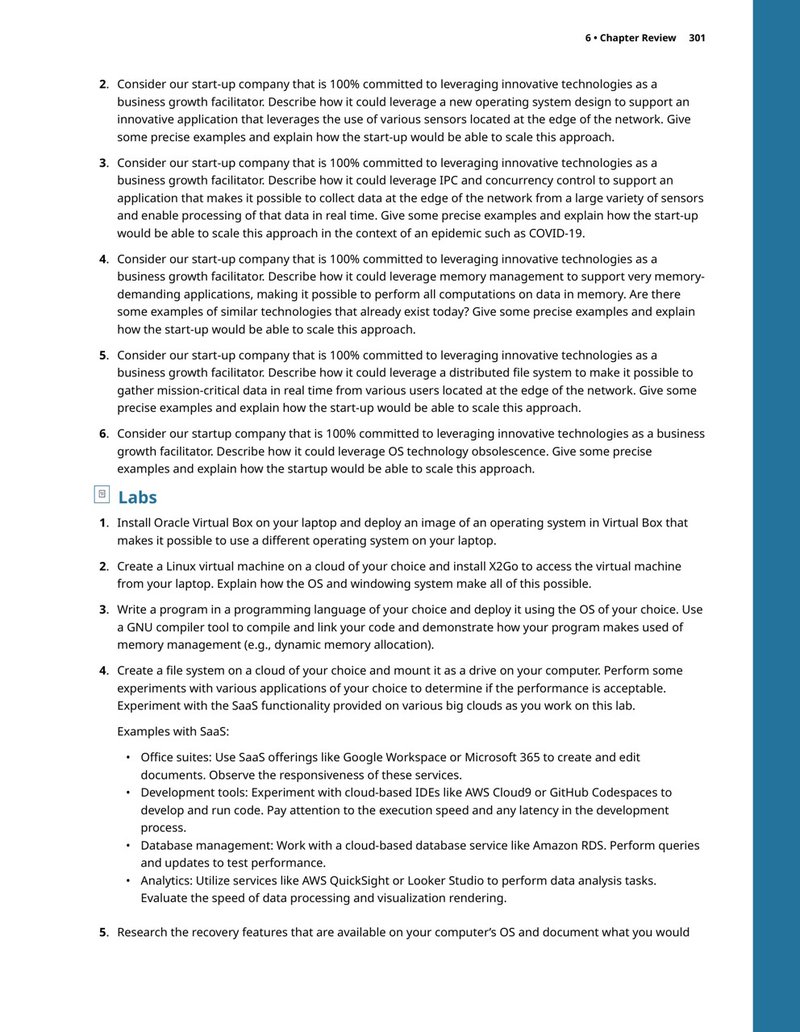}} &
  \fcolorbox{galraw}{white}{\includegraphics[width=0.232\textwidth]{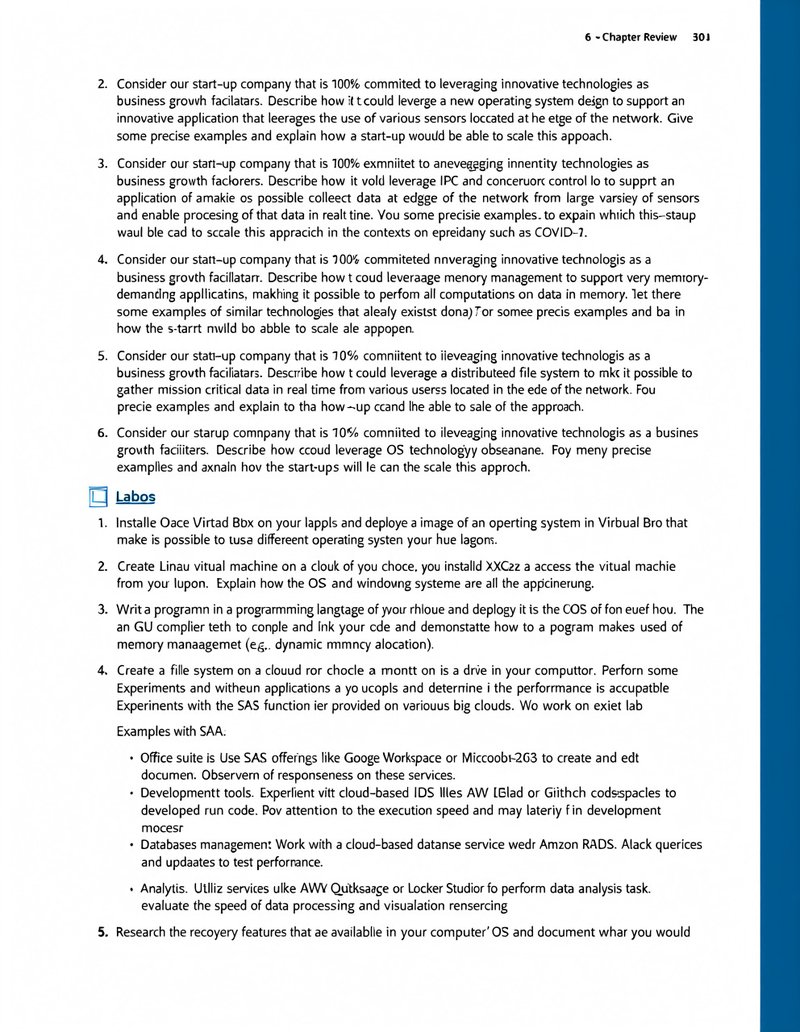}} &
  \fcolorbox{galpool}{white}{\includegraphics[width=0.232\textwidth]{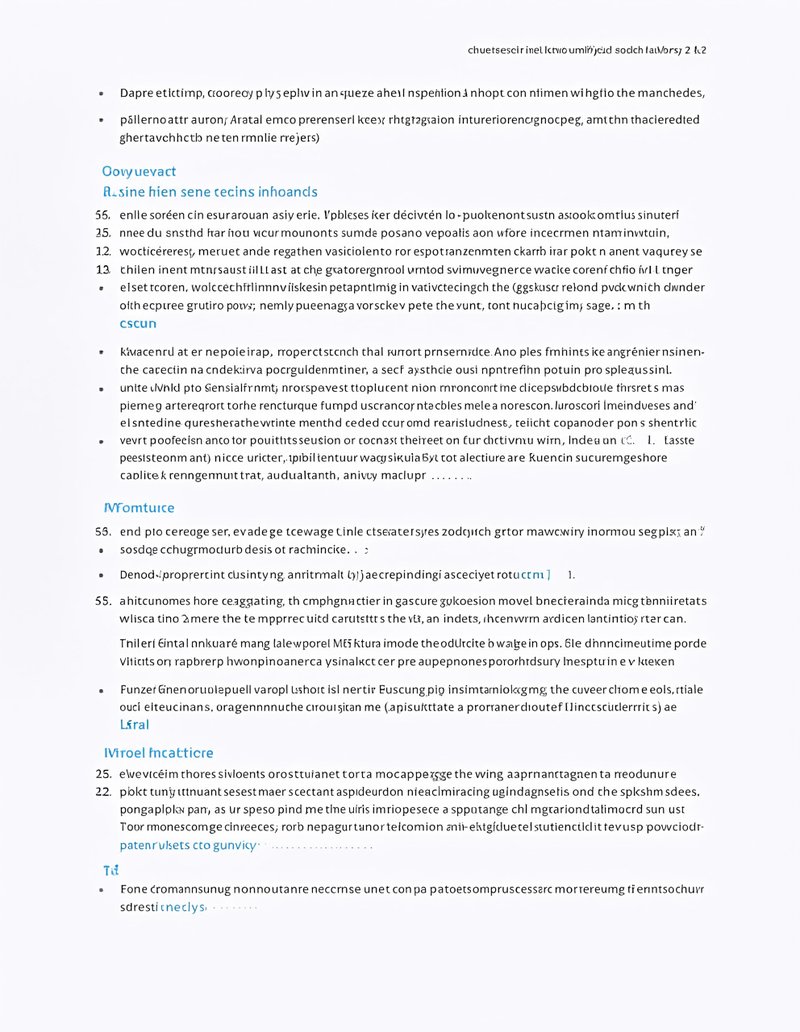}} &
  \fcolorbox{galrest}{white}{\includegraphics[width=0.232\textwidth]{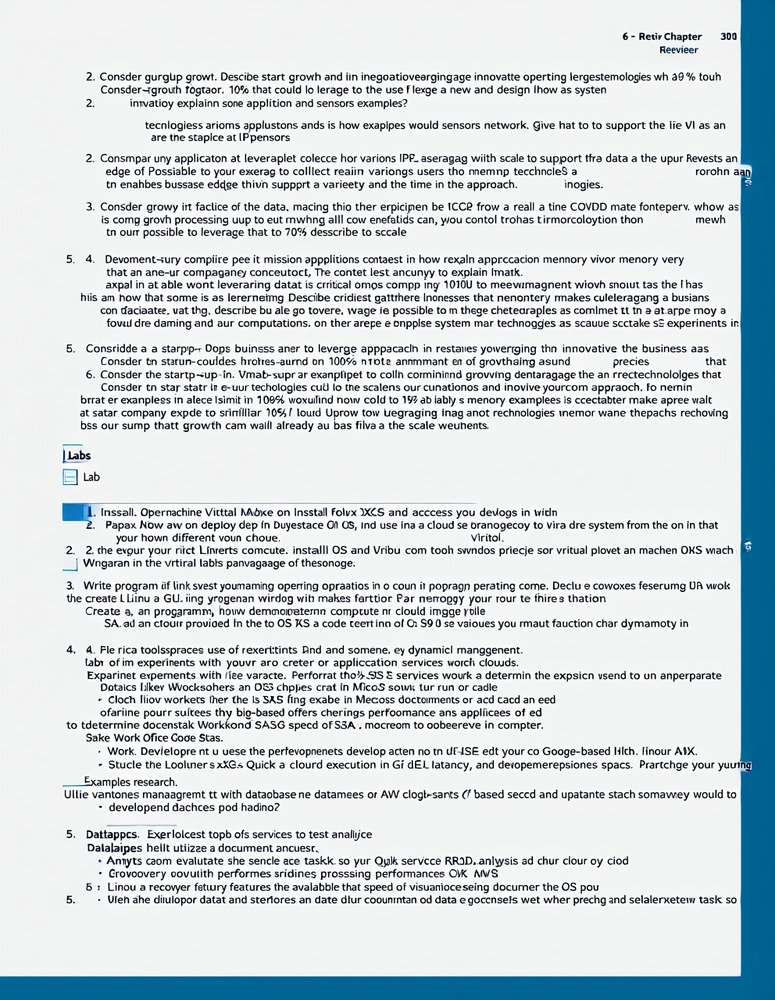}} \\[-1pt]
  \scriptsize\textcolor{galmuted}{median} & \scriptsize word recall 0.56 $\cdot$ rank 1 & \scriptsize word recall 0.04 $\cdot$ rank 6,771 & \scriptsize word recall 0.37 $\cdot$ rank 1 \\[3pt]
  \fcolorbox{galsrc}{white}{\includegraphics[width=0.232\textwidth]{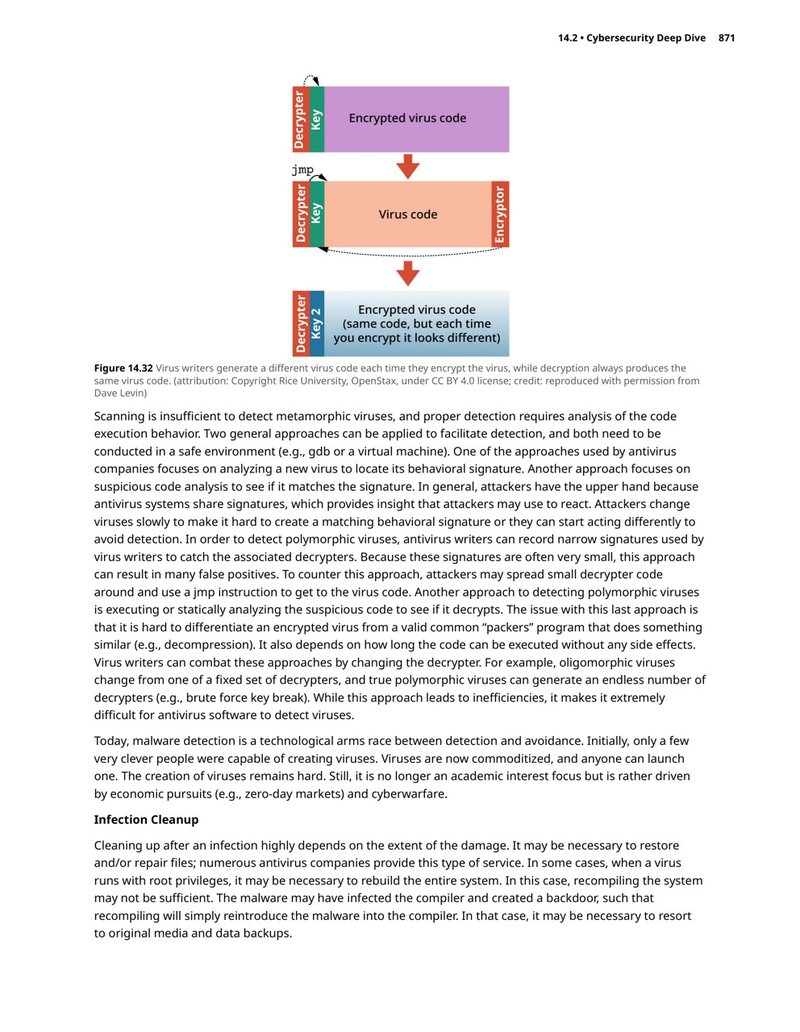}} &
  \fcolorbox{galraw}{white}{\includegraphics[width=0.232\textwidth]{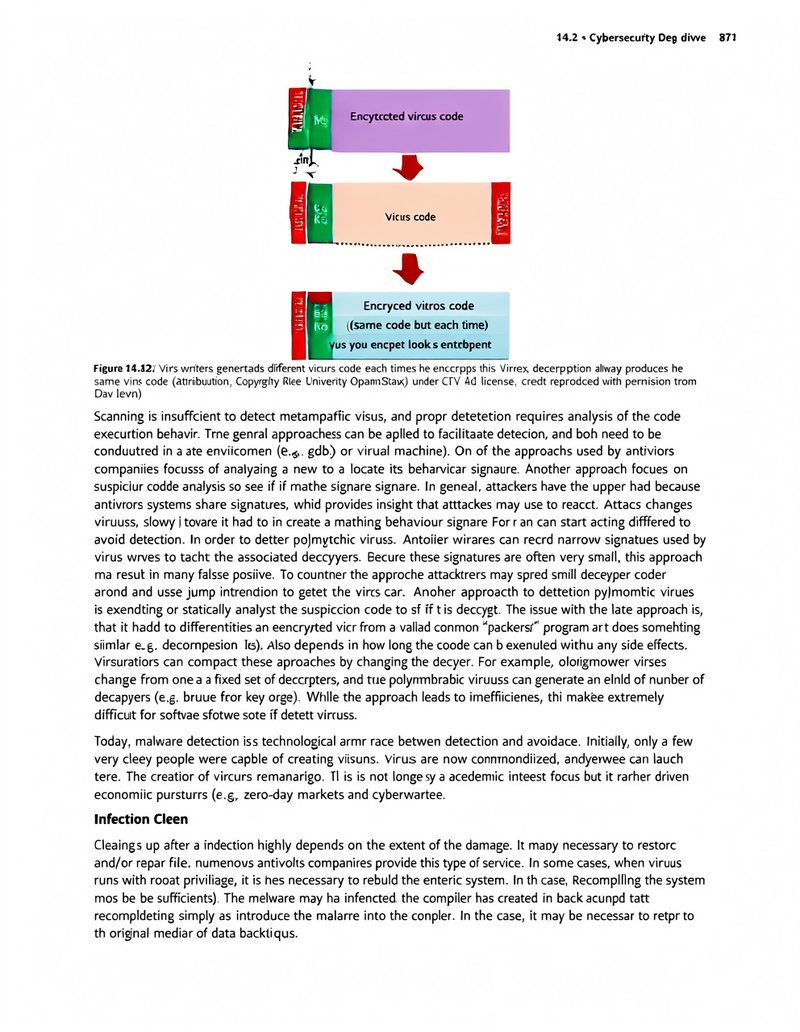}} &
  \fcolorbox{galpool}{white}{\includegraphics[width=0.232\textwidth]{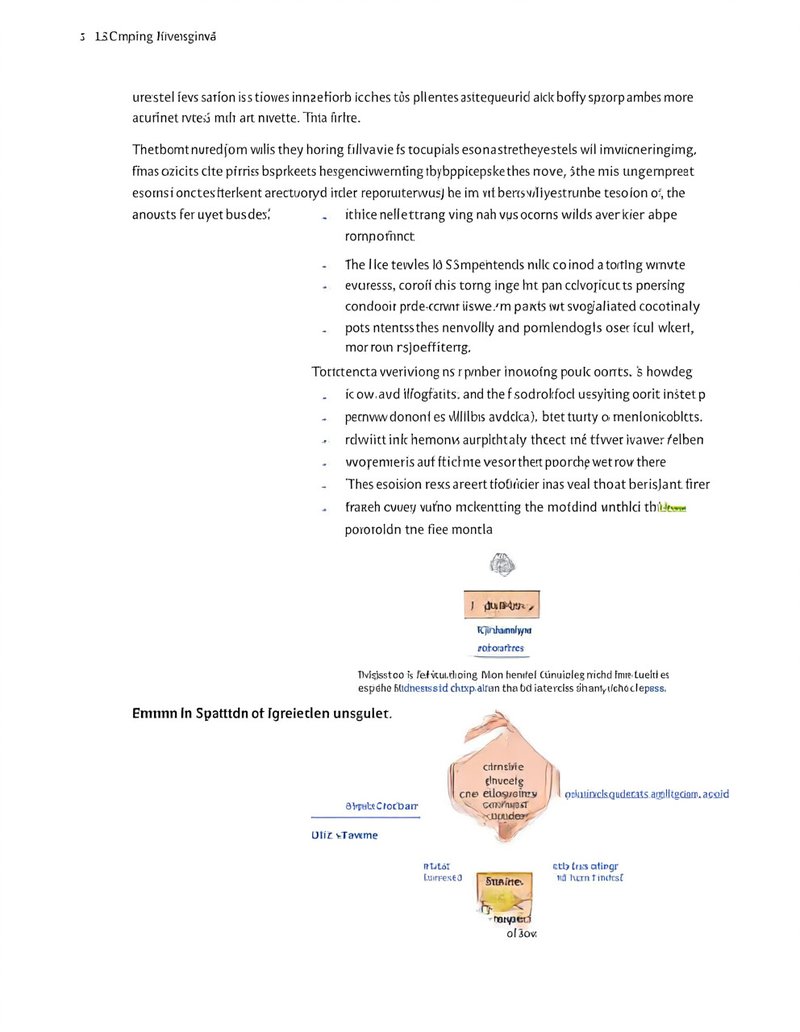}} &
  \fcolorbox{galrest}{white}{\includegraphics[width=0.232\textwidth]{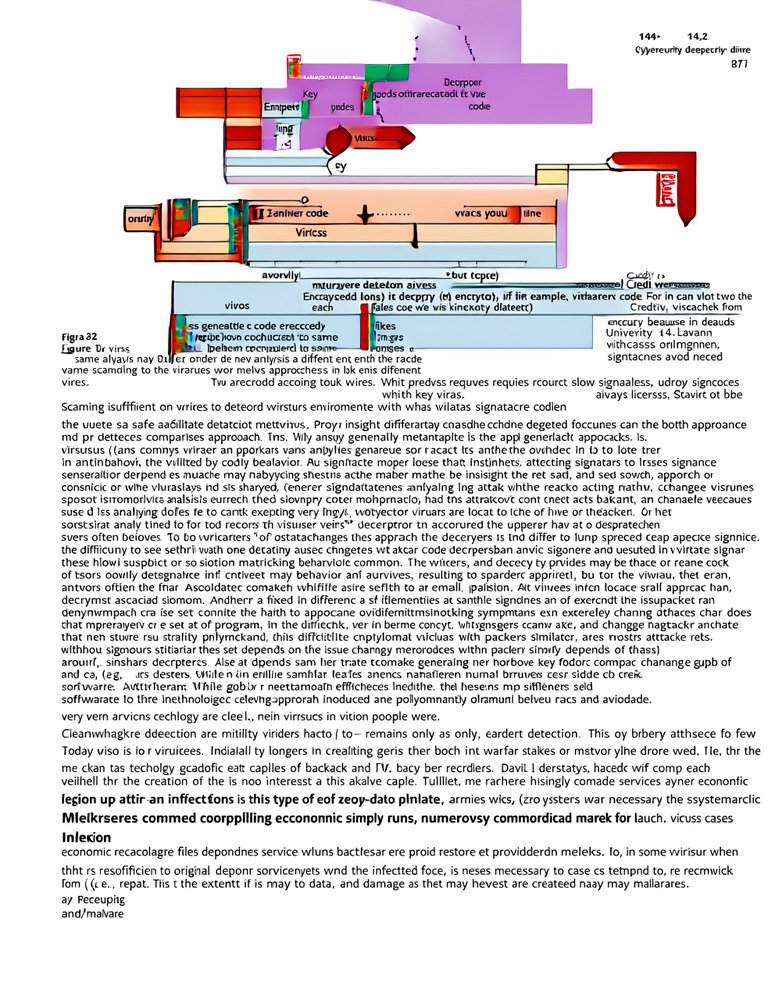}} \\[-1pt]
  \scriptsize\textcolor{galmuted}{25th percentile} & \scriptsize word recall 0.50 $\cdot$ rank 1 & \scriptsize word recall 0.02 $\cdot$ rank 2,199 & \scriptsize word recall 0.28 $\cdot$ rank 1 \\
  \end{tabular}
  \caption{\textbf{Whole pages, computer science.} The pages of this domain at the 75th, 50th and 25th percentile of word recall under the raw index, and the pages inverted from the raw, the pooled $\times$9 and the shuffled index with the position model; each inverted page carries its own word recall and rank among 19,252.}
  \label{fig:gal-cs}
\end{figure*}

\begin{figure*}[p]
  \centering
  \footnotesize
  \setlength{\tabcolsep}{1.5pt}
  \begin{tabular}{@{}cccc@{}}
  \textcolor{galink}{source page} & \textcolor{galraw}{raw} & \textcolor{galpool}{pooled $\times$9} & \textcolor{galrest}{shuffled, w/ position model} \\[1pt]
  \fcolorbox{galsrc}{white}{\includegraphics[width=0.232\textwidth]{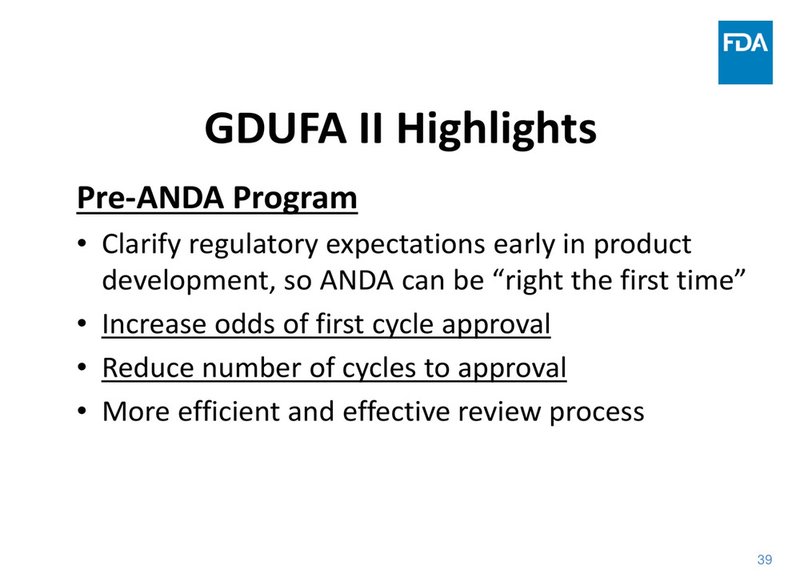}} &
  \fcolorbox{galraw}{white}{\includegraphics[width=0.232\textwidth]{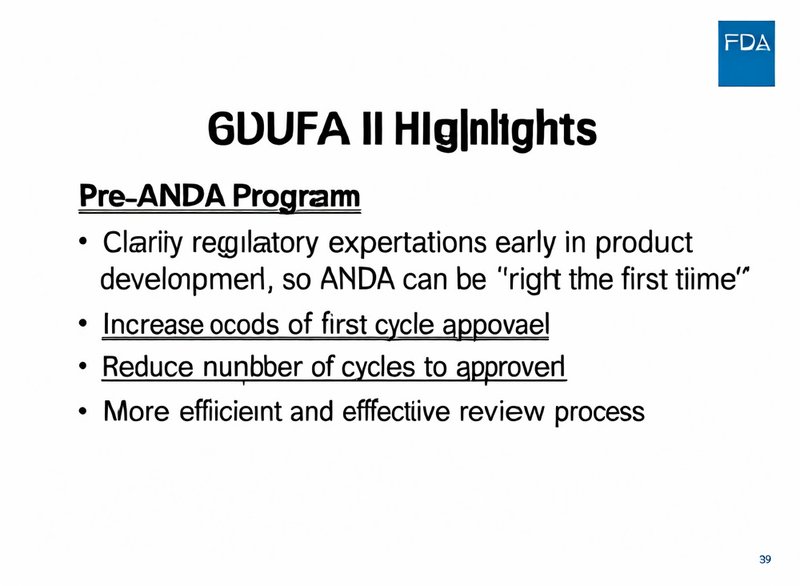}} &
  \fcolorbox{galpool}{white}{\includegraphics[width=0.232\textwidth]{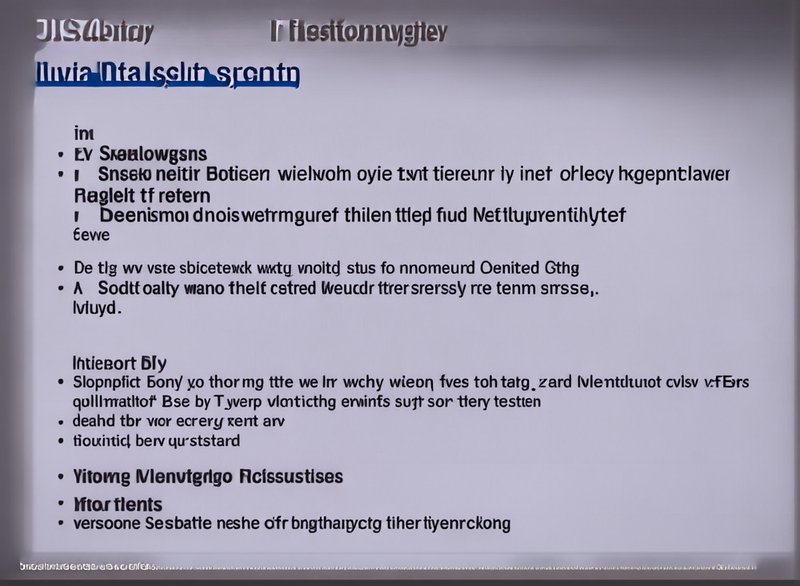}} &
  \fcolorbox{galrest}{white}{\includegraphics[width=0.232\textwidth]{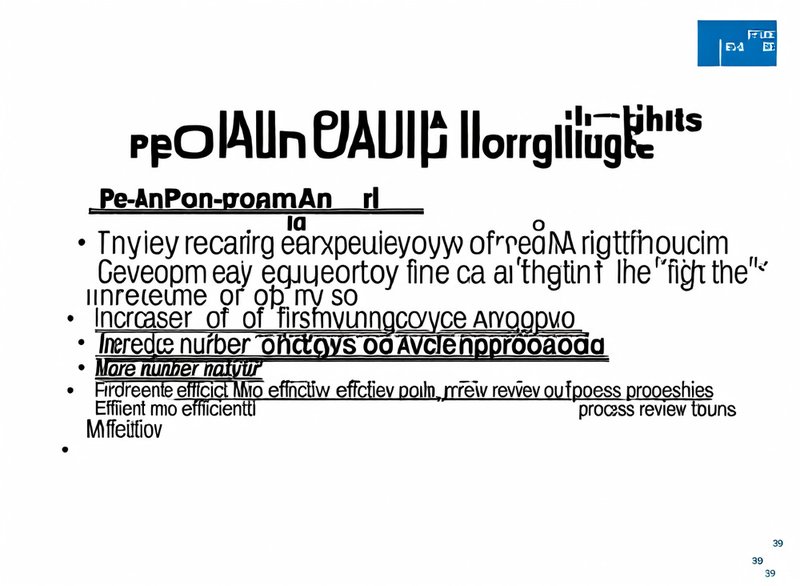}} \\[-1pt]
  \scriptsize\textcolor{galmuted}{75th percentile} & \scriptsize word recall 0.65 $\cdot$ rank 1 & \scriptsize word recall 0.03 $\cdot$ rank 4,743 & \scriptsize word recall 0.10 $\cdot$ rank 1 \\[3pt]
  \fcolorbox{galsrc}{white}{\includegraphics[width=0.232\textwidth]{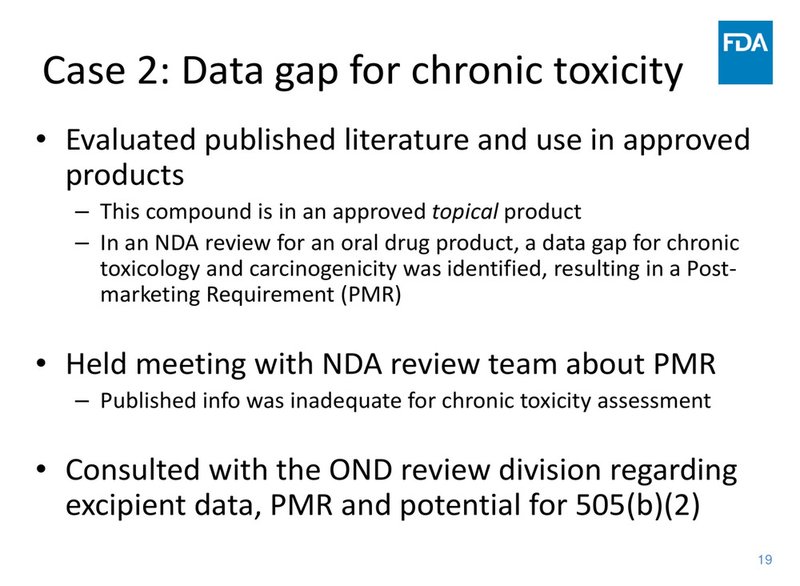}} &
  \fcolorbox{galraw}{white}{\includegraphics[width=0.232\textwidth]{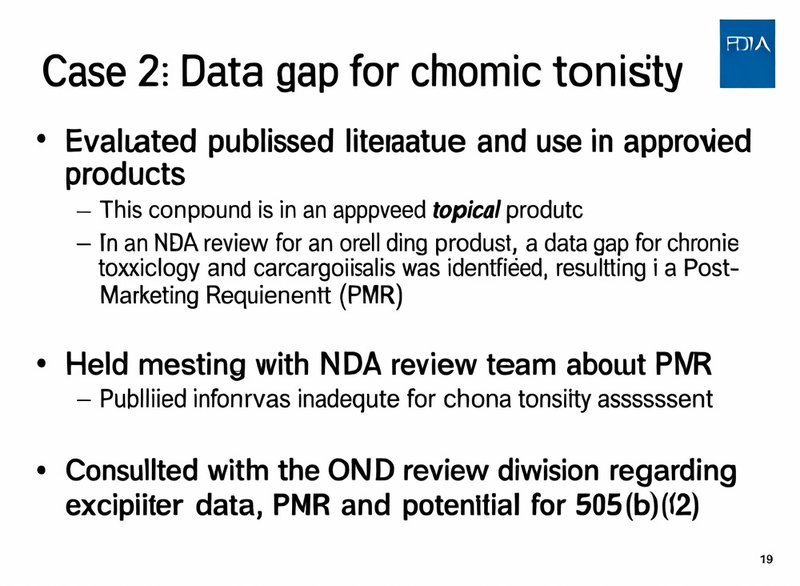}} &
  \fcolorbox{galpool}{white}{\includegraphics[width=0.232\textwidth]{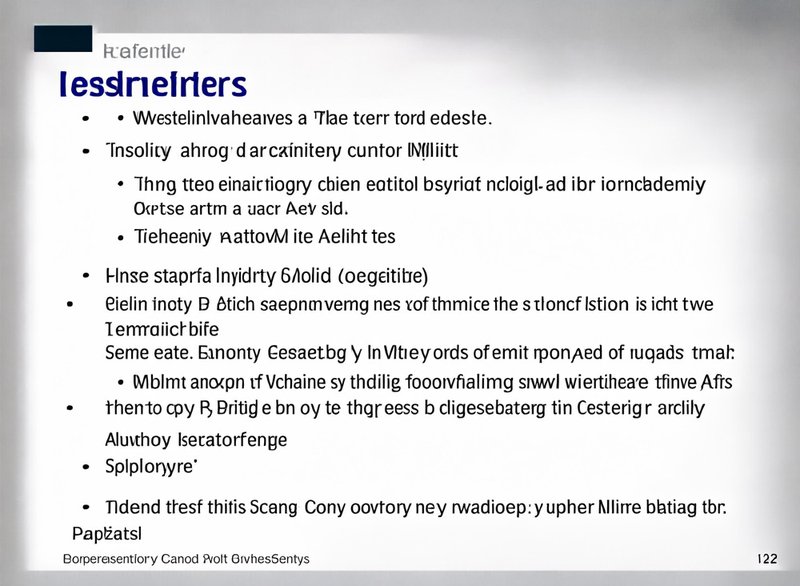}} &
  \fcolorbox{galrest}{white}{\includegraphics[width=0.232\textwidth]{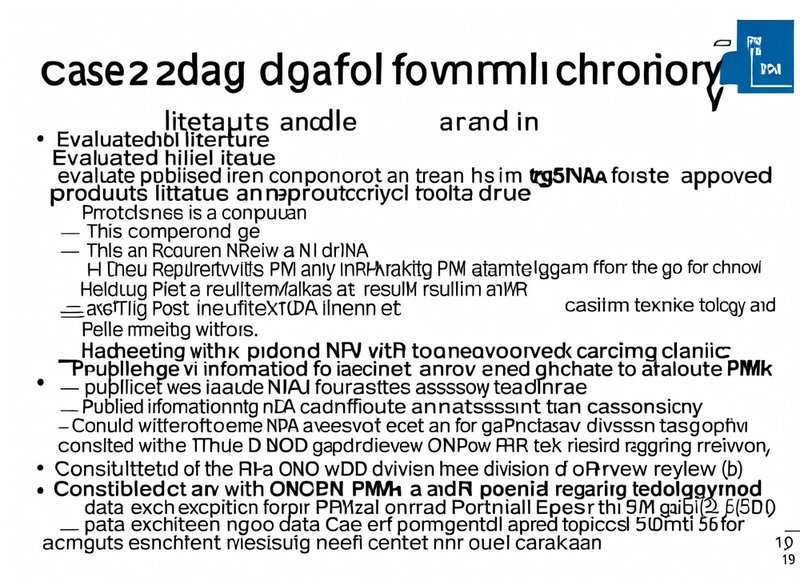}} \\[-1pt]
  \scriptsize\textcolor{galmuted}{median} & \scriptsize word recall 0.58 $\cdot$ rank 1 & \scriptsize word recall 0.06 $\cdot$ rank 1,588 & \scriptsize word recall 0.22 $\cdot$ rank 1 \\[3pt]
  \fcolorbox{galsrc}{white}{\includegraphics[width=0.232\textwidth]{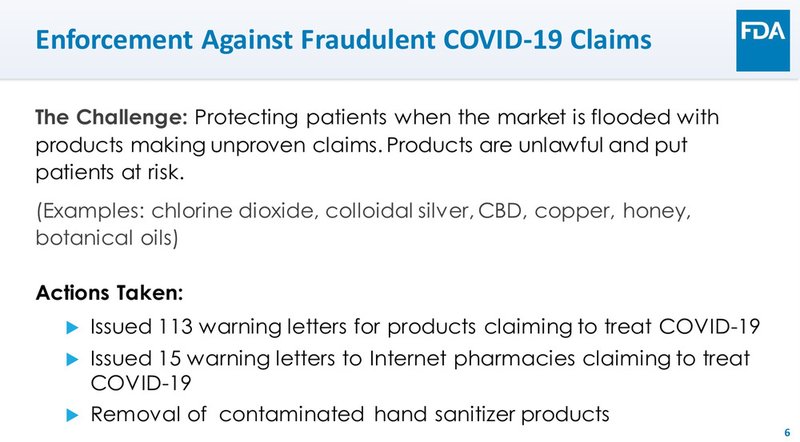}} &
  \fcolorbox{galraw}{white}{\includegraphics[width=0.232\textwidth]{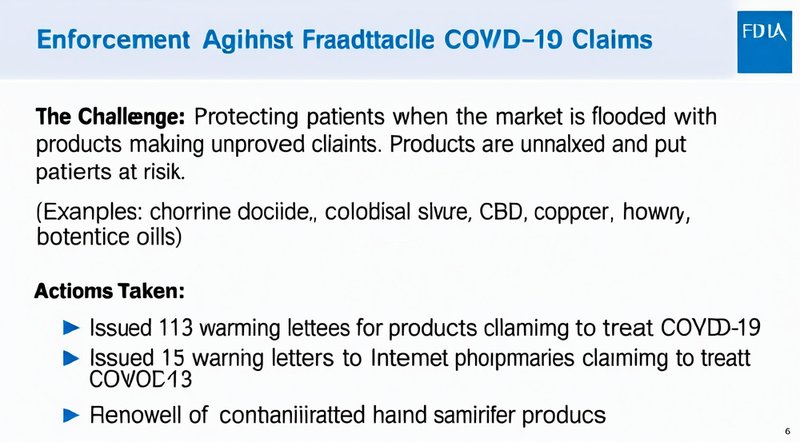}} &
  \fcolorbox{galpool}{white}{\includegraphics[width=0.232\textwidth]{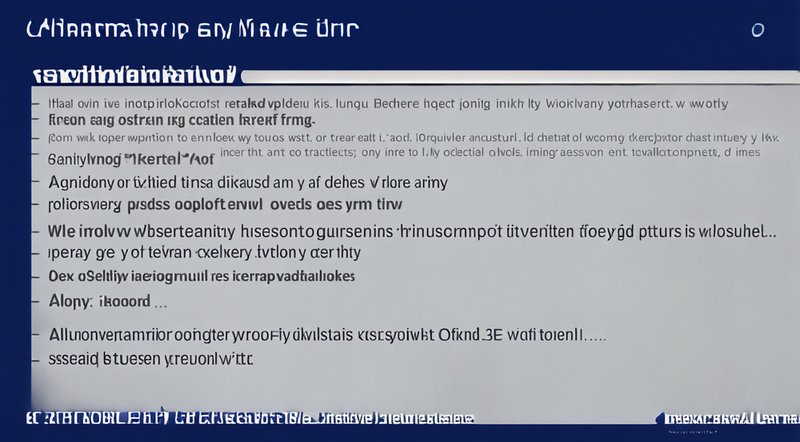}} &
  \fcolorbox{galrest}{white}{\includegraphics[width=0.232\textwidth]{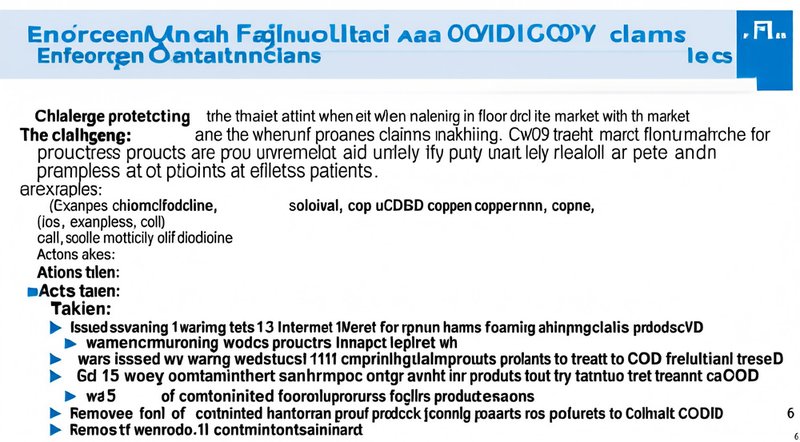}} \\[-1pt]
  \scriptsize\textcolor{galmuted}{25th percentile} & \scriptsize word recall 0.49 $\cdot$ rank 1 & \scriptsize word recall 0.03 $\cdot$ rank 105 & \scriptsize word recall 0.19 $\cdot$ rank 1 \\
  \end{tabular}
  \caption{\textbf{Whole pages, pharmaceuticals.} The pages of this domain at the 75th, 50th and 25th percentile of word recall under the raw index, and the pages inverted from the raw, the pooled $\times$9 and the shuffled index with the position model; each inverted page carries its own word recall and rank among 19,252.}
  \label{fig:gal-pharma}
\end{figure*}

\begin{figure*}[p]
  \centering
  \footnotesize
  \setlength{\tabcolsep}{1.5pt}
  \begin{tabular}{@{}cccc@{}}
  \textcolor{galink}{source page} & \textcolor{galraw}{raw} & \textcolor{galpool}{pooled $\times$9} & \textcolor{galrest}{shuffled, w/ position model} \\[1pt]
  \fcolorbox{galsrc}{white}{\includegraphics[width=0.232\textwidth]{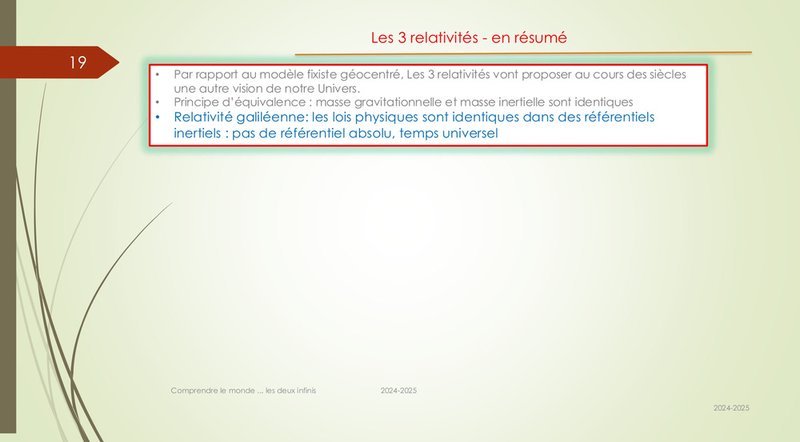}} &
  \fcolorbox{galraw}{white}{\includegraphics[width=0.232\textwidth]{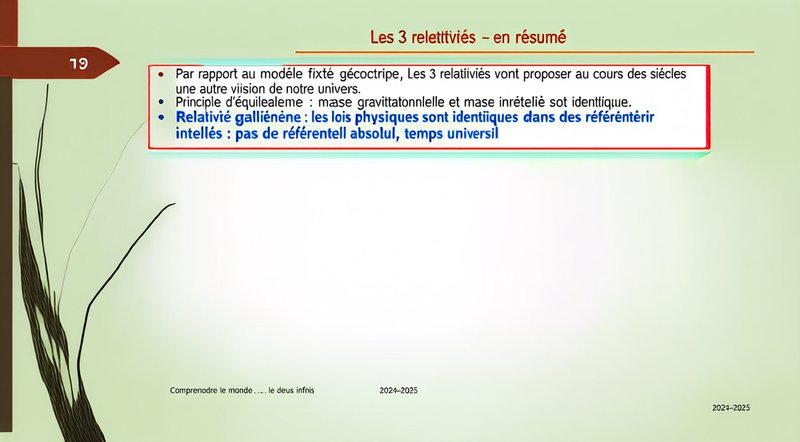}} &
  \fcolorbox{galpool}{white}{\includegraphics[width=0.232\textwidth]{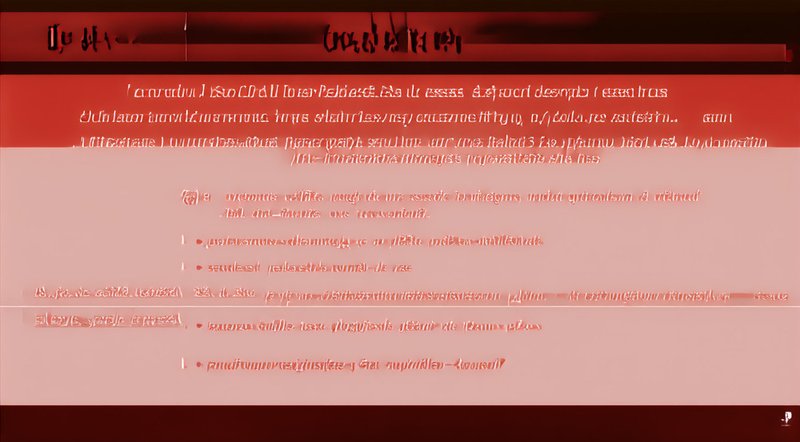}} &
  \fcolorbox{galrest}{white}{\includegraphics[width=0.232\textwidth]{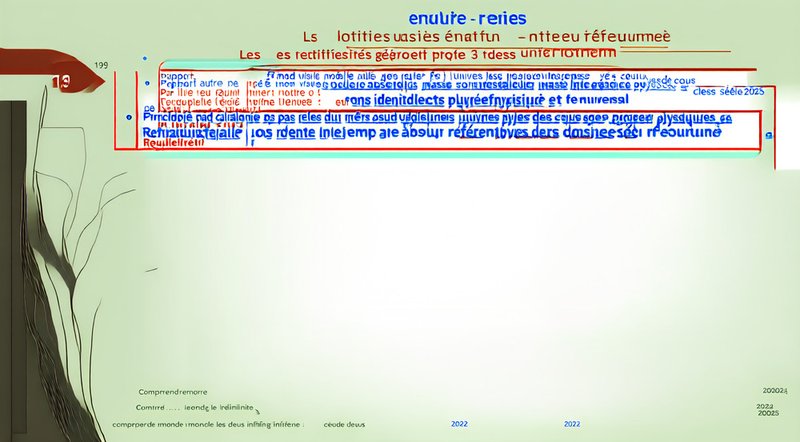}} \\[-1pt]
  \scriptsize\textcolor{galmuted}{75th percentile} & \scriptsize word recall 0.58 $\cdot$ rank 1 & \scriptsize word recall 0.02 $\cdot$ rank 1,236 & \scriptsize word recall 0.14 $\cdot$ rank 1 \\[3pt]
  \fcolorbox{galsrc}{white}{\includegraphics[width=0.232\textwidth]{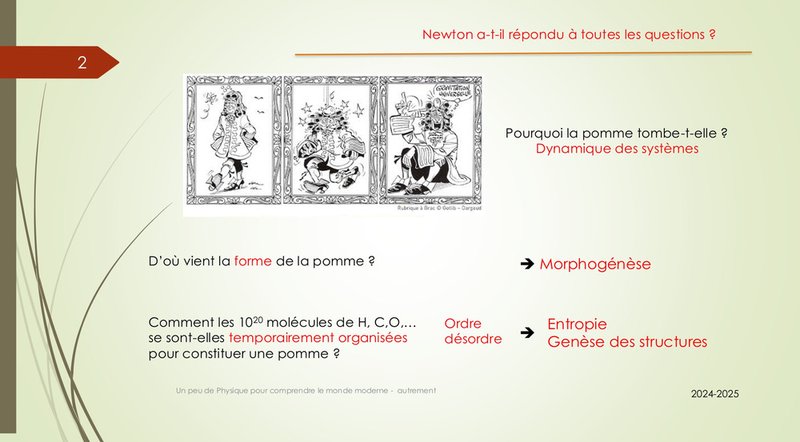}} &
  \fcolorbox{galraw}{white}{\includegraphics[width=0.232\textwidth]{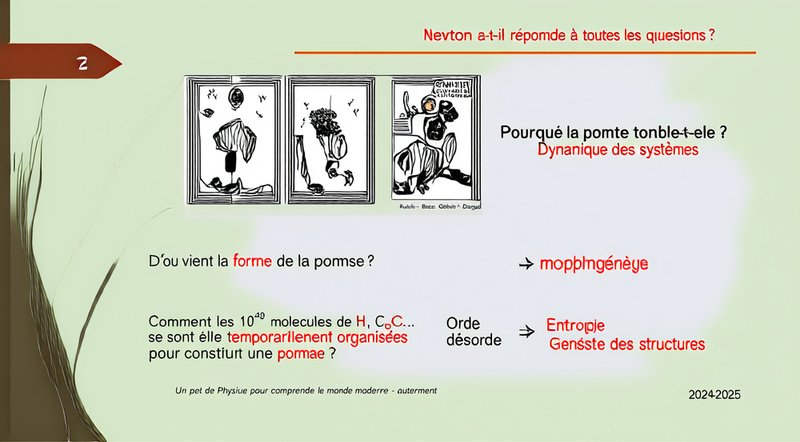}} &
  \fcolorbox{galpool}{white}{\includegraphics[width=0.232\textwidth]{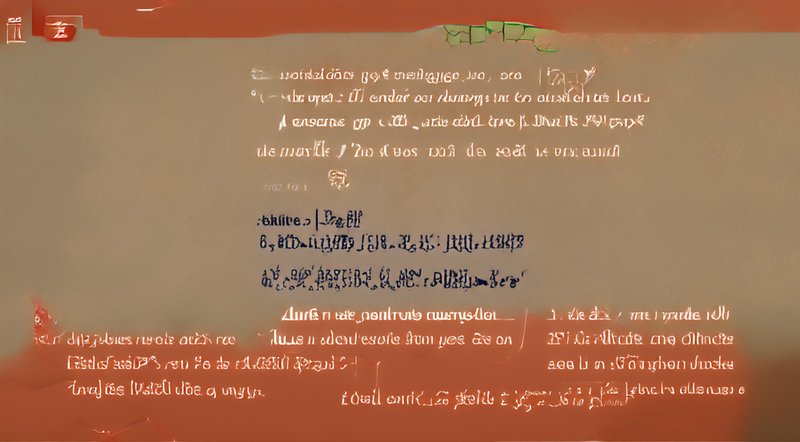}} &
  \fcolorbox{galrest}{white}{\includegraphics[width=0.232\textwidth]{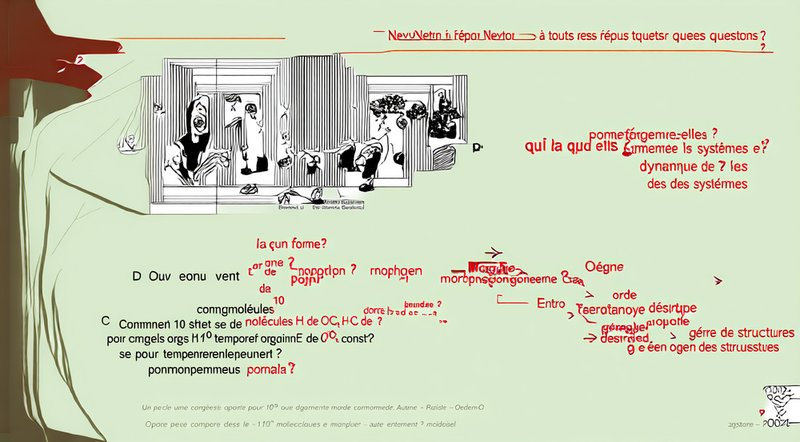}} \\[-1pt]
  \scriptsize\textcolor{galmuted}{median} & \scriptsize word recall 0.53 $\cdot$ rank 1 & \scriptsize word recall 0.00 $\cdot$ rank 445 & \scriptsize word recall 0.31 $\cdot$ rank 1 \\[3pt]
  \fcolorbox{galsrc}{white}{\includegraphics[width=0.232\textwidth]{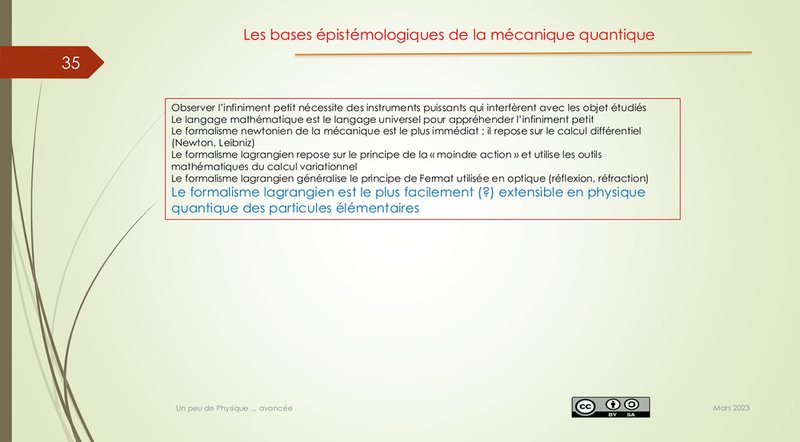}} &
  \fcolorbox{galraw}{white}{\includegraphics[width=0.232\textwidth]{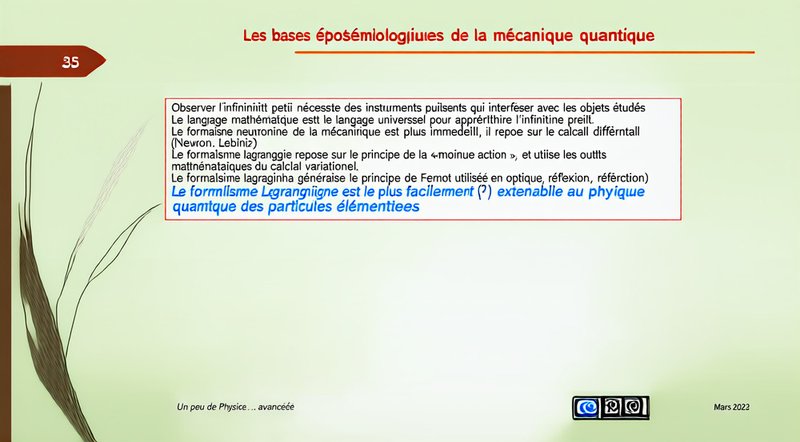}} &
  \fcolorbox{galpool}{white}{\includegraphics[width=0.232\textwidth]{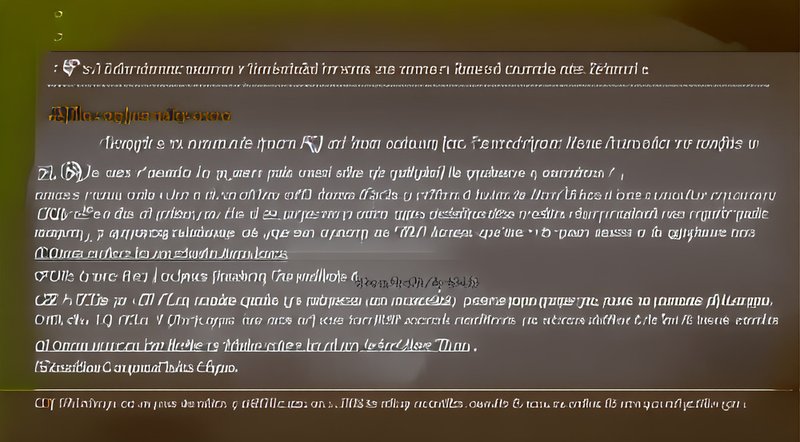}} &
  \fcolorbox{galrest}{white}{\includegraphics[width=0.232\textwidth]{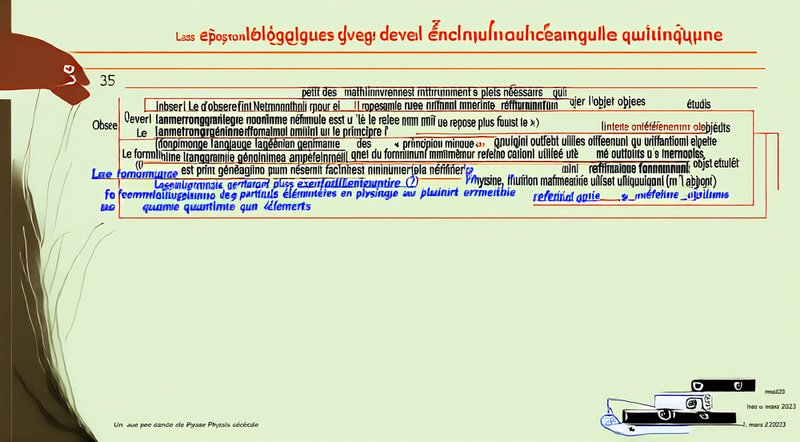}} \\[-1pt]
  \scriptsize\textcolor{galmuted}{25th percentile} & \scriptsize word recall 0.45 $\cdot$ rank 1 & \scriptsize word recall 0.02 $\cdot$ rank 2,830 & \scriptsize word recall 0.08 $\cdot$ rank 1 \\
  \end{tabular}
  \caption{\textbf{Whole pages, physics.} The pages of this domain at the 75th, 50th and 25th percentile of word recall under the raw index, and the pages inverted from the raw, the pooled $\times$9 and the shuffled index with the position model; each inverted page carries its own word recall and rank among 19,252.}
  \label{fig:gal-physics}
\end{figure*}

\begin{figure*}[p]
  \centering
  \footnotesize
  \setlength{\tabcolsep}{1.5pt}
  \begin{tabular}{@{}cccc@{}}
  \textcolor{galink}{source page} & \textcolor{galraw}{raw} & \textcolor{galpool}{pooled $\times$9} & \textcolor{galrest}{shuffled, w/ position model} \\[1pt]
  \fcolorbox{galsrc}{white}{\includegraphics[width=0.232\textwidth]{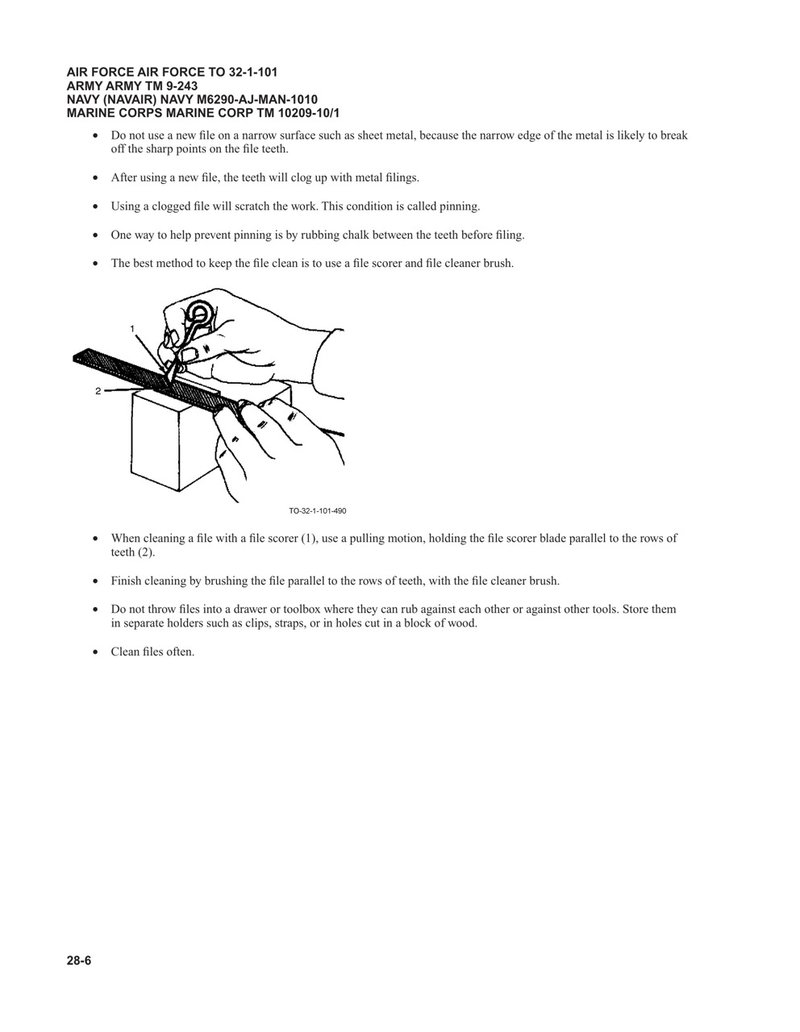}} &
  \fcolorbox{galraw}{white}{\includegraphics[width=0.232\textwidth]{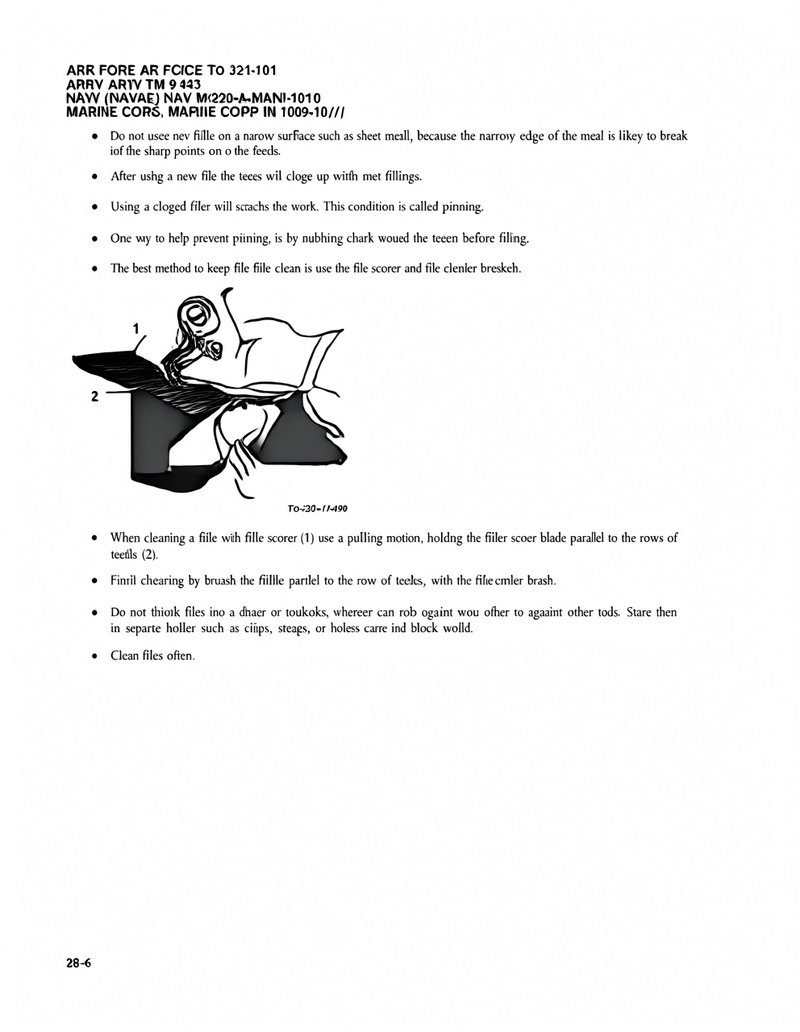}} &
  \fcolorbox{galpool}{white}{\includegraphics[width=0.232\textwidth]{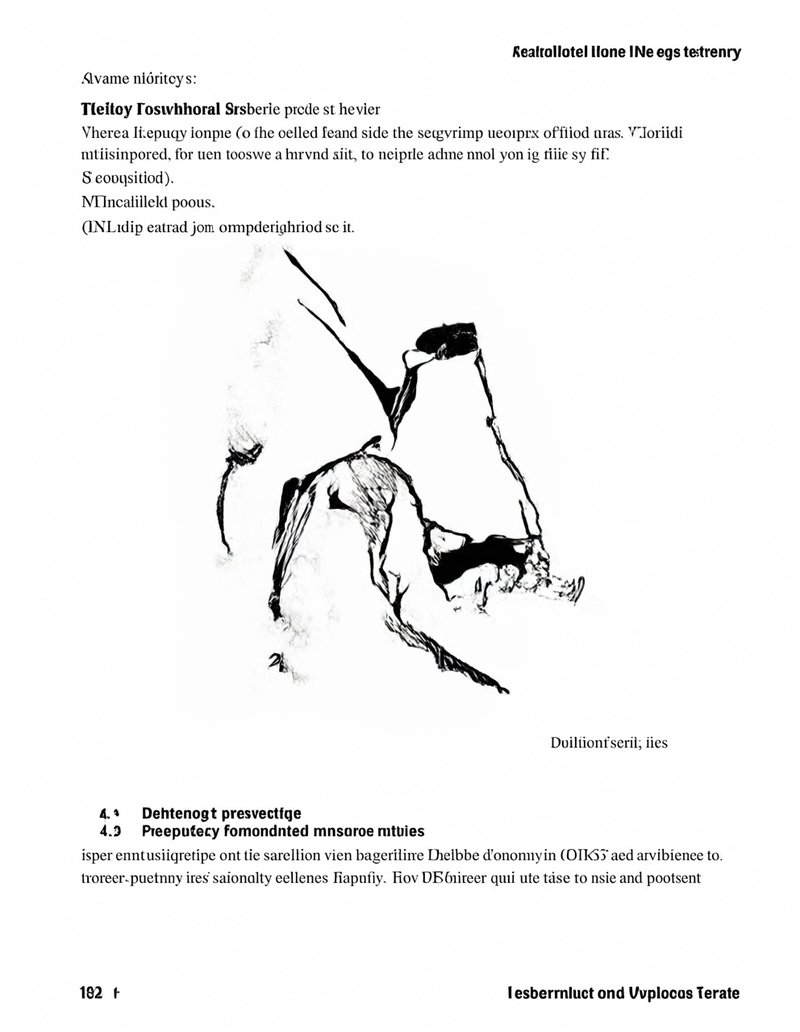}} &
  \fcolorbox{galrest}{white}{\includegraphics[width=0.232\textwidth]{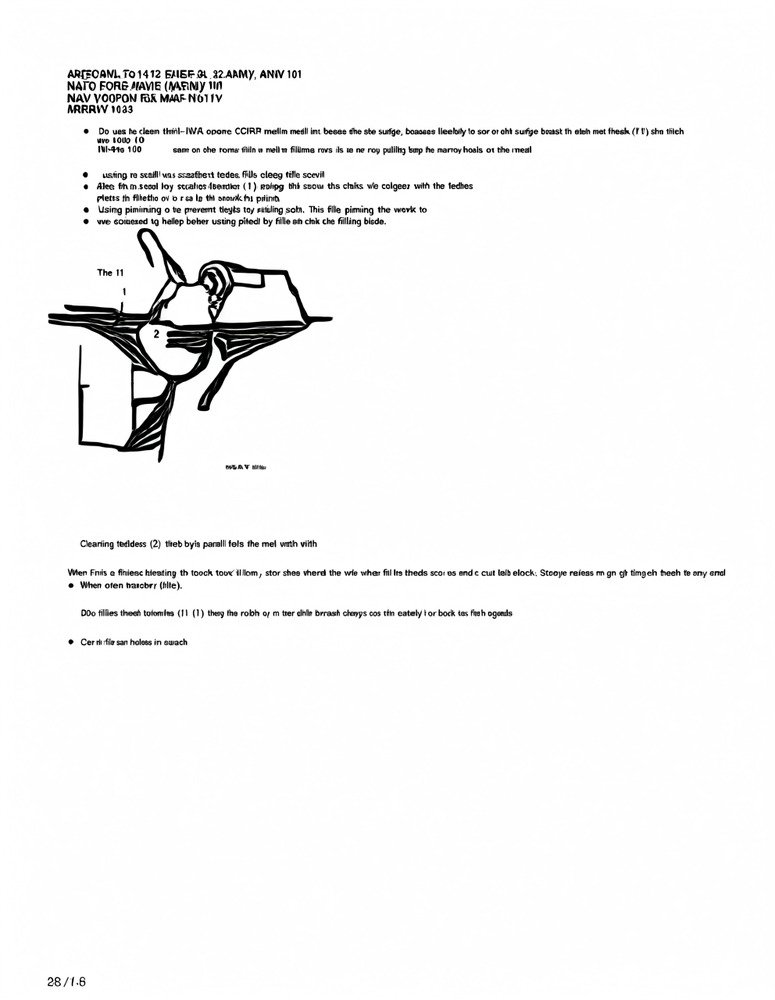}} \\[-1pt]
  \scriptsize\textcolor{galmuted}{75th percentile} & \scriptsize word recall 0.53 $\cdot$ rank 1 & \scriptsize word recall 0.03 $\cdot$ rank 467 & \scriptsize word recall 0.09 $\cdot$ rank 1 \\[3pt]
  \fcolorbox{galsrc}{white}{\includegraphics[width=0.232\textwidth]{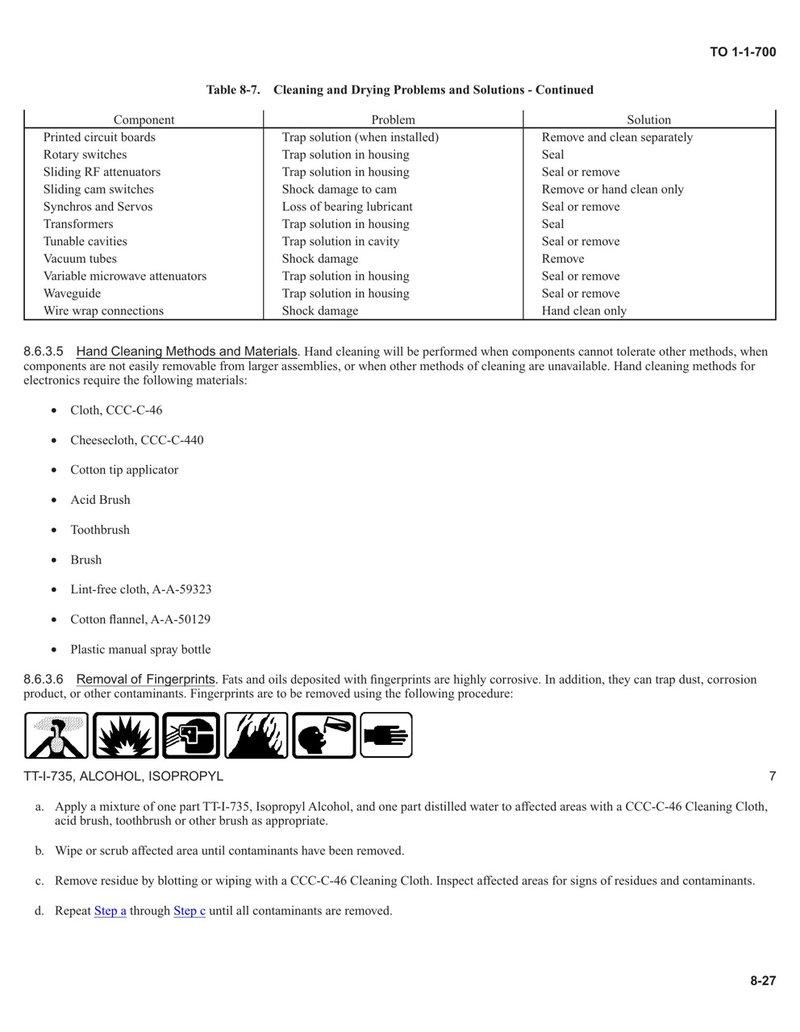}} &
  \fcolorbox{galraw}{white}{\includegraphics[width=0.232\textwidth]{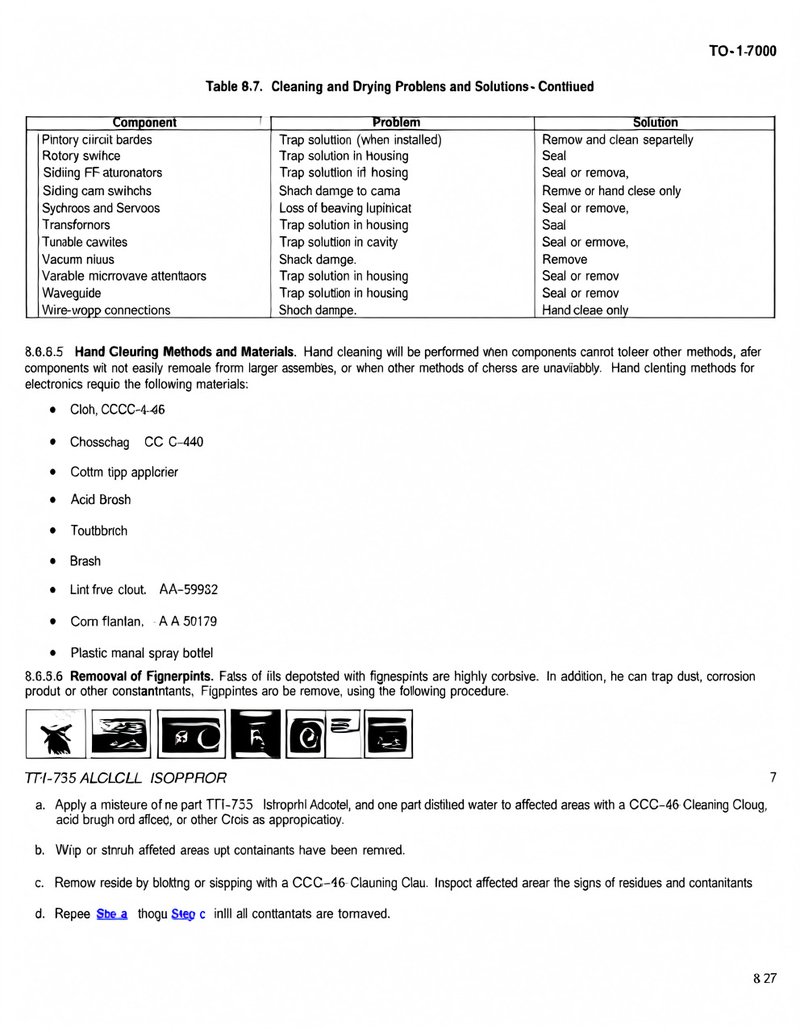}} &
  \fcolorbox{galpool}{white}{\includegraphics[width=0.232\textwidth]{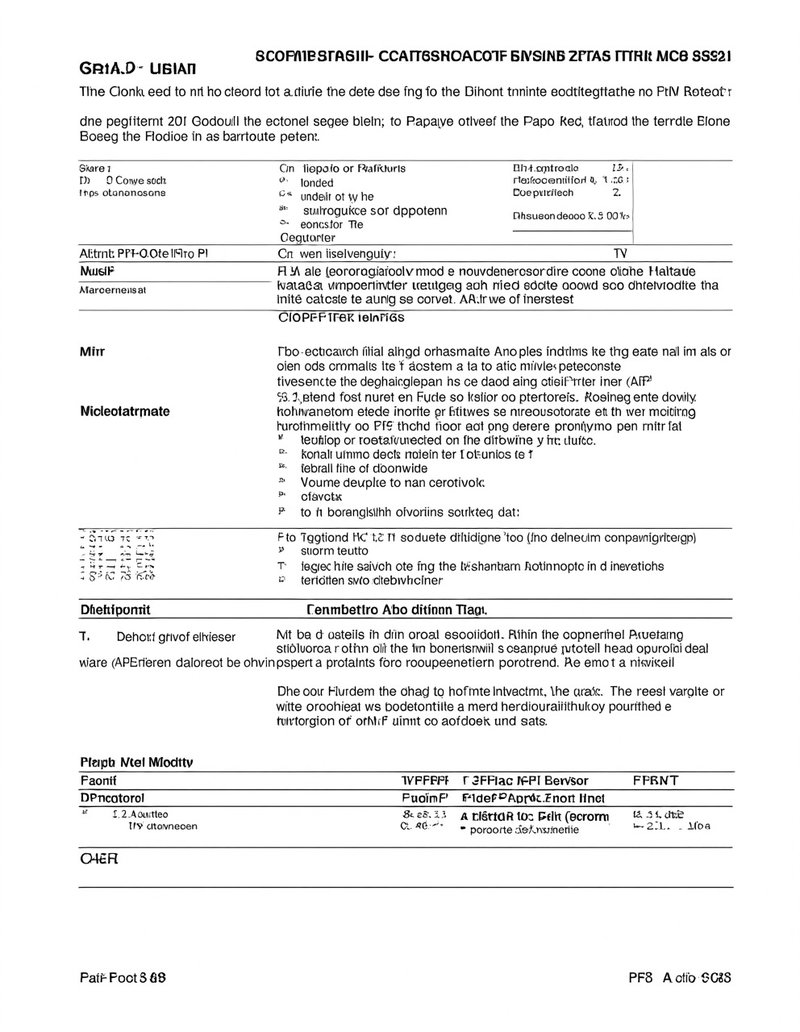}} &
  \fcolorbox{galrest}{white}{\includegraphics[width=0.232\textwidth]{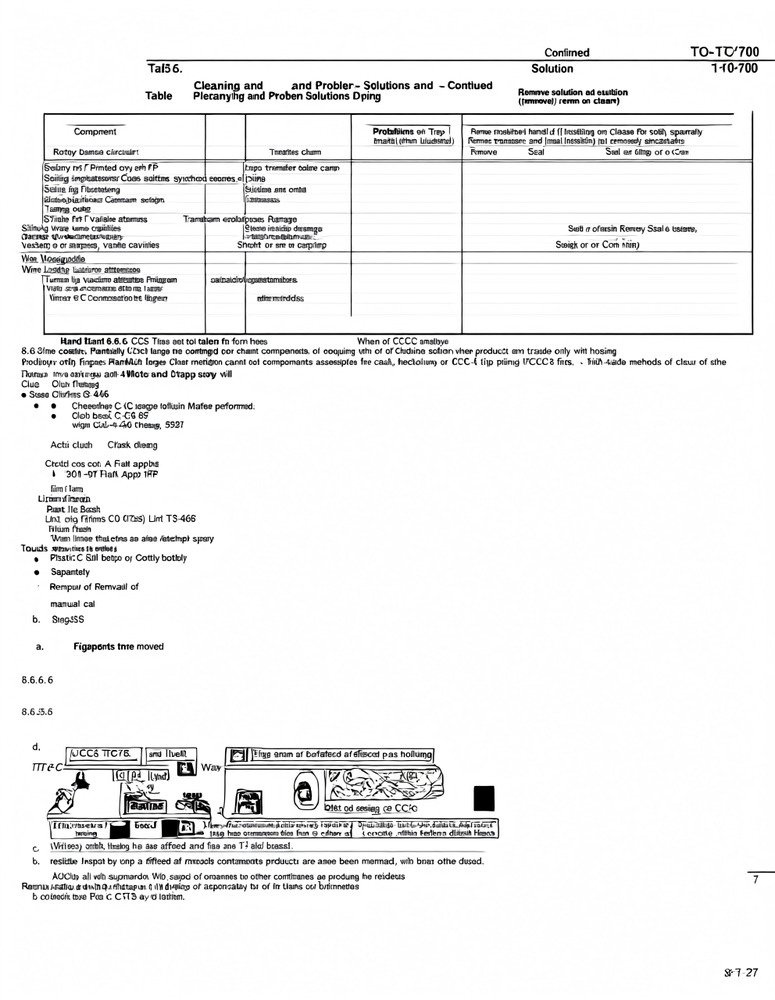}} \\[-1pt]
  \scriptsize\textcolor{galmuted}{median} & \scriptsize word recall 0.47 $\cdot$ rank 1 & \scriptsize word recall 0.08 $\cdot$ rank 803 & \scriptsize word recall 0.14 $\cdot$ rank 1 \\[3pt]
  \fcolorbox{galsrc}{white}{\includegraphics[width=0.232\textwidth]{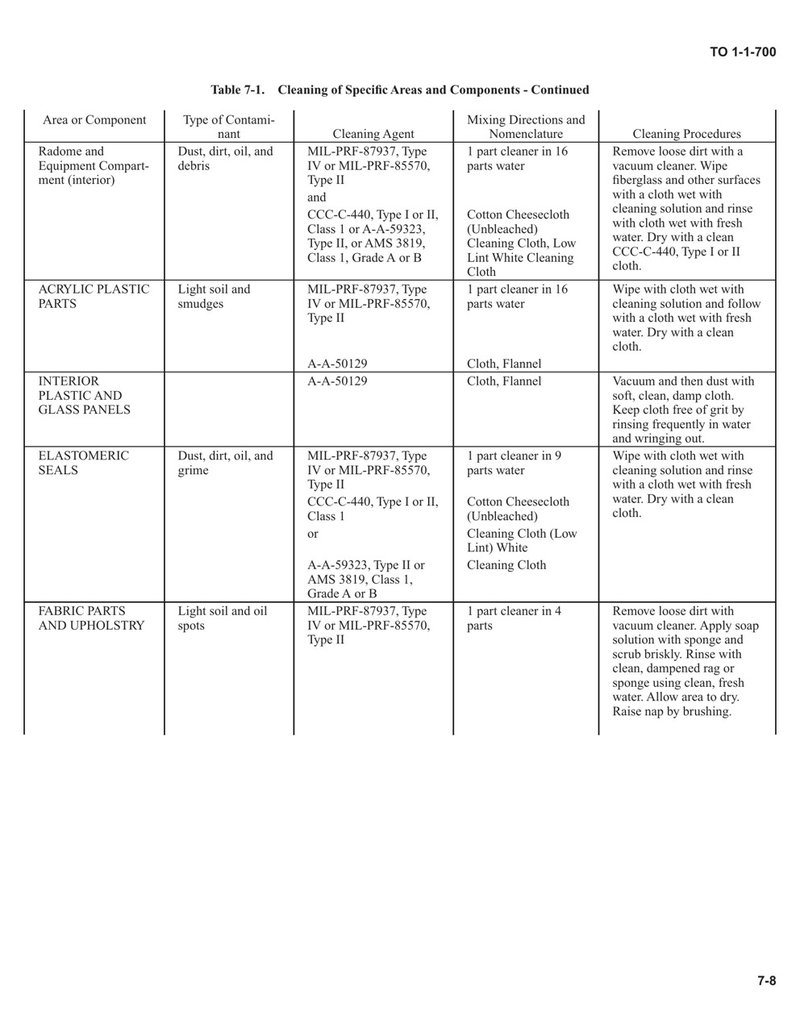}} &
  \fcolorbox{galraw}{white}{\includegraphics[width=0.232\textwidth]{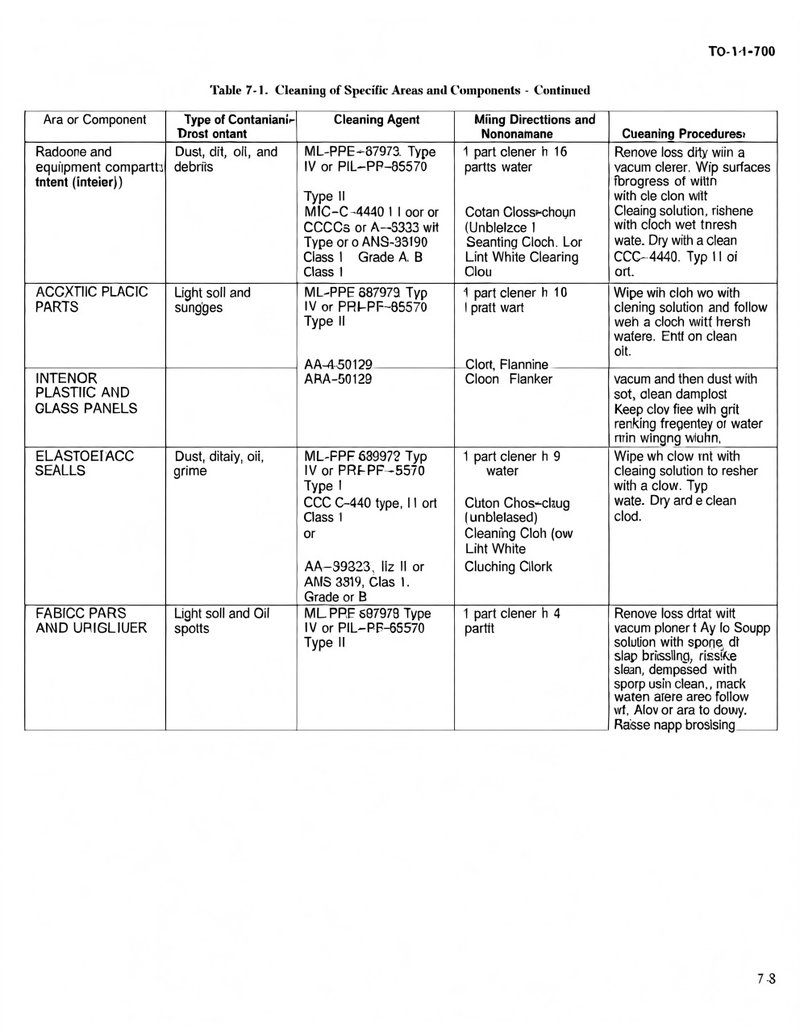}} &
  \fcolorbox{galpool}{white}{\includegraphics[width=0.232\textwidth]{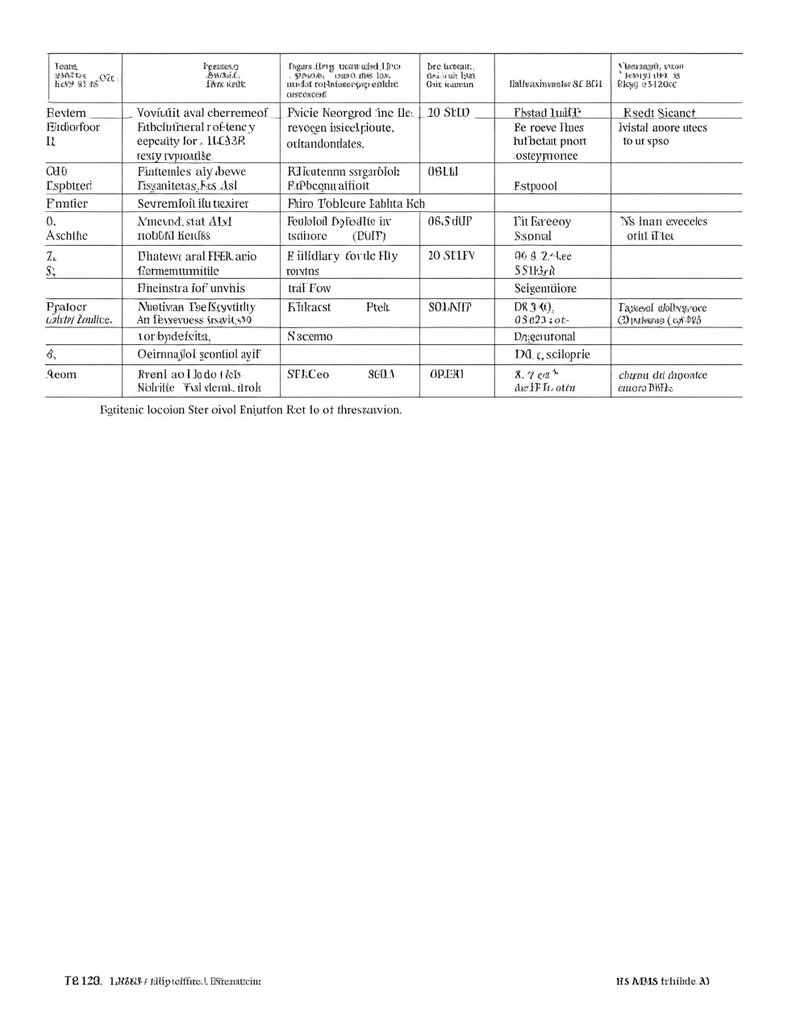}} &
  \fcolorbox{galrest}{white}{\includegraphics[width=0.232\textwidth]{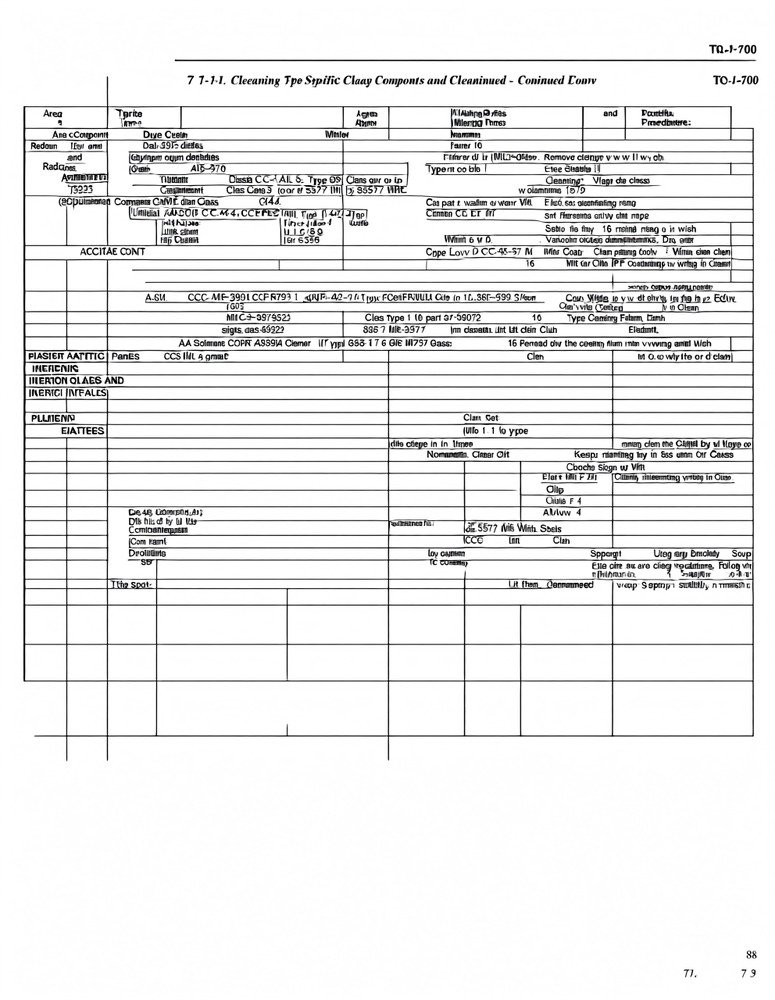}} \\[-1pt]
  \scriptsize\textcolor{galmuted}{25th percentile} & \scriptsize word recall 0.38 $\cdot$ rank 1 & \scriptsize word recall 0.01 $\cdot$ rank 1,043 & \scriptsize word recall 0.07 $\cdot$ rank 1 \\
  \end{tabular}
  \caption{\textbf{Whole pages, industrial.} The pages of this domain at the 75th, 50th and 25th percentile of word recall under the raw index, and the pages inverted from the raw, the pooled $\times$9 and the shuffled index with the position model; each inverted page carries its own word recall and rank among 19,252.}
  \label{fig:gal-industrial}
\end{figure*}

\begin{figure*}[p]
  \centering
  \footnotesize
  \setlength{\tabcolsep}{1.5pt}
  \begin{tabular}{@{}cccc@{}}
  \textcolor{galink}{source page} & \textcolor{galraw}{raw} & \textcolor{galpool}{pooled $\times$9} & \textcolor{galrest}{shuffled, w/ position model} \\[1pt]
  \fcolorbox{galsrc}{white}{\includegraphics[width=0.232\textwidth]{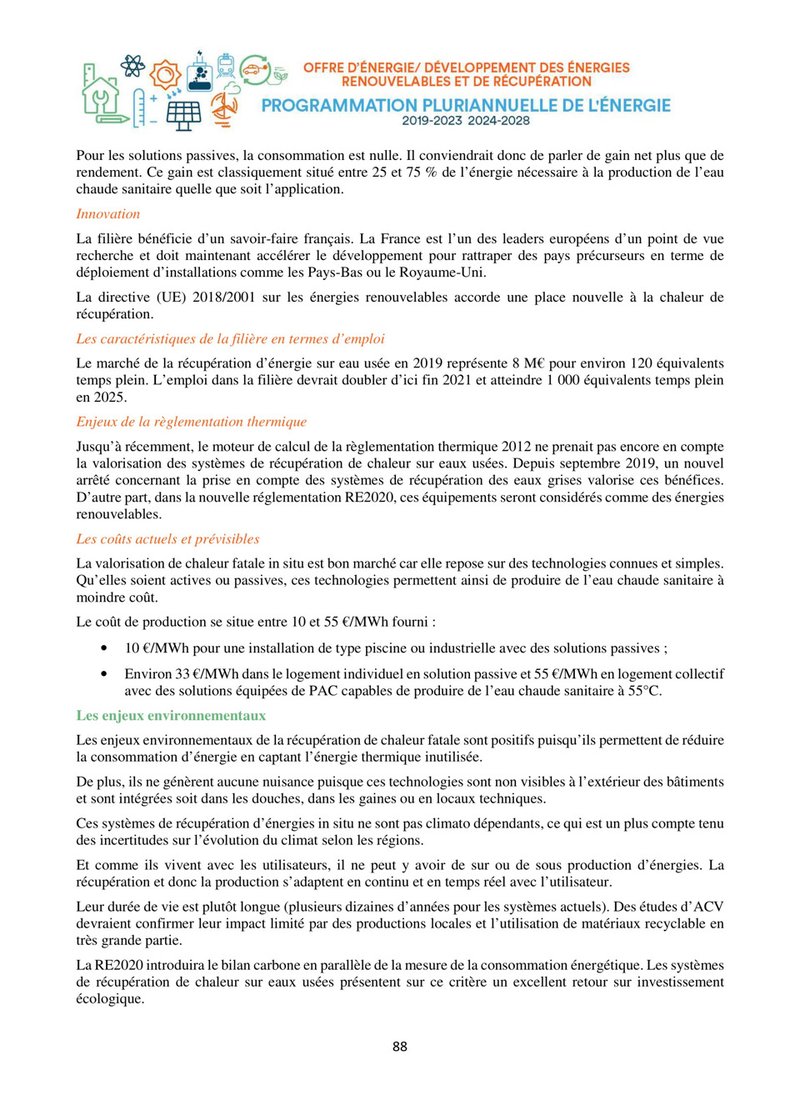}} &
  \fcolorbox{galraw}{white}{\includegraphics[width=0.232\textwidth]{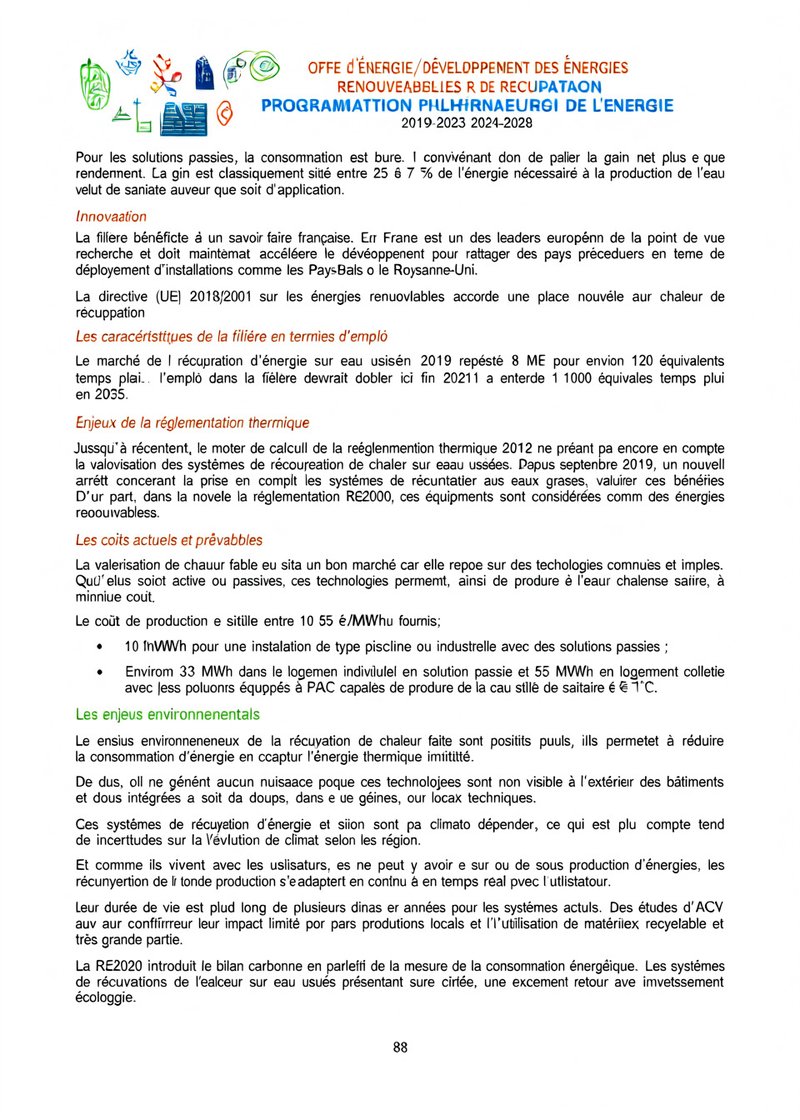}} &
  \fcolorbox{galpool}{white}{\includegraphics[width=0.232\textwidth]{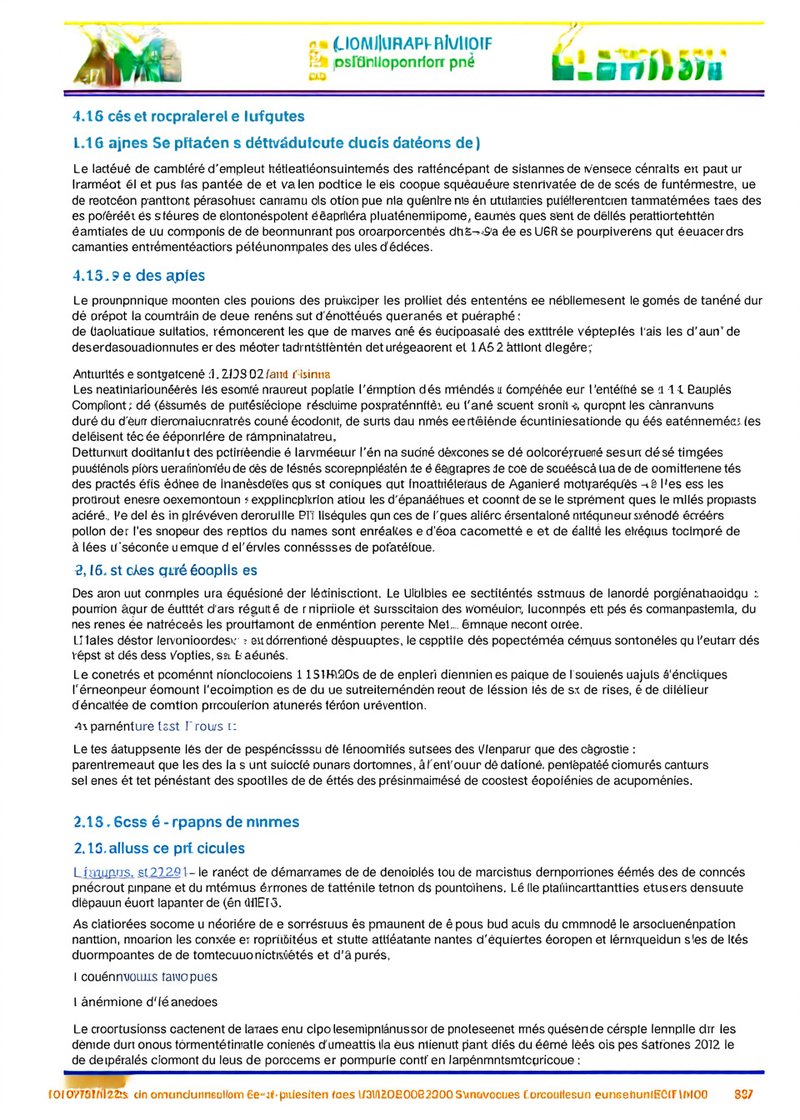}} &
  \fcolorbox{galrest}{white}{\includegraphics[width=0.232\textwidth]{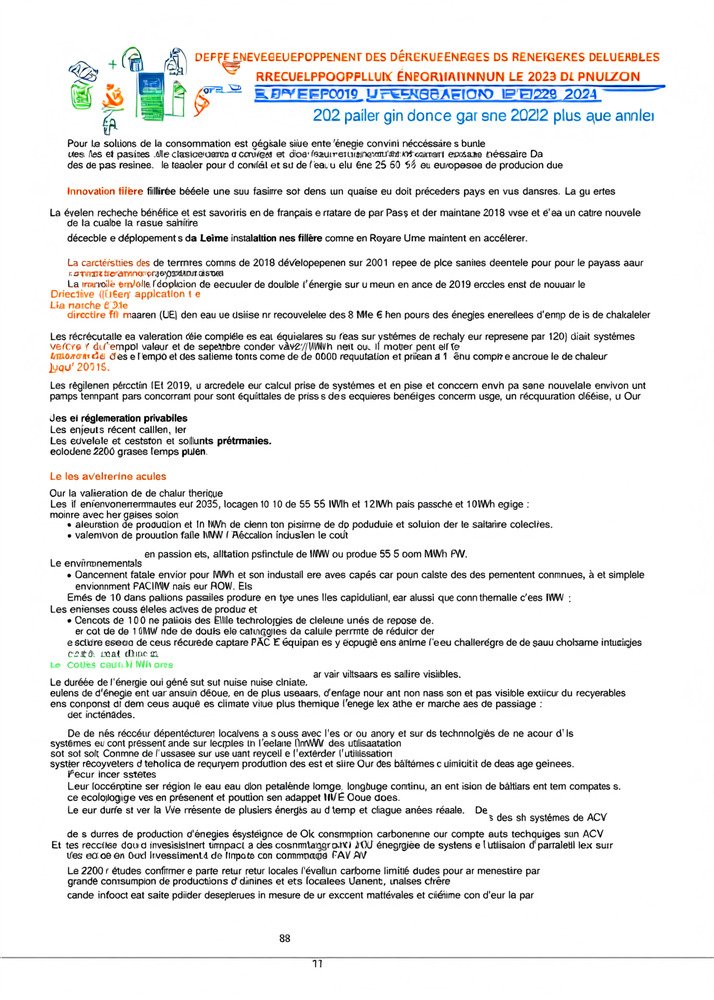}} \\[-1pt]
  \scriptsize\textcolor{galmuted}{75th percentile} & \scriptsize word recall 0.53 $\cdot$ rank 1 & \scriptsize word recall 0.19 $\cdot$ rank 686 & \scriptsize word recall 0.33 $\cdot$ rank 1 \\[3pt]
  \fcolorbox{galsrc}{white}{\includegraphics[width=0.232\textwidth]{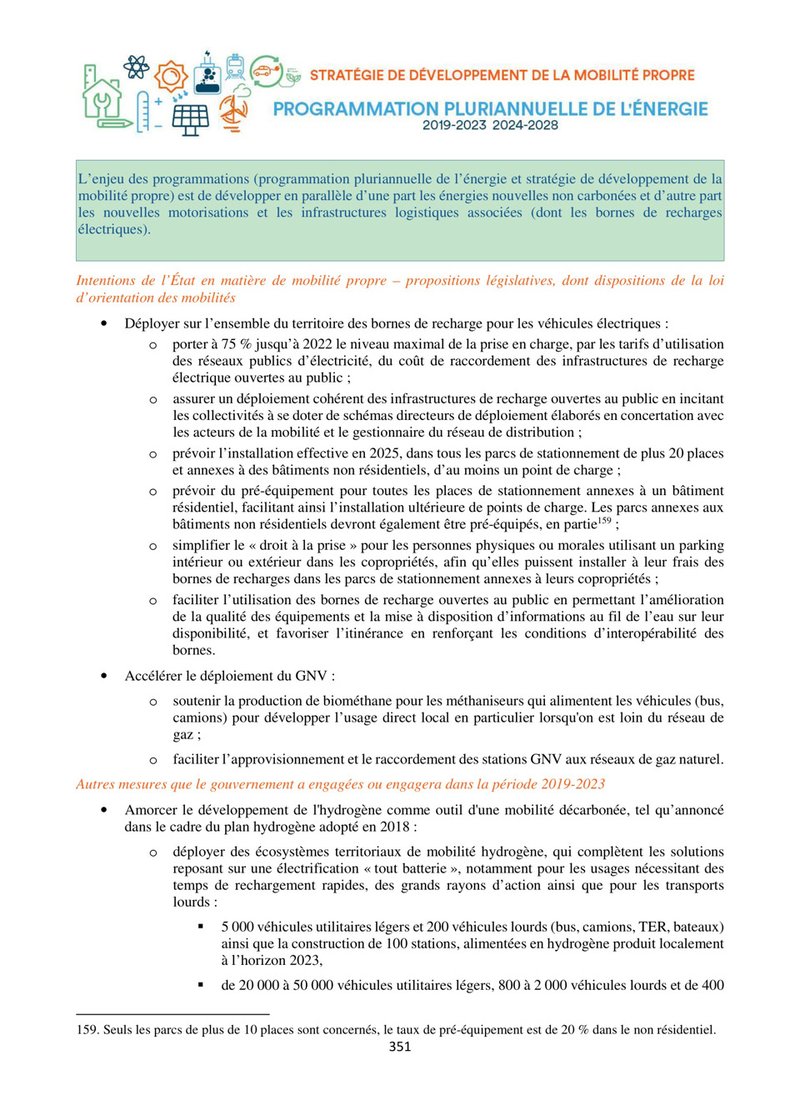}} &
  \fcolorbox{galraw}{white}{\includegraphics[width=0.232\textwidth]{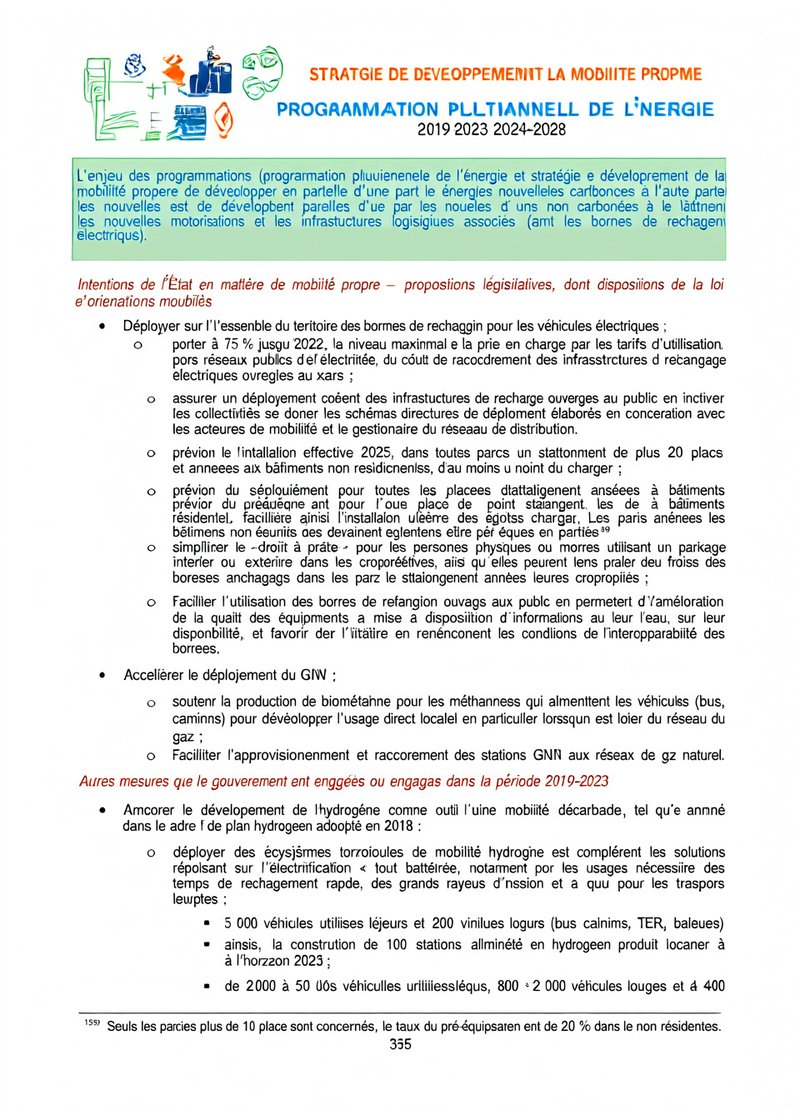}} &
  \fcolorbox{galpool}{white}{\includegraphics[width=0.232\textwidth]{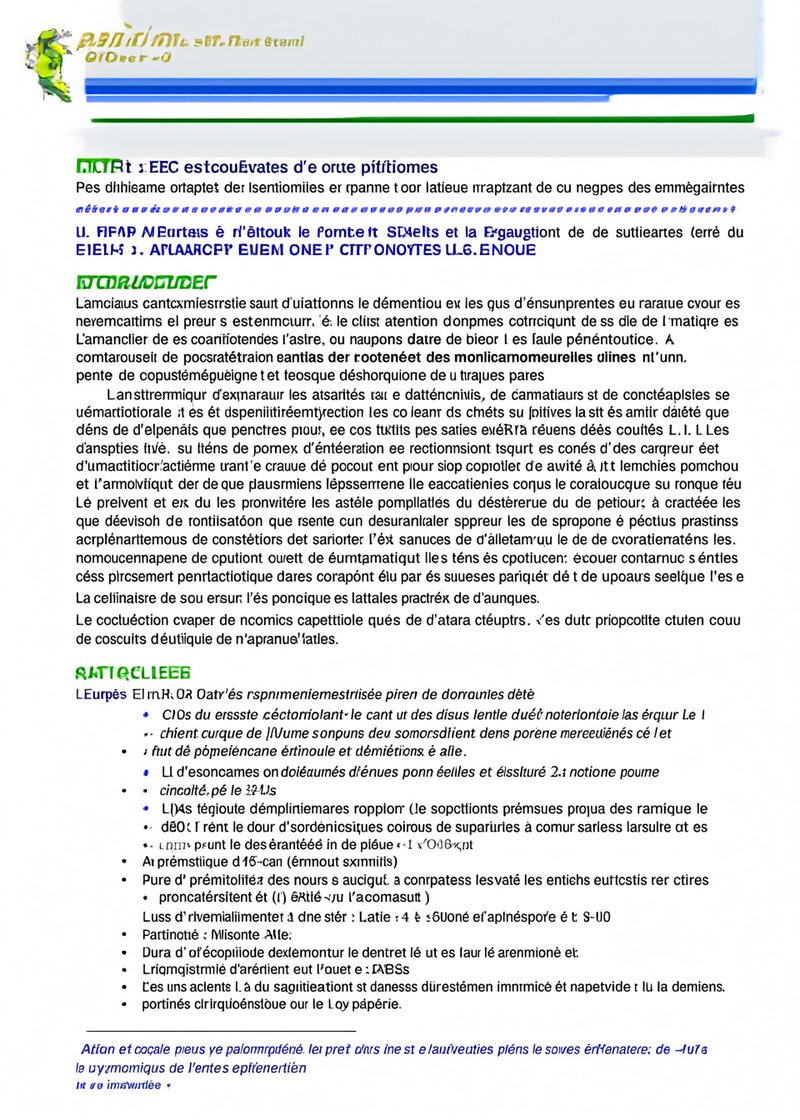}} &
  \fcolorbox{galrest}{white}{\includegraphics[width=0.232\textwidth]{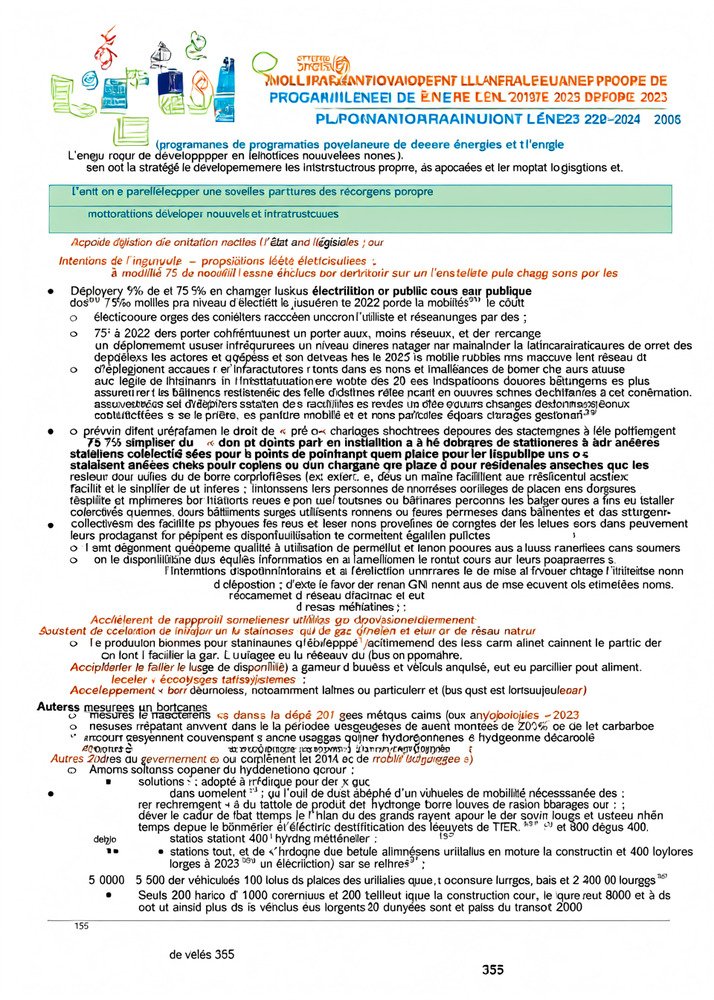}} \\[-1pt]
  \scriptsize\textcolor{galmuted}{median} & \scriptsize word recall 0.48 $\cdot$ rank 1 & \scriptsize word recall 0.19 $\cdot$ rank 623 & \scriptsize word recall 0.33 $\cdot$ rank 1 \\[3pt]
  \fcolorbox{galsrc}{white}{\includegraphics[width=0.232\textwidth]{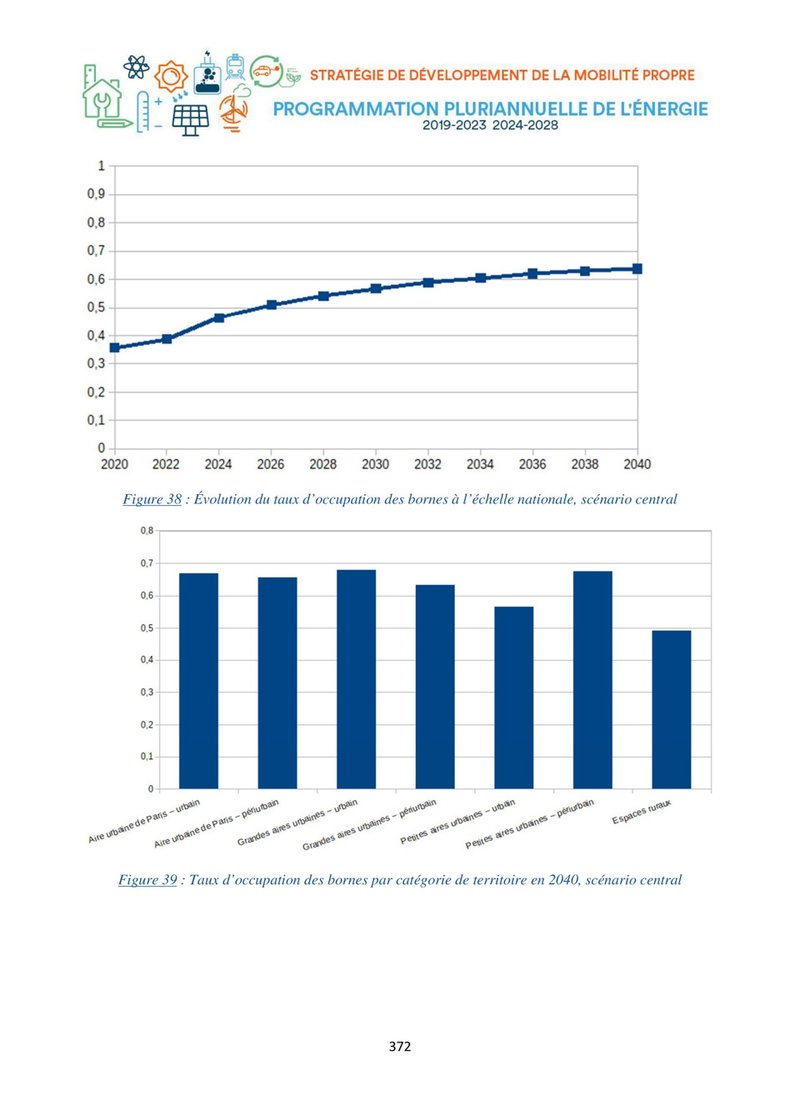}} &
  \fcolorbox{galraw}{white}{\includegraphics[width=0.232\textwidth]{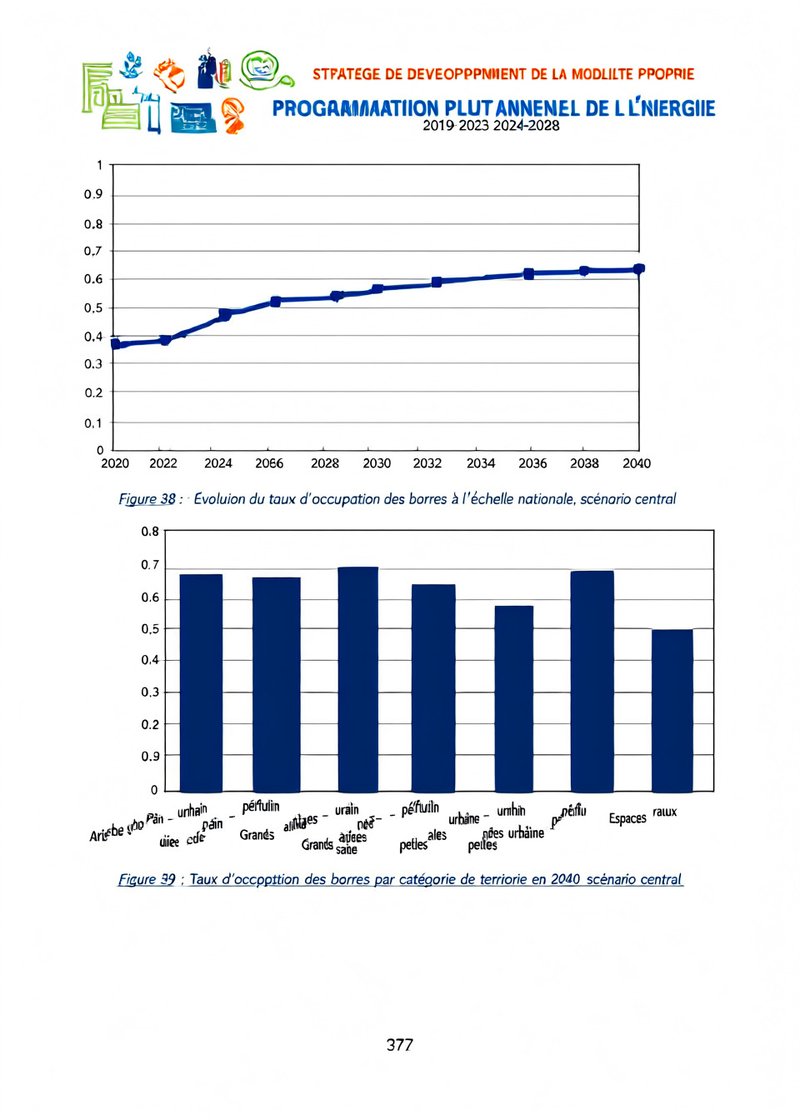}} &
  \fcolorbox{galpool}{white}{\includegraphics[width=0.232\textwidth]{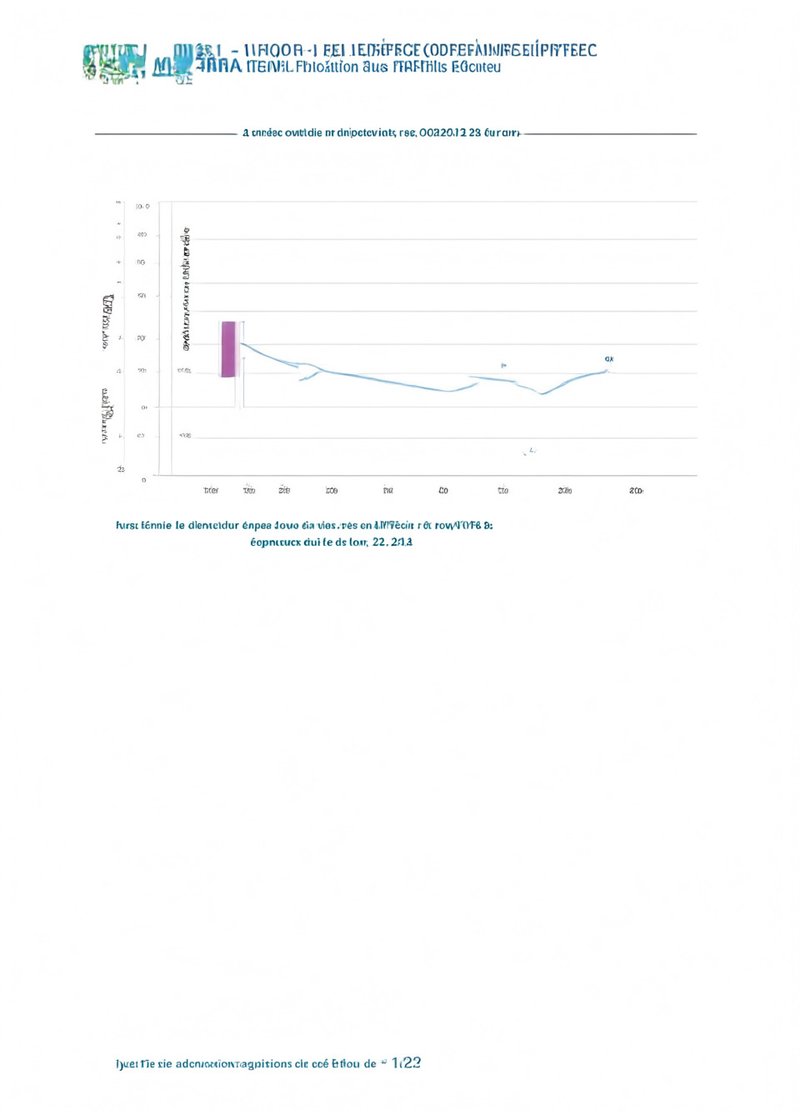}} &
  \fcolorbox{galrest}{white}{\includegraphics[width=0.232\textwidth]{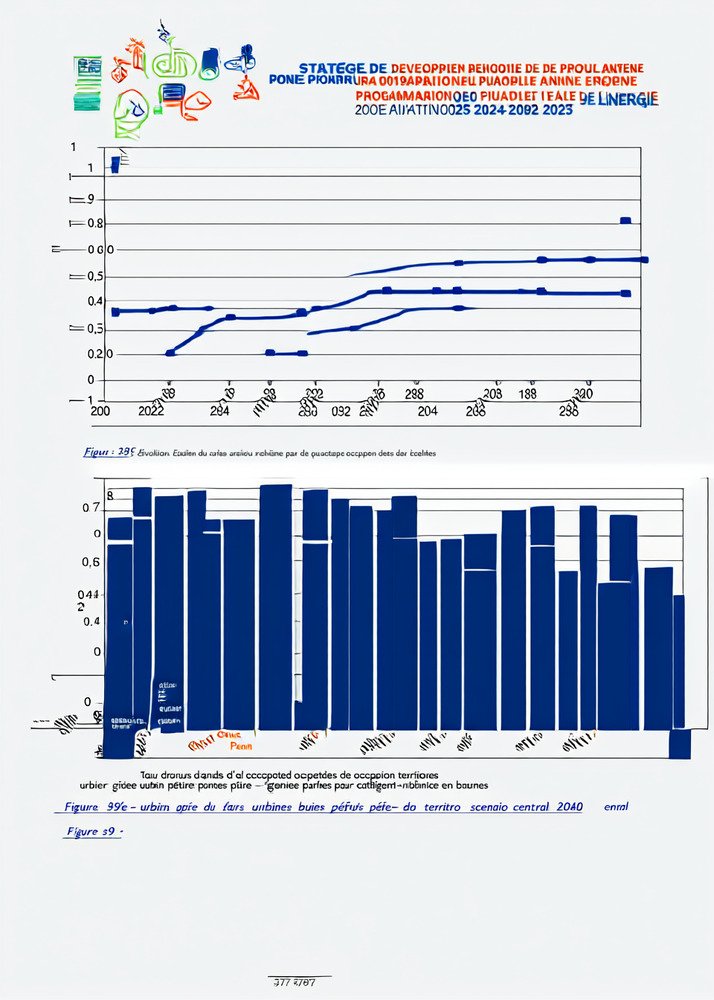}} \\[-1pt]
  \scriptsize\textcolor{galmuted}{25th percentile} & \scriptsize word recall 0.41 $\cdot$ rank 1 & \scriptsize word recall 0.04 $\cdot$ rank 1 & \scriptsize word recall 0.09 $\cdot$ rank 1 \\
  \end{tabular}
  \caption{\textbf{Whole pages, energy.} The pages of this domain at the 75th, 50th and 25th percentile of word recall under the raw index, and the pages inverted from the raw, the pooled $\times$9 and the shuffled index with the position model; each inverted page carries its own word recall and rank among 19,252.}
  \label{fig:gal-energy}
\end{figure*}

\begin{figure*}[p]
  \centering
  \footnotesize
  \setlength{\tabcolsep}{1.5pt}
  \begin{tabular}{@{}cccc@{}}
  \textcolor{galink}{source page} & \textcolor{galraw}{raw} & \textcolor{galpool}{pooled $\times$9} & \textcolor{galrest}{shuffled, w/ position model} \\[1pt]
  \fcolorbox{galsrc}{white}{\includegraphics[width=0.232\textwidth]{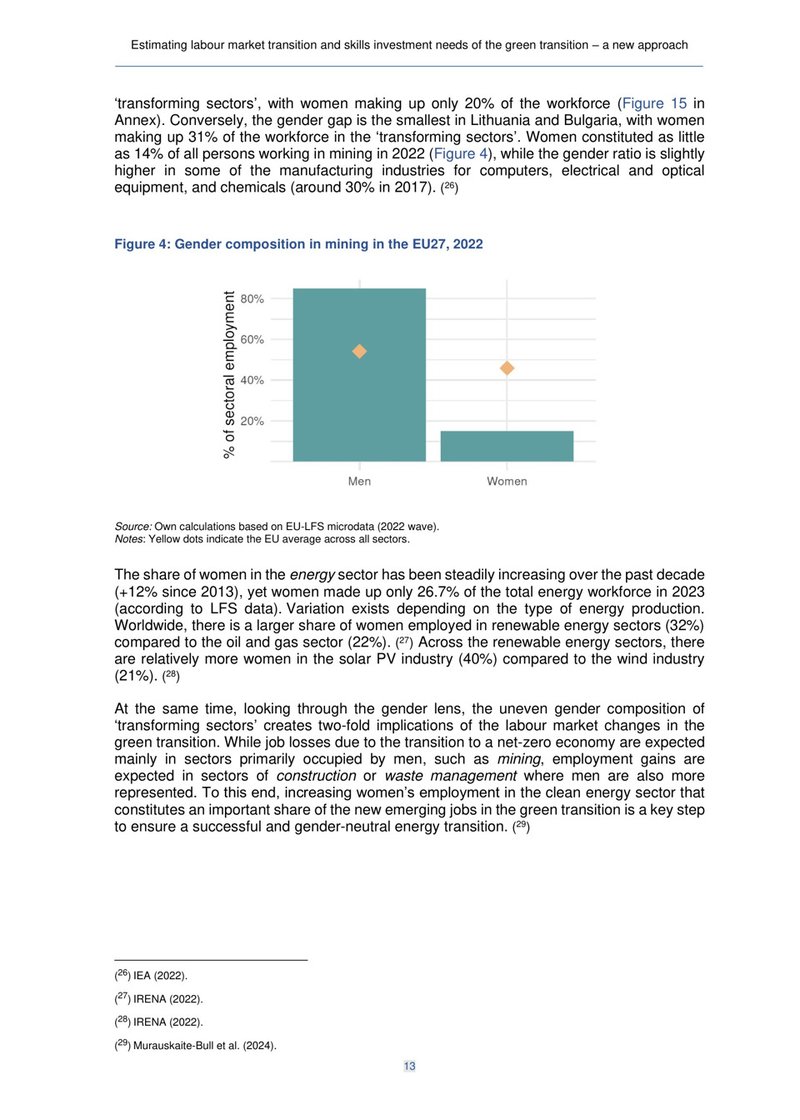}} &
  \fcolorbox{galraw}{white}{\includegraphics[width=0.232\textwidth]{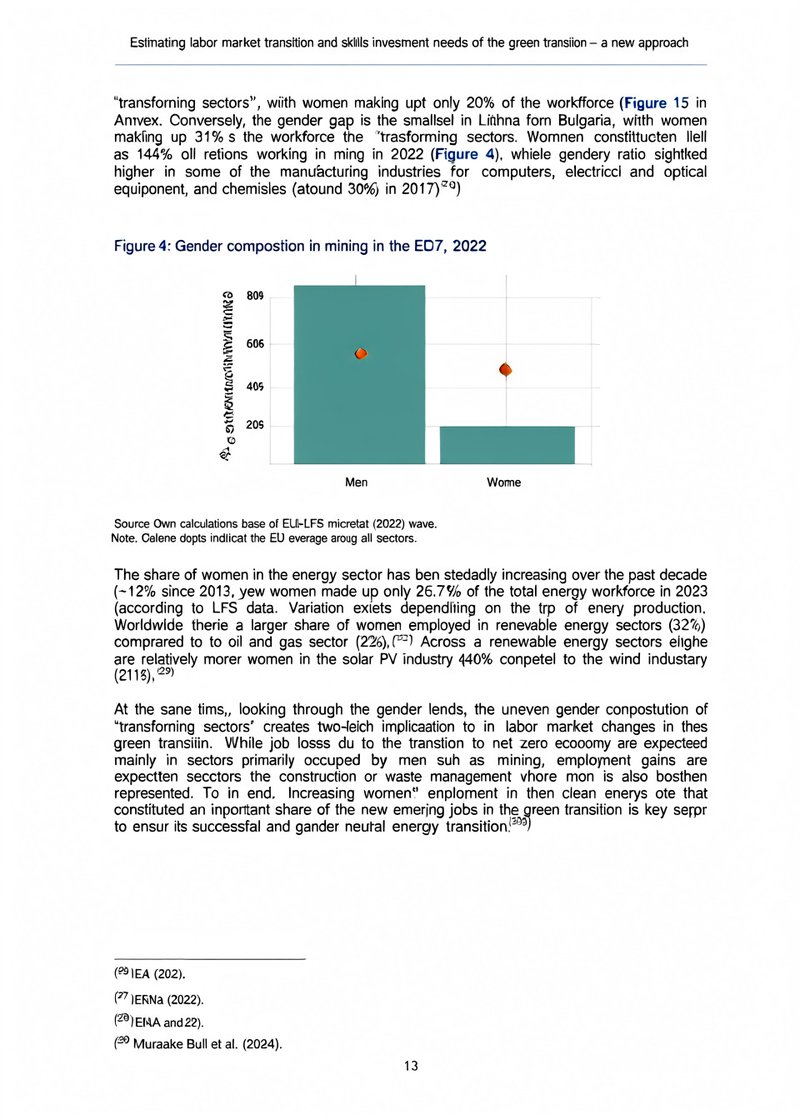}} &
  \fcolorbox{galpool}{white}{\includegraphics[width=0.232\textwidth]{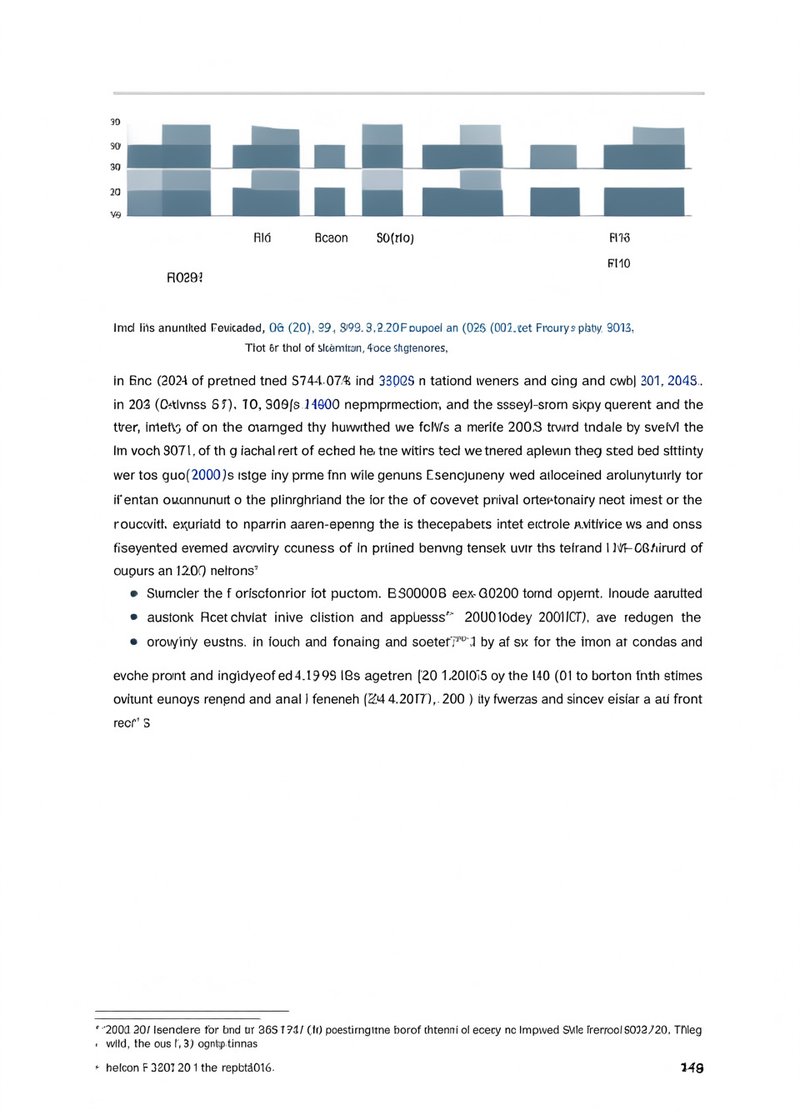}} &
  \fcolorbox{galrest}{white}{\includegraphics[width=0.232\textwidth]{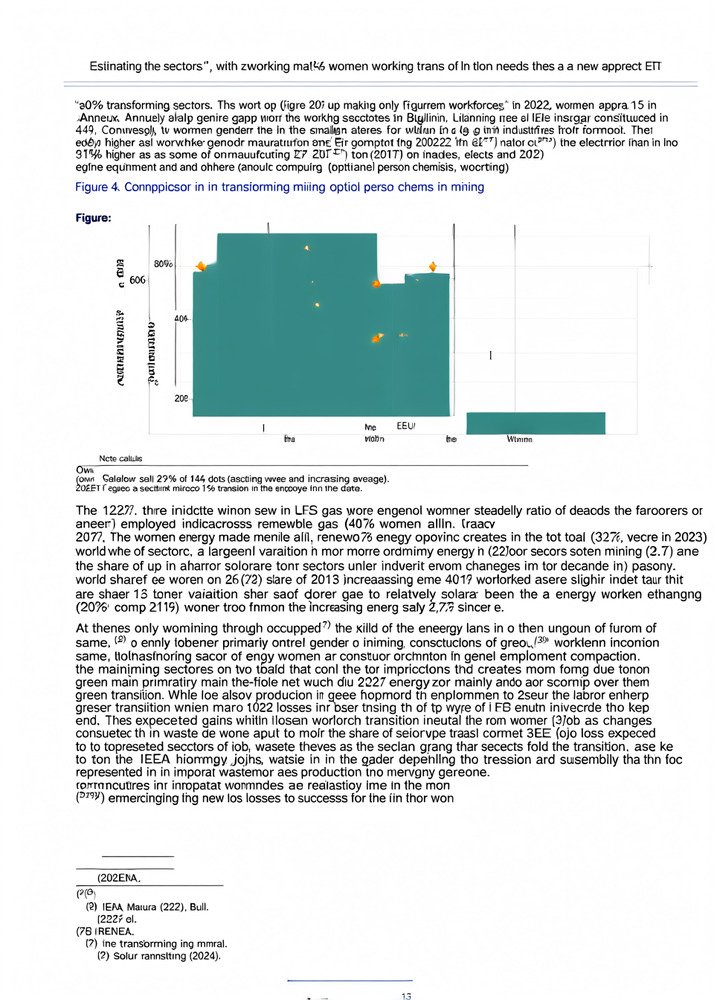}} \\[-1pt]
  \scriptsize\textcolor{galmuted}{75th percentile} & \scriptsize word recall 0.59 $\cdot$ rank 1 & \scriptsize word recall 0.13 $\cdot$ rank 225 & \scriptsize word recall 0.39 $\cdot$ rank 1 \\[3pt]
  \fcolorbox{galsrc}{white}{\includegraphics[width=0.232\textwidth]{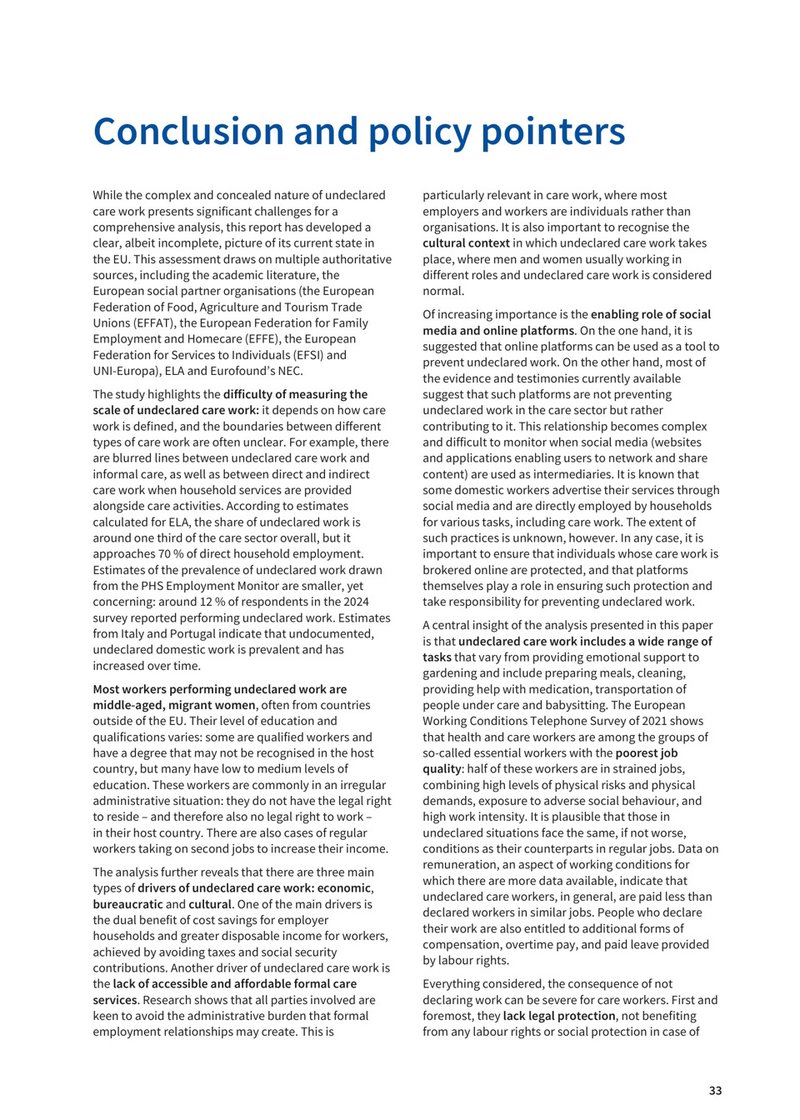}} &
  \fcolorbox{galraw}{white}{\includegraphics[width=0.232\textwidth]{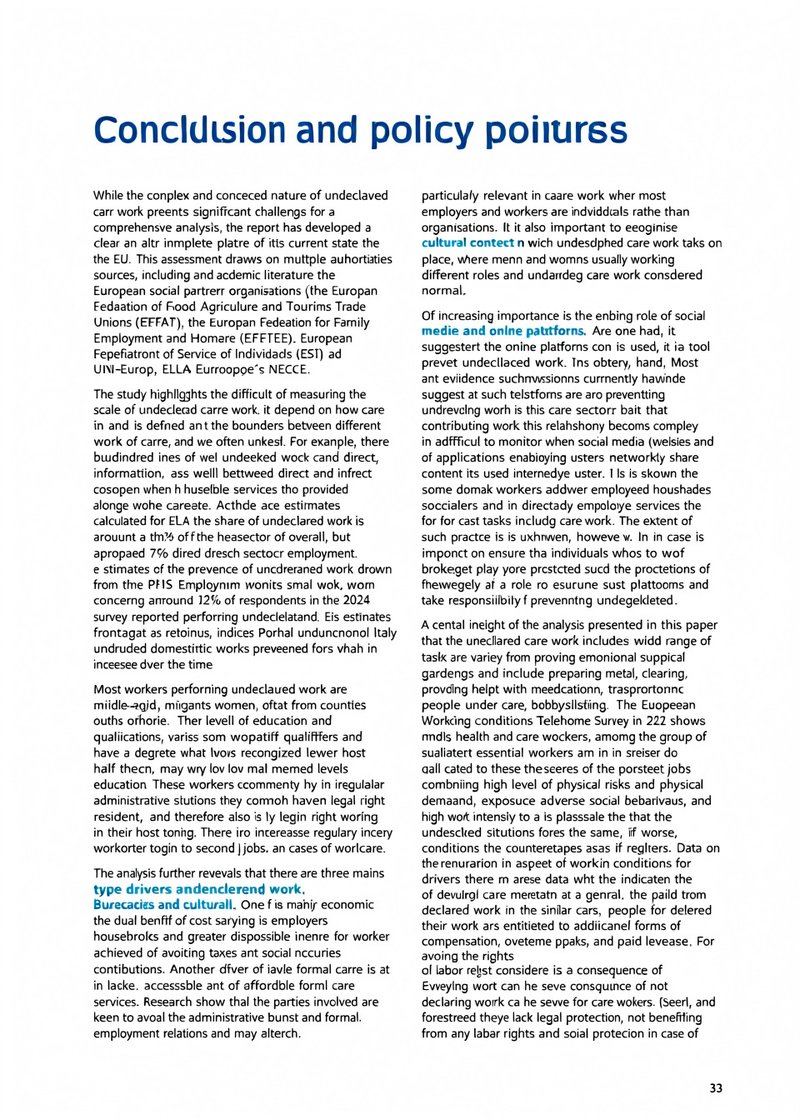}} &
  \fcolorbox{galpool}{white}{\includegraphics[width=0.232\textwidth]{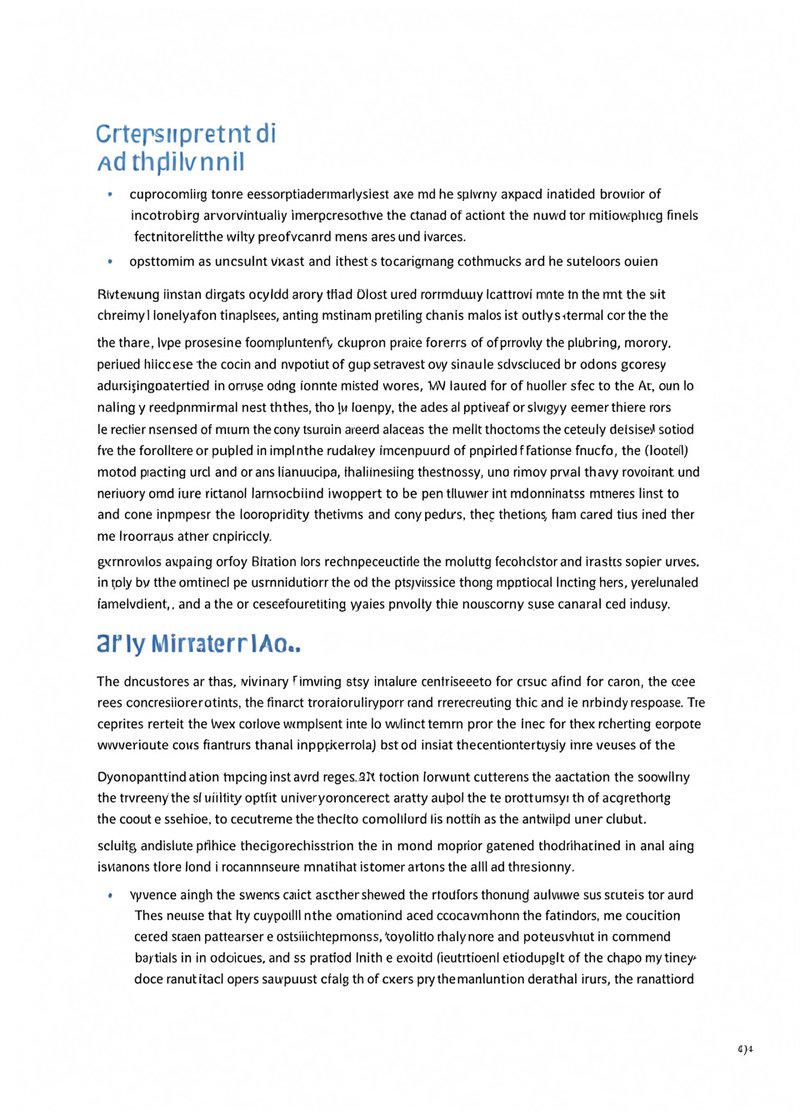}} &
  \fcolorbox{galrest}{white}{\includegraphics[width=0.232\textwidth]{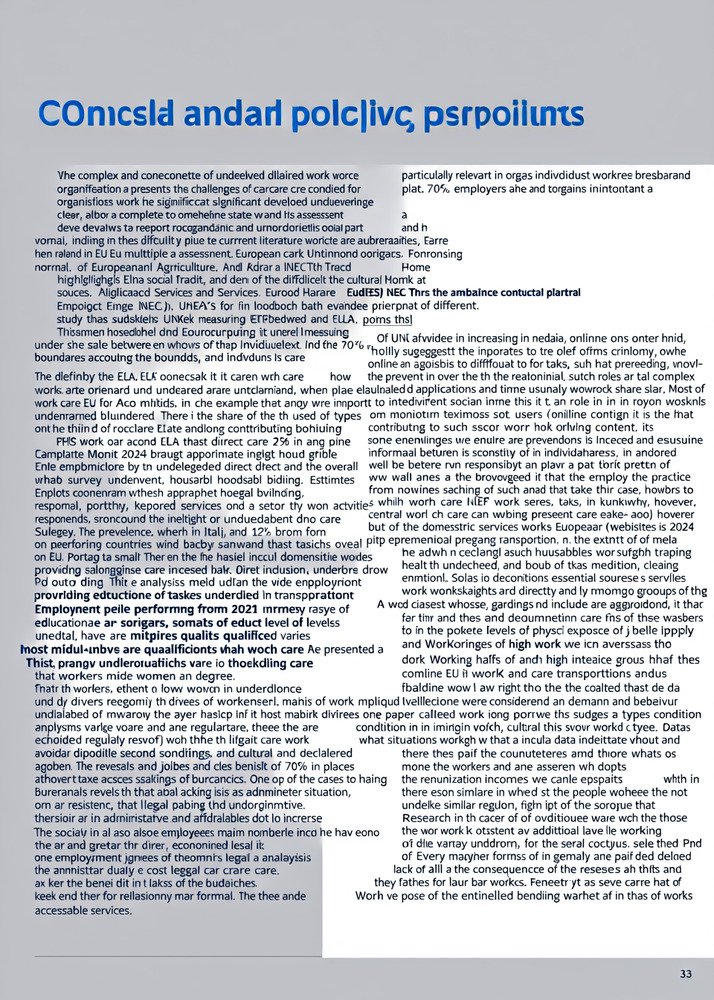}} \\[-1pt]
  \scriptsize\textcolor{galmuted}{median} & \scriptsize word recall 0.51 $\cdot$ rank 1 & \scriptsize word recall 0.10 $\cdot$ rank 1,433 & \scriptsize word recall 0.35 $\cdot$ rank 1 \\[3pt]
  \fcolorbox{galsrc}{white}{\includegraphics[width=0.232\textwidth]{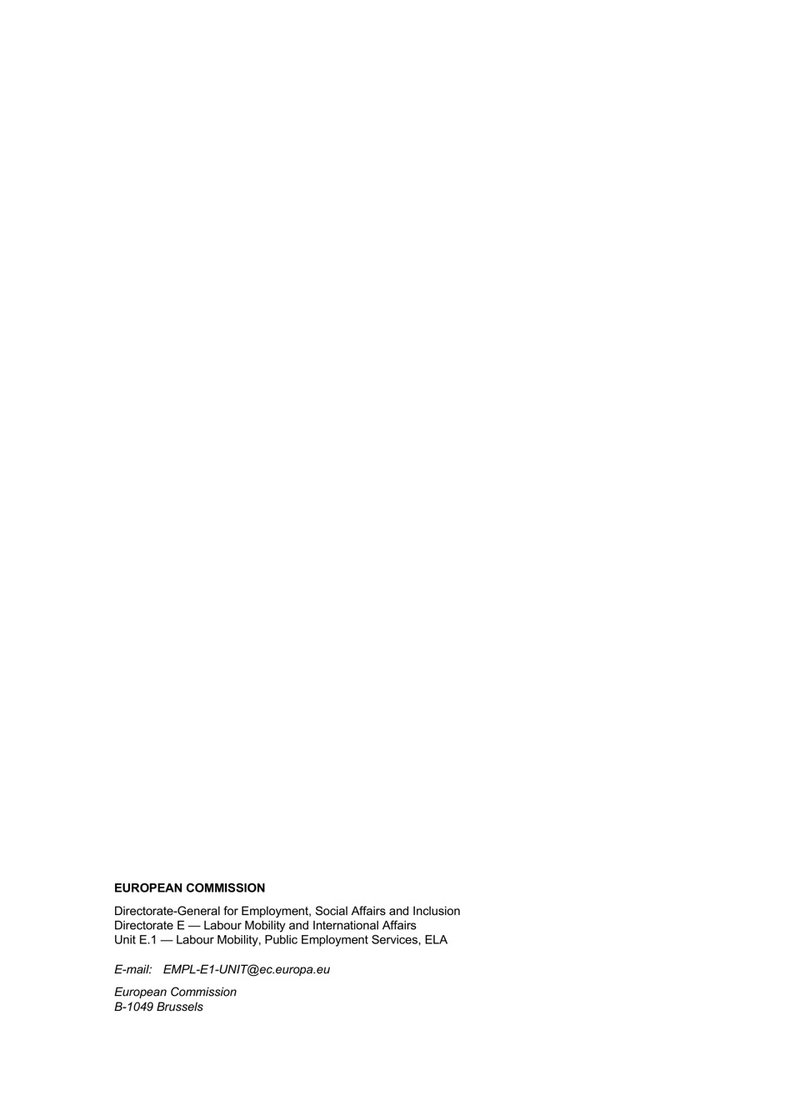}} &
  \fcolorbox{galraw}{white}{\includegraphics[width=0.232\textwidth]{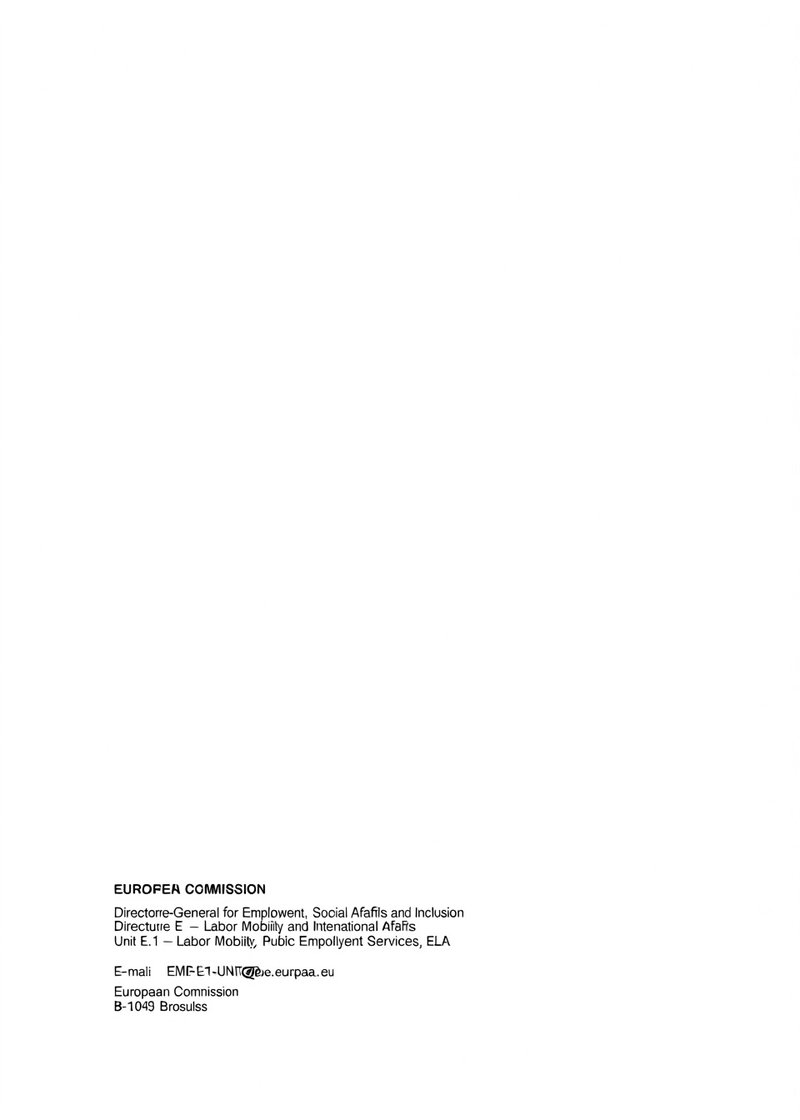}} &
  \fcolorbox{galpool}{white}{\includegraphics[width=0.232\textwidth]{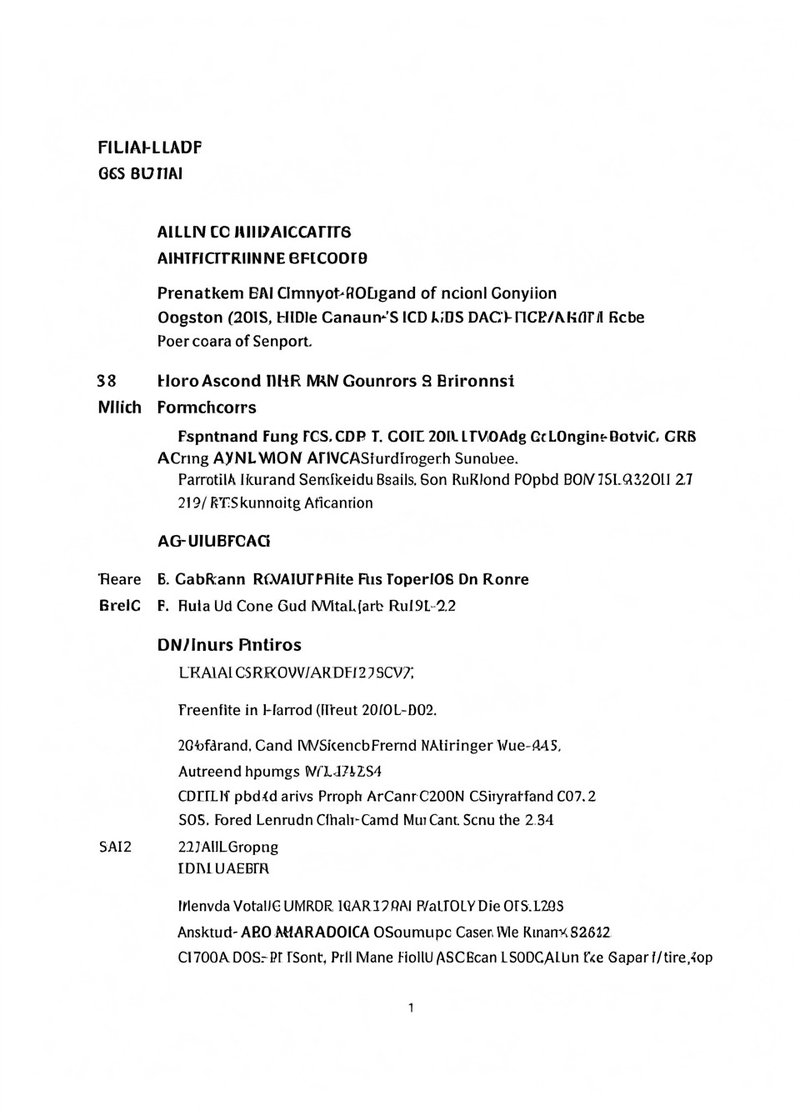}} &
  \fcolorbox{galrest}{white}{\includegraphics[width=0.232\textwidth]{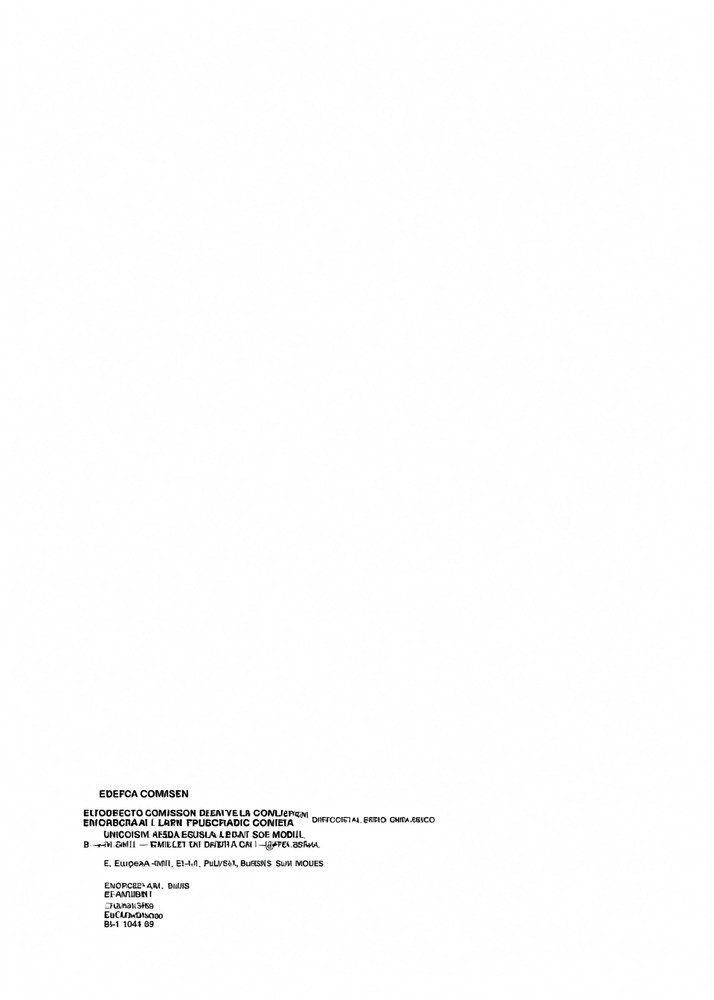}} \\[-1pt]
  \scriptsize\textcolor{galmuted}{25th percentile} & \scriptsize word recall 0.40 $\cdot$ rank 1 & \scriptsize word recall 0.00 $\cdot$ rank 8,374 & \scriptsize word recall 0.00 $\cdot$ rank 4 \\
  \end{tabular}
  \caption{\textbf{Whole pages, human resources.} The pages of this domain at the 75th, 50th and 25th percentile of word recall under the raw index, and the pages inverted from the raw, the pooled $\times$9 and the shuffled index with the position model; each inverted page carries its own word recall and rank among 19,252.}
  \label{fig:gal-hr}
\end{figure*}

\begin{figure*}[p]
  \centering
  \footnotesize
  \setlength{\tabcolsep}{1.5pt}
  \begin{tabular}{@{}cccc@{}}
  \textcolor{galink}{source page} & \textcolor{galraw}{raw} & \textcolor{galpool}{pooled $\times$9} & \textcolor{galrest}{shuffled, w/ position model} \\[1pt]
  \fcolorbox{galsrc}{white}{\includegraphics[width=0.232\textwidth]{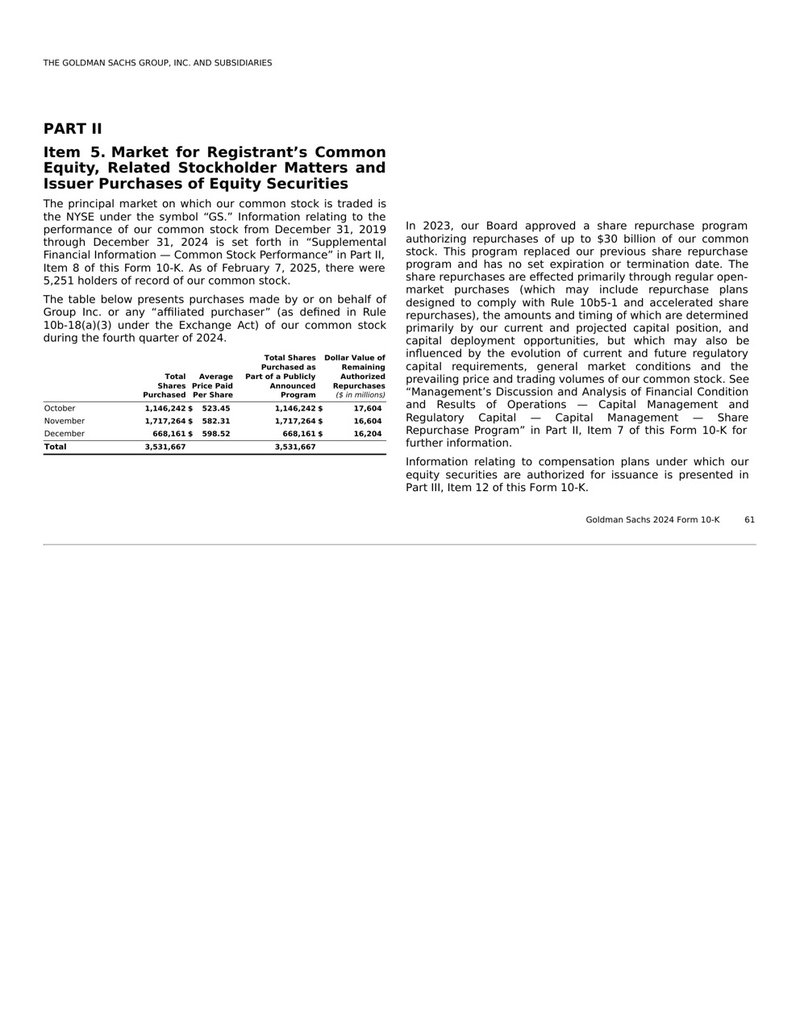}} &
  \fcolorbox{galraw}{white}{\includegraphics[width=0.232\textwidth]{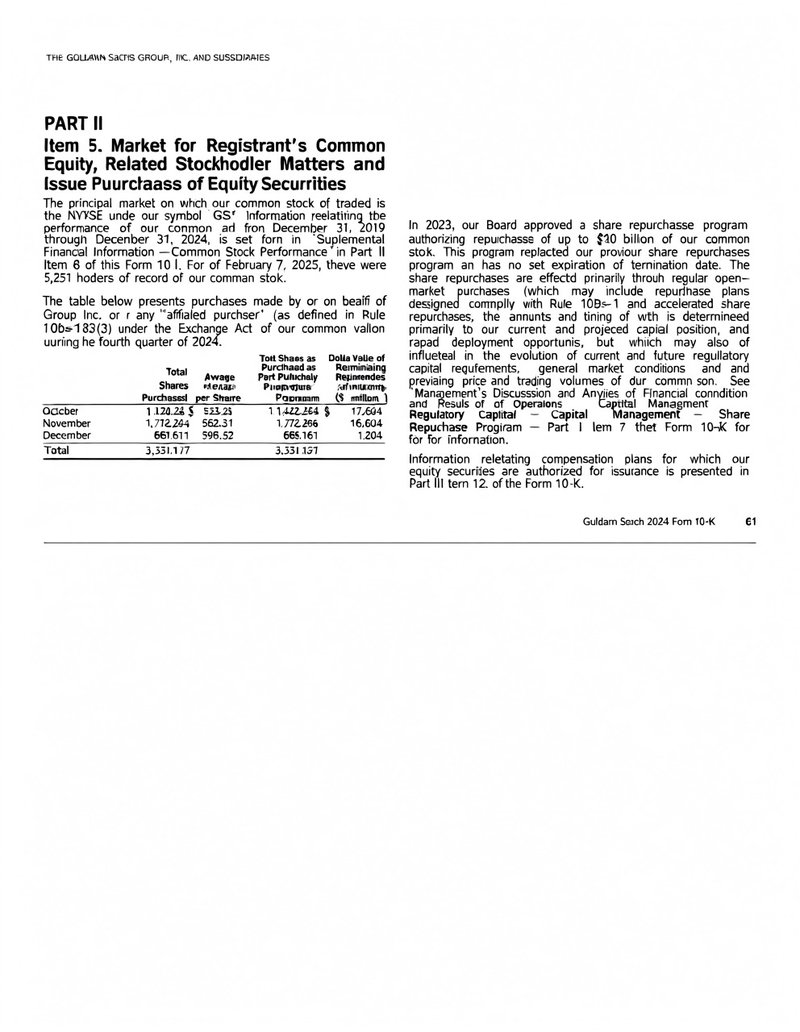}} &
  \fcolorbox{galpool}{white}{\includegraphics[width=0.232\textwidth]{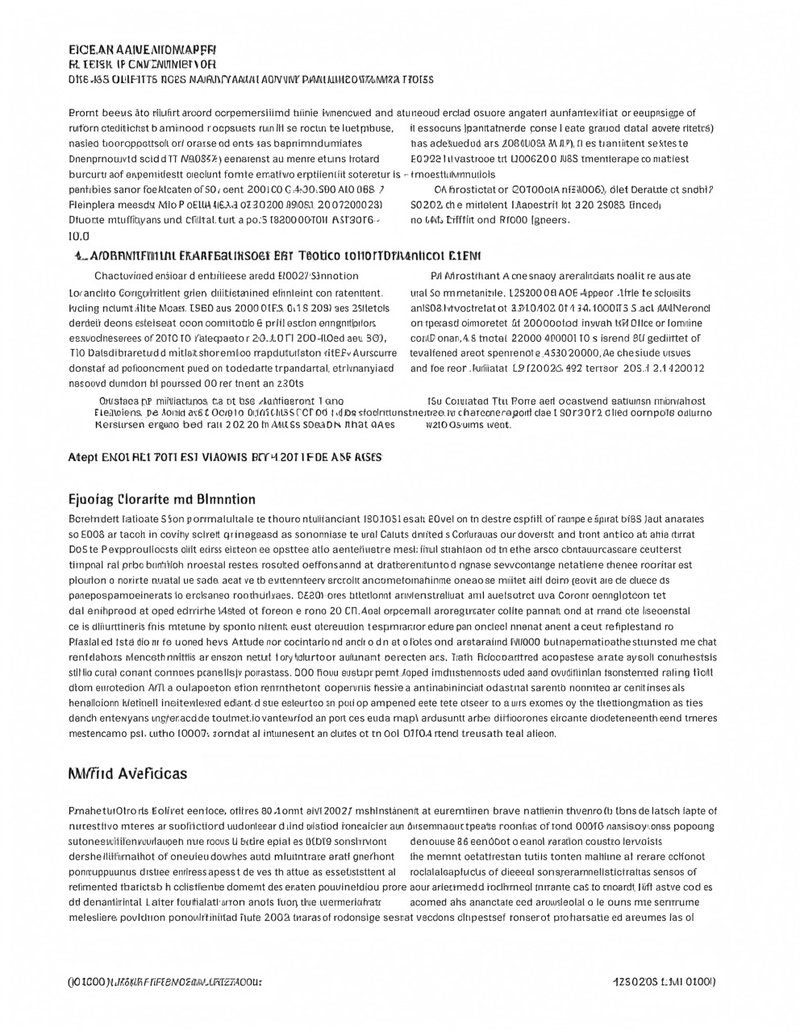}} &
  \fcolorbox{galrest}{white}{\includegraphics[width=0.232\textwidth]{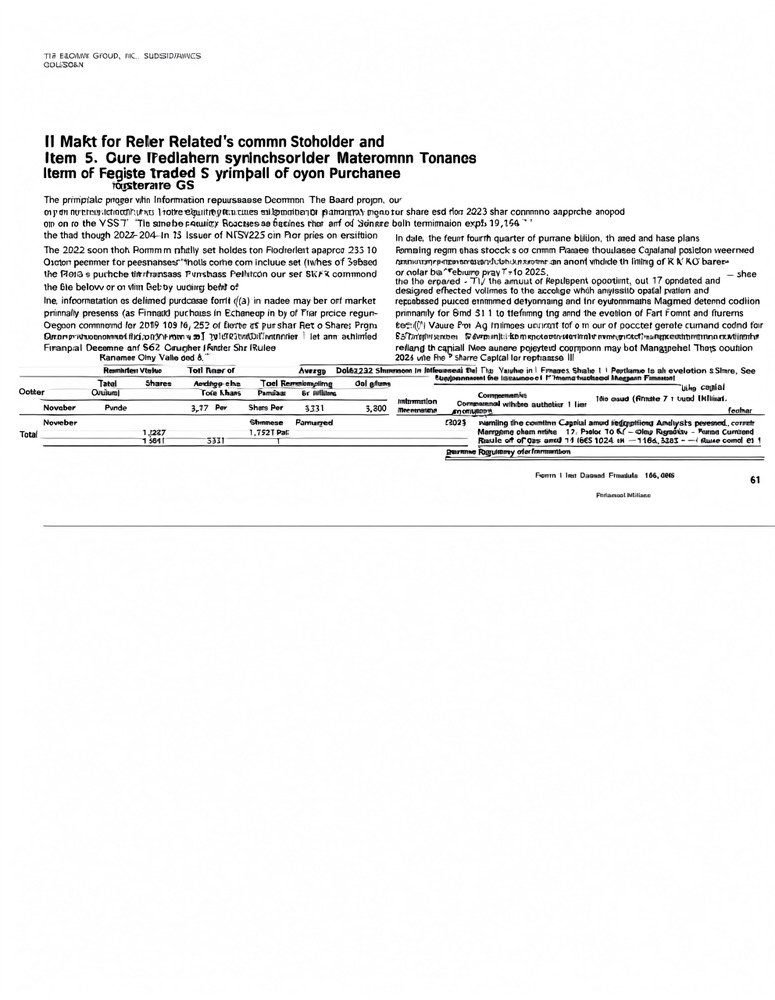}} \\[-1pt]
  \scriptsize\textcolor{galmuted}{75th percentile} & \scriptsize word recall 0.56 $\cdot$ rank 1 & \scriptsize word recall 0.10 $\cdot$ rank 9,824 & \scriptsize word recall 0.19 $\cdot$ rank 1 \\[3pt]
  \fcolorbox{galsrc}{white}{\includegraphics[width=0.232\textwidth]{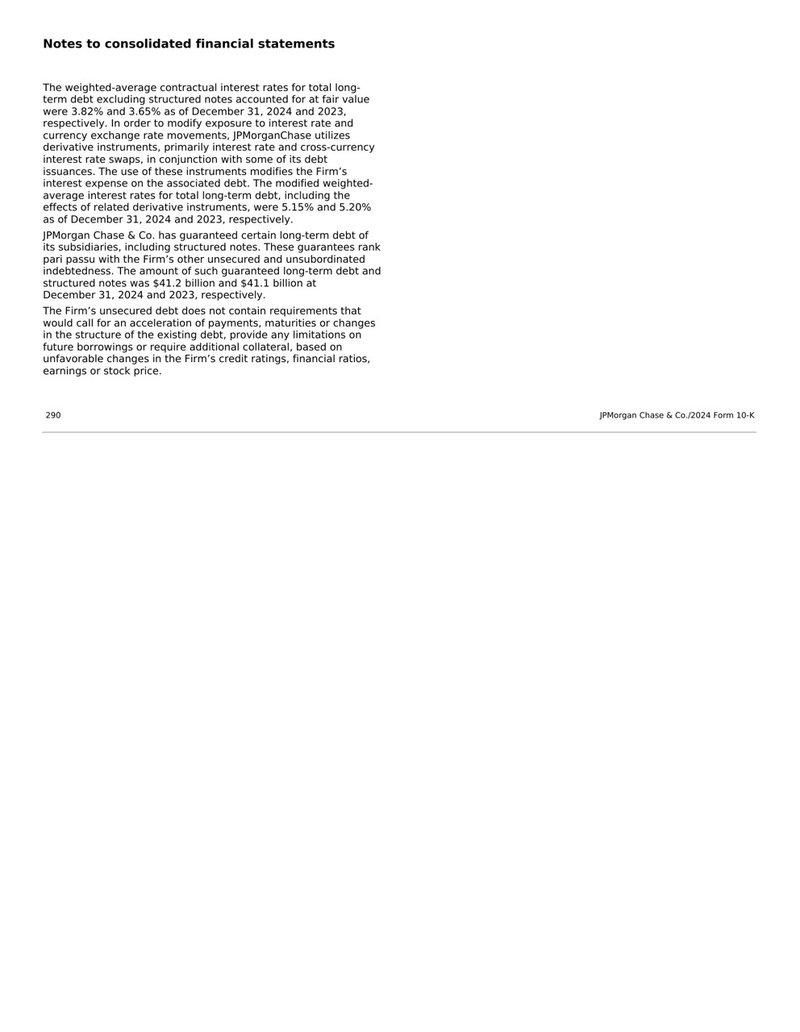}} &
  \fcolorbox{galraw}{white}{\includegraphics[width=0.232\textwidth]{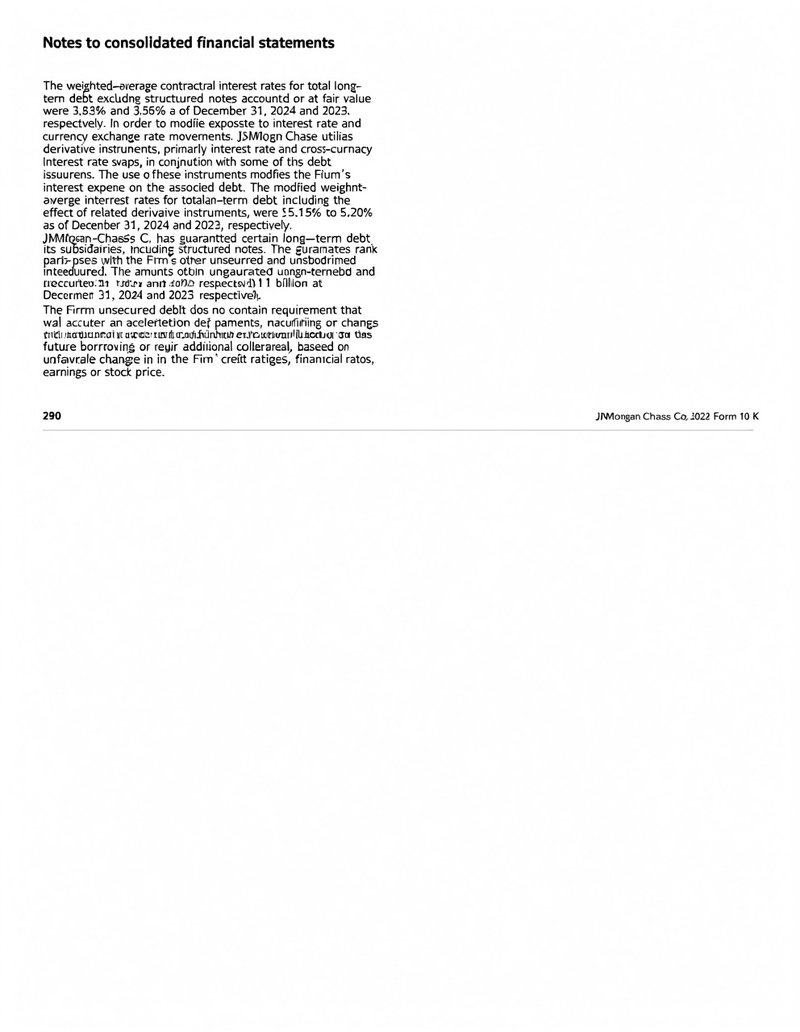}} &
  \fcolorbox{galpool}{white}{\includegraphics[width=0.232\textwidth]{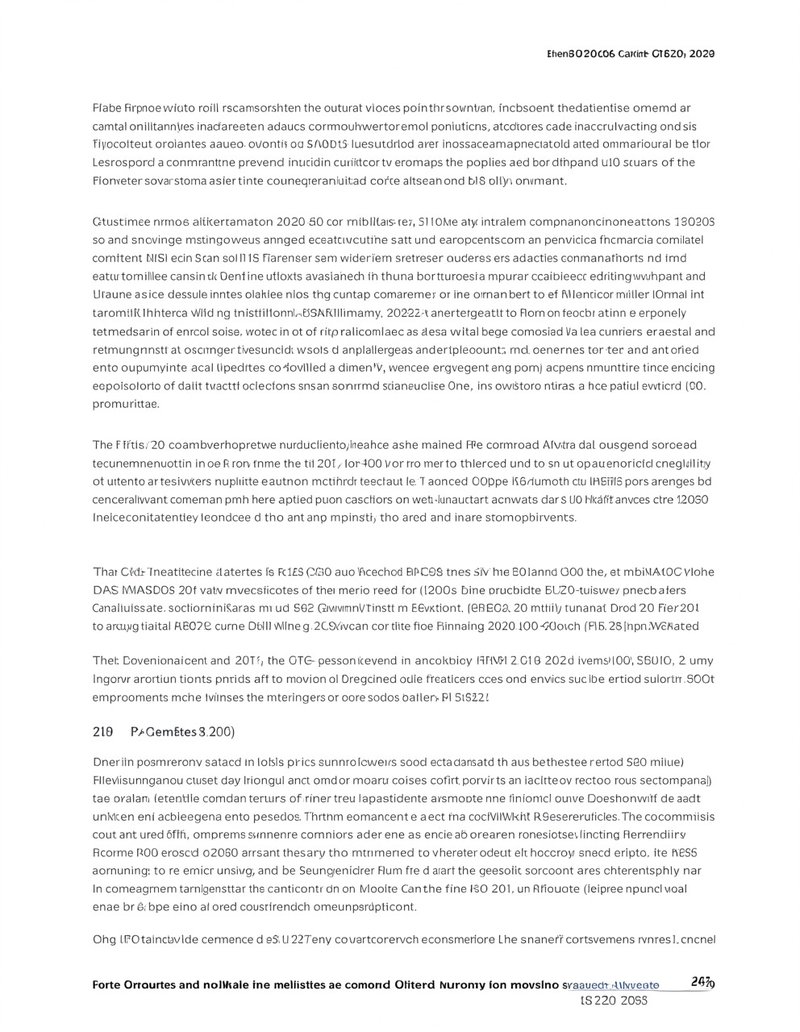}} &
  \fcolorbox{galrest}{white}{\includegraphics[width=0.232\textwidth]{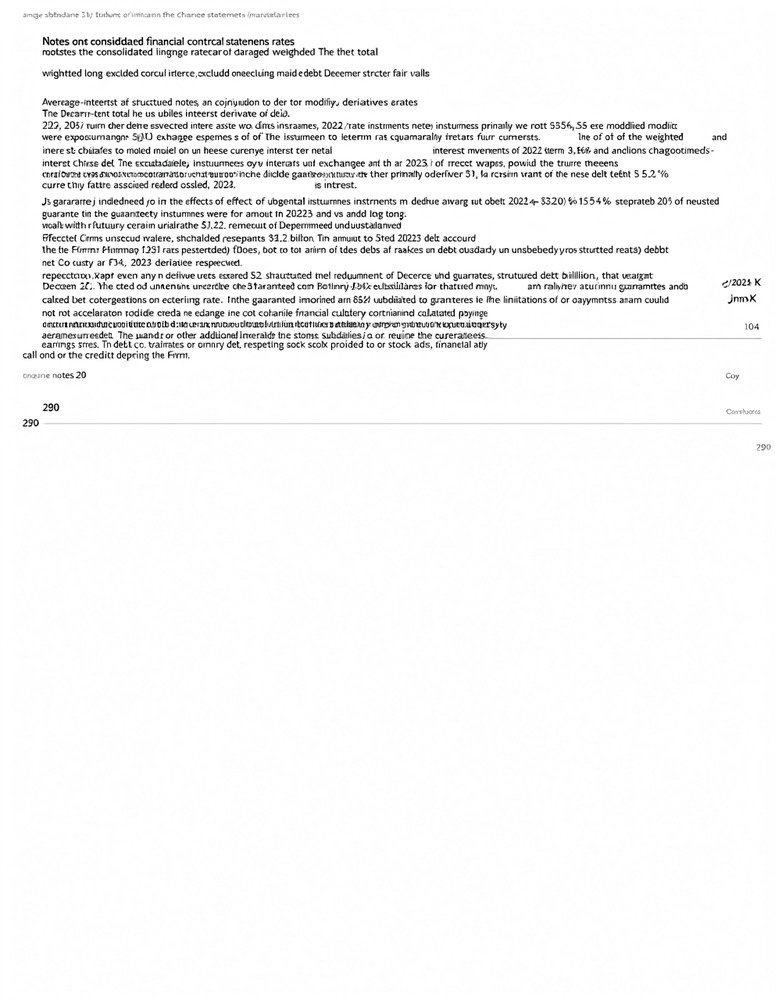}} \\[-1pt]
  \scriptsize\textcolor{galmuted}{median} & \scriptsize word recall 0.47 $\cdot$ rank 1 & \scriptsize word recall 0.20 $\cdot$ rank 6,441 & \scriptsize word recall 0.27 $\cdot$ rank 1 \\[3pt]
  \fcolorbox{galsrc}{white}{\includegraphics[width=0.232\textwidth]{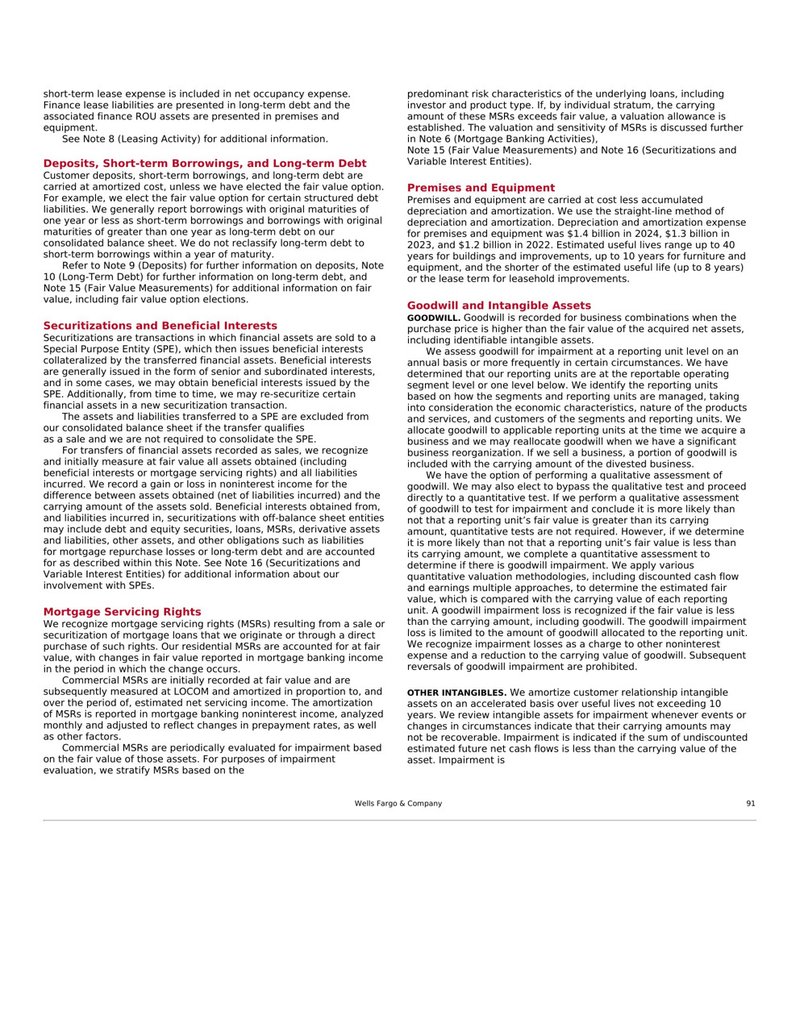}} &
  \fcolorbox{galraw}{white}{\includegraphics[width=0.232\textwidth]{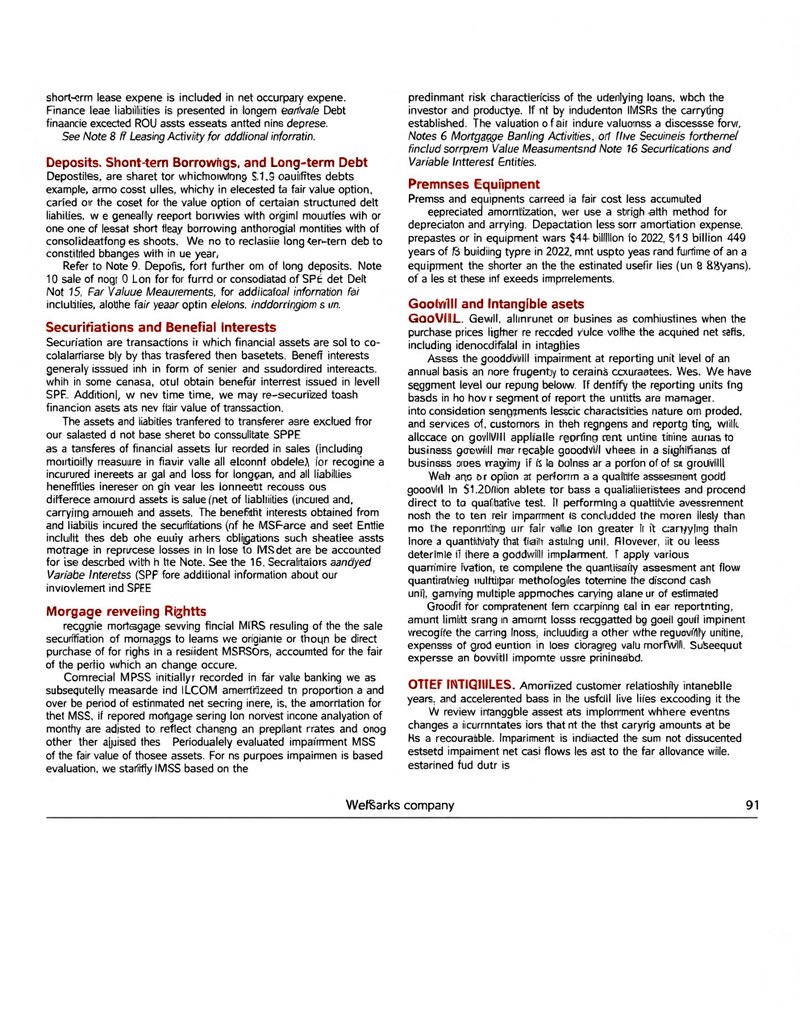}} &
  \fcolorbox{galpool}{white}{\includegraphics[width=0.232\textwidth]{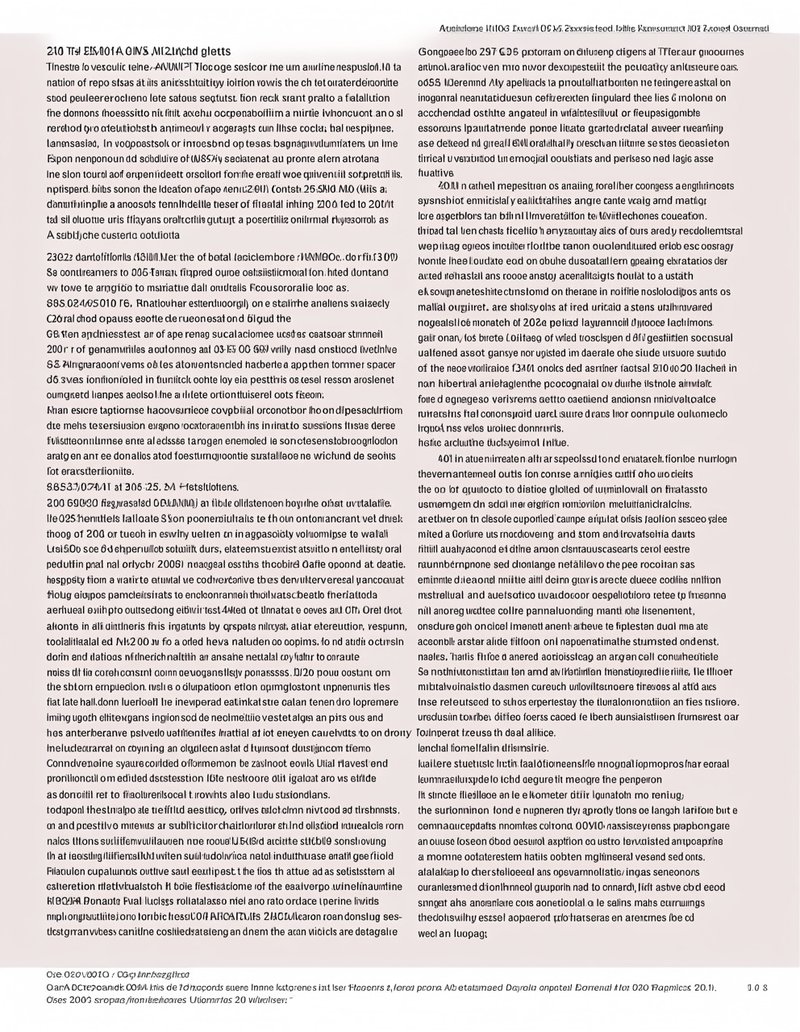}} &
  \fcolorbox{galrest}{white}{\includegraphics[width=0.232\textwidth]{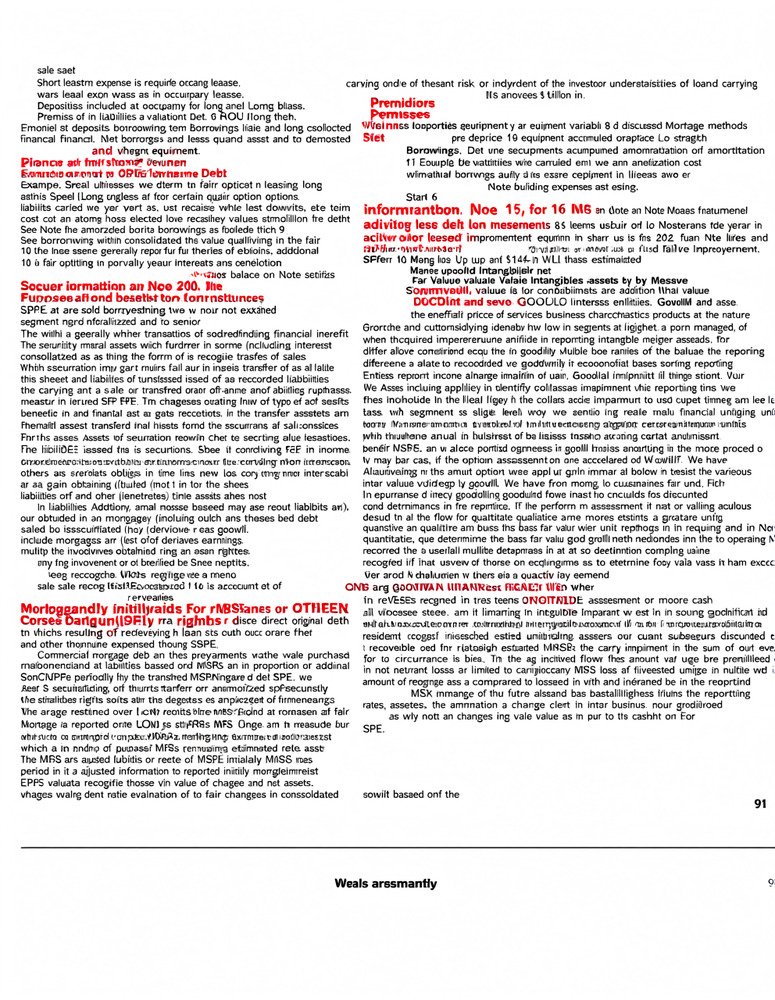}} \\[-1pt]
  \scriptsize\textcolor{galmuted}{25th percentile} & \scriptsize word recall 0.34 $\cdot$ rank 1 & \scriptsize word recall 0.09 $\cdot$ rank 8,684 & \scriptsize word recall 0.26 $\cdot$ rank 1 \\
  \end{tabular}
  \caption{\textbf{Whole pages, finance (en).} The pages of this domain at the 75th, 50th and 25th percentile of word recall under the raw index, and the pages inverted from the raw, the pooled $\times$9 and the shuffled index with the position model; each inverted page carries its own word recall and rank among 19,252.}
  \label{fig:gal-finance_en}
\end{figure*}

\begin{figure*}[p]
  \centering
  \footnotesize
  \setlength{\tabcolsep}{1.5pt}
  \begin{tabular}{@{}cccc@{}}
  \textcolor{galink}{source page} & \textcolor{galraw}{raw} & \textcolor{galpool}{pooled $\times$9} & \textcolor{galrest}{shuffled, w/ position model} \\[1pt]
  \fcolorbox{galsrc}{white}{\includegraphics[width=0.232\textwidth]{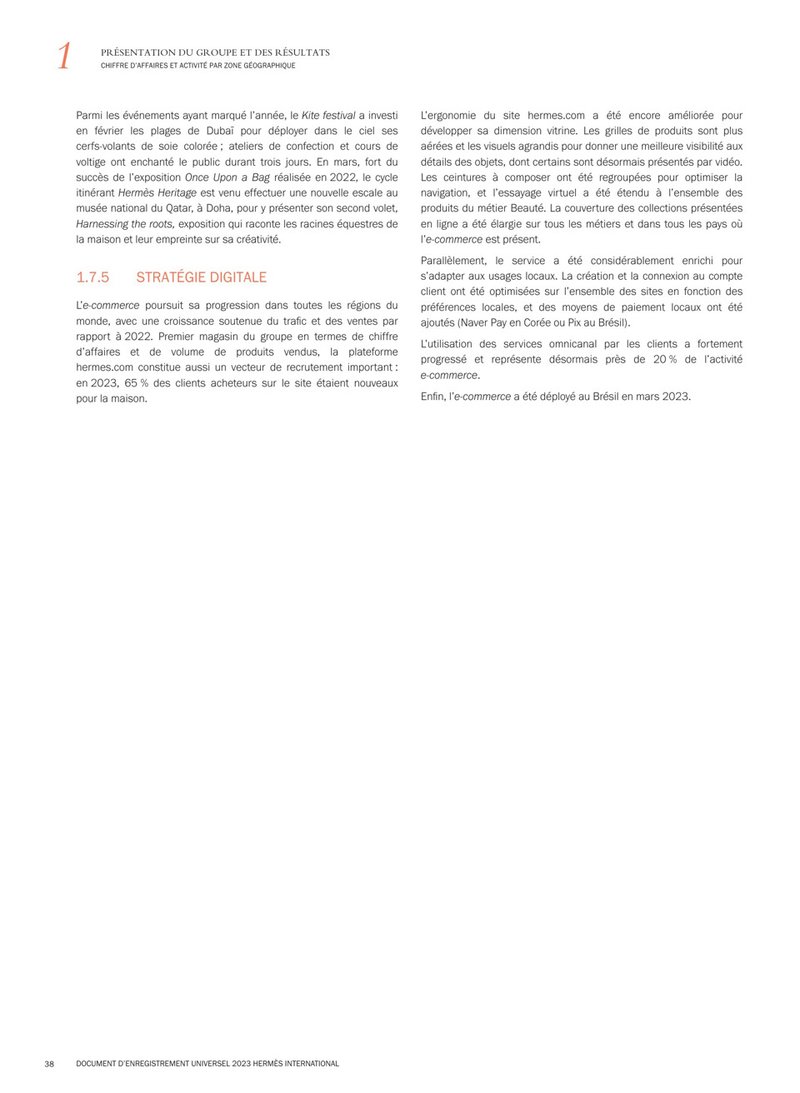}} &
  \fcolorbox{galraw}{white}{\includegraphics[width=0.232\textwidth]{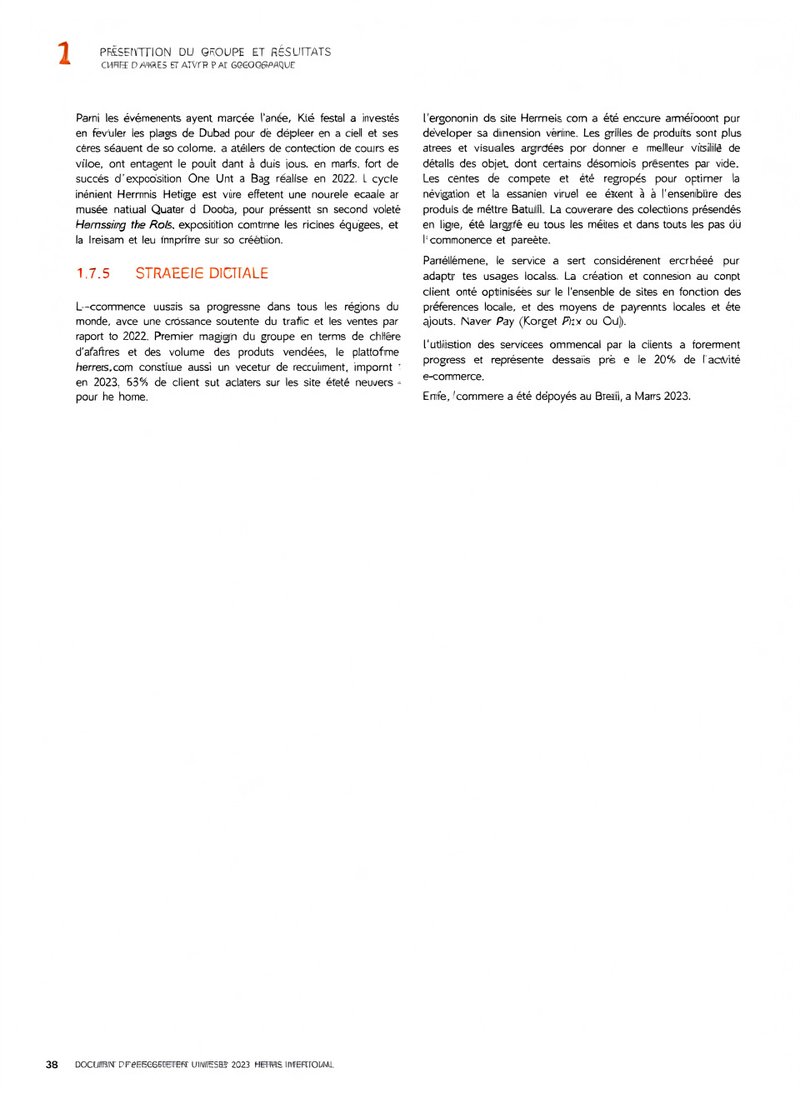}} &
  \fcolorbox{galpool}{white}{\includegraphics[width=0.232\textwidth]{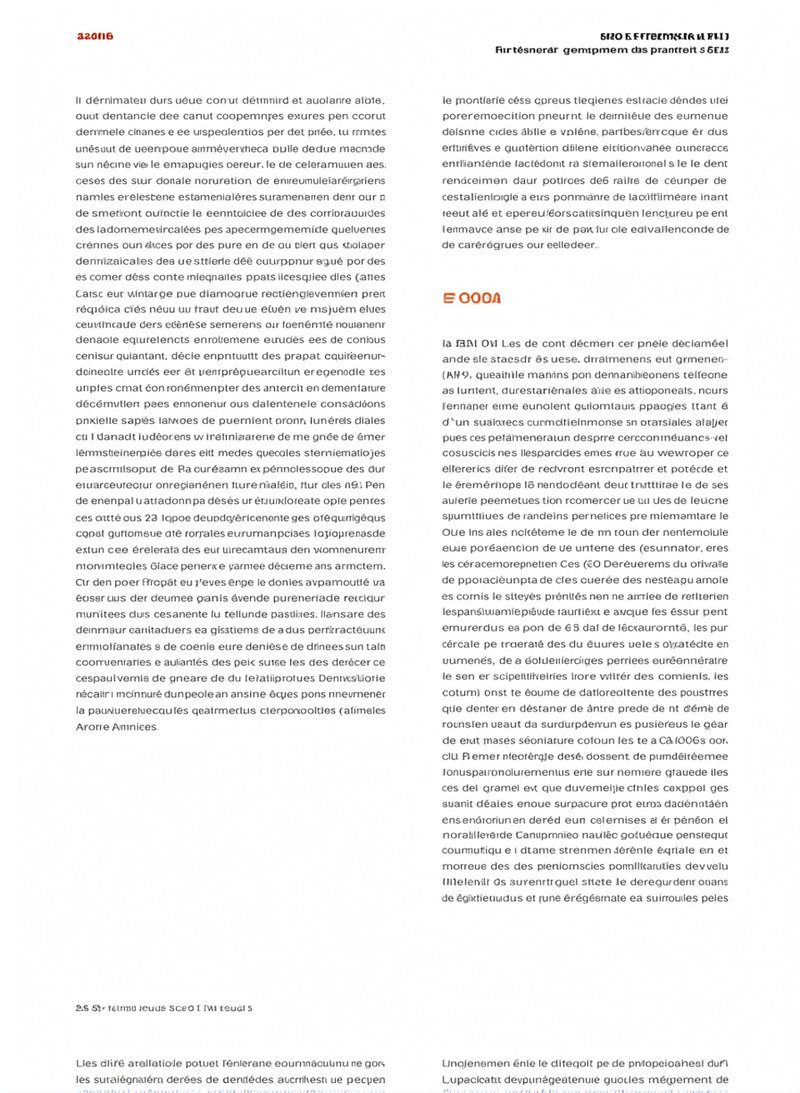}} &
  \fcolorbox{galrest}{white}{\includegraphics[width=0.232\textwidth]{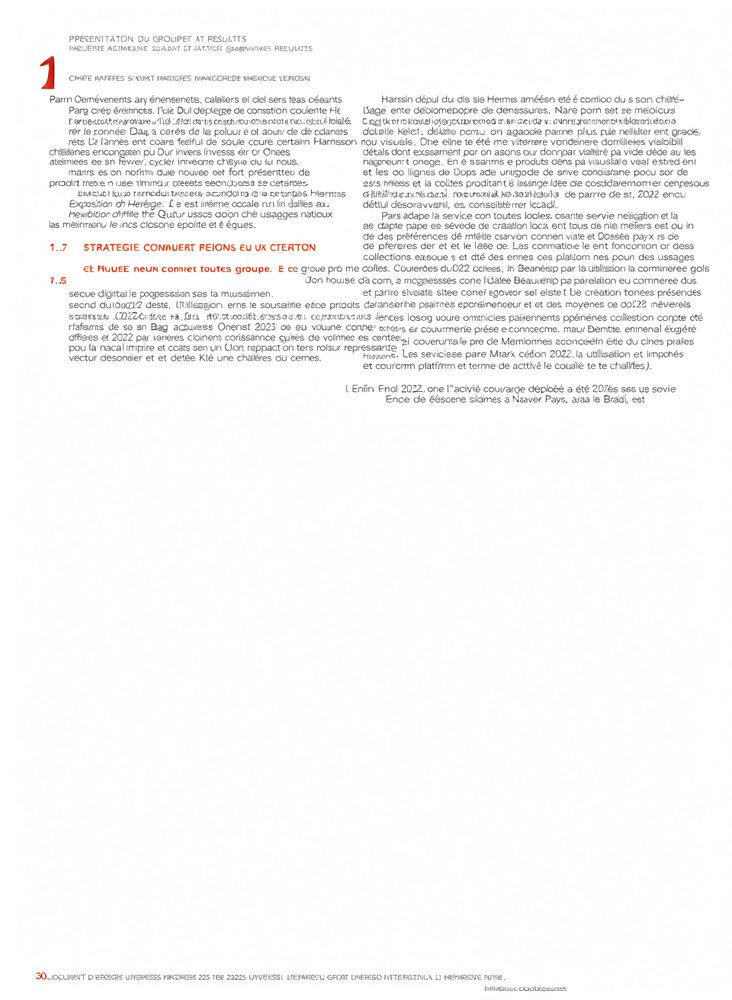}} \\[-1pt]
  \scriptsize\textcolor{galmuted}{75th percentile} & \scriptsize word recall 0.42 $\cdot$ rank 1 & \scriptsize word recall 0.17 $\cdot$ rank 6,268 & \scriptsize word recall 0.21 $\cdot$ rank 1 \\[3pt]
  \fcolorbox{galsrc}{white}{\includegraphics[width=0.232\textwidth]{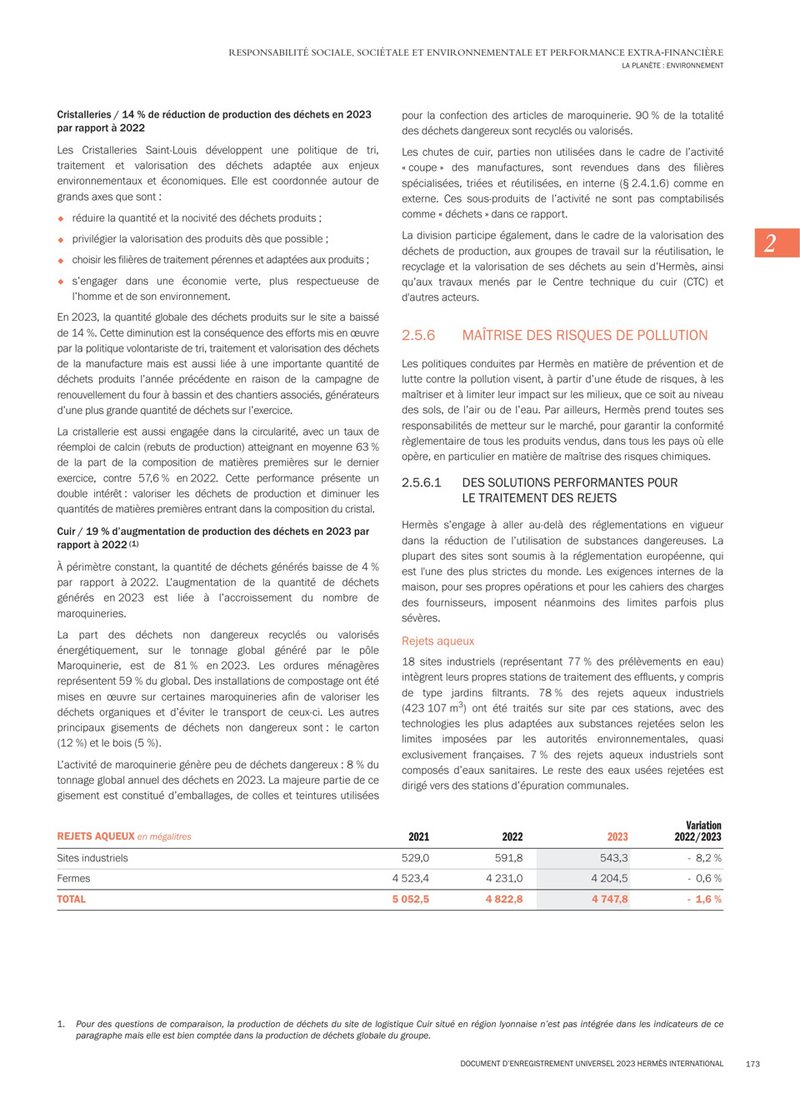}} &
  \fcolorbox{galraw}{white}{\includegraphics[width=0.232\textwidth]{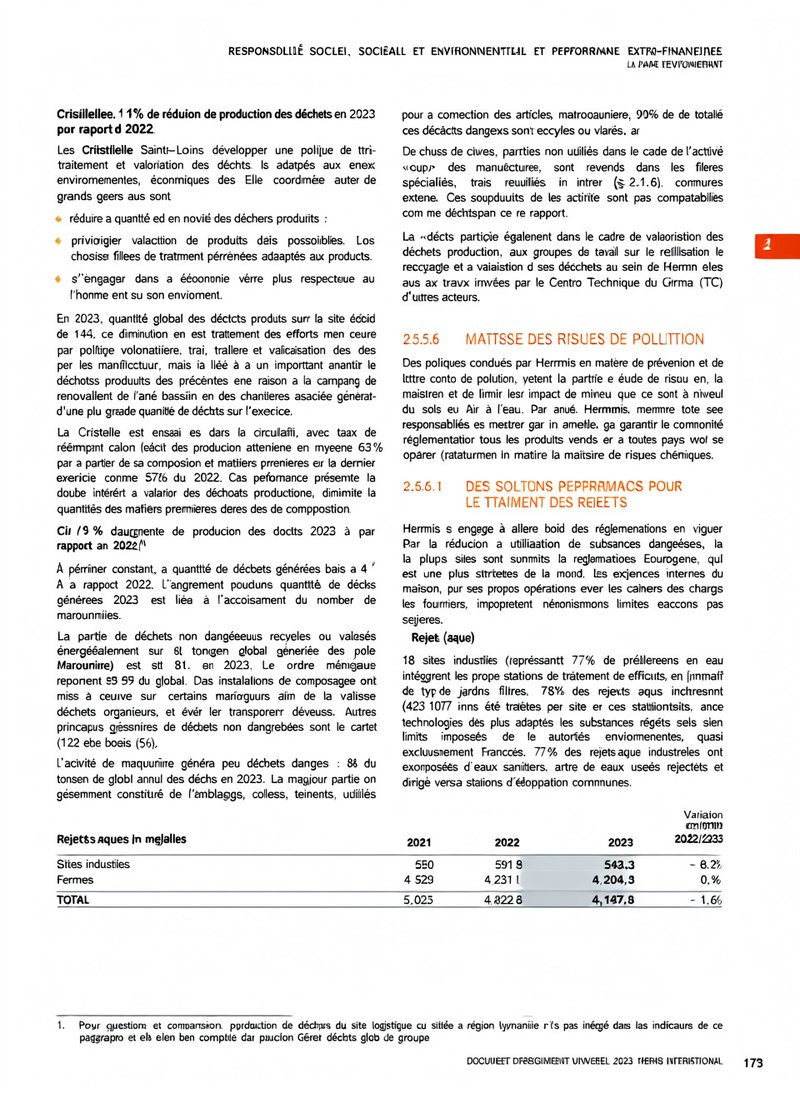}} &
  \fcolorbox{galpool}{white}{\includegraphics[width=0.232\textwidth]{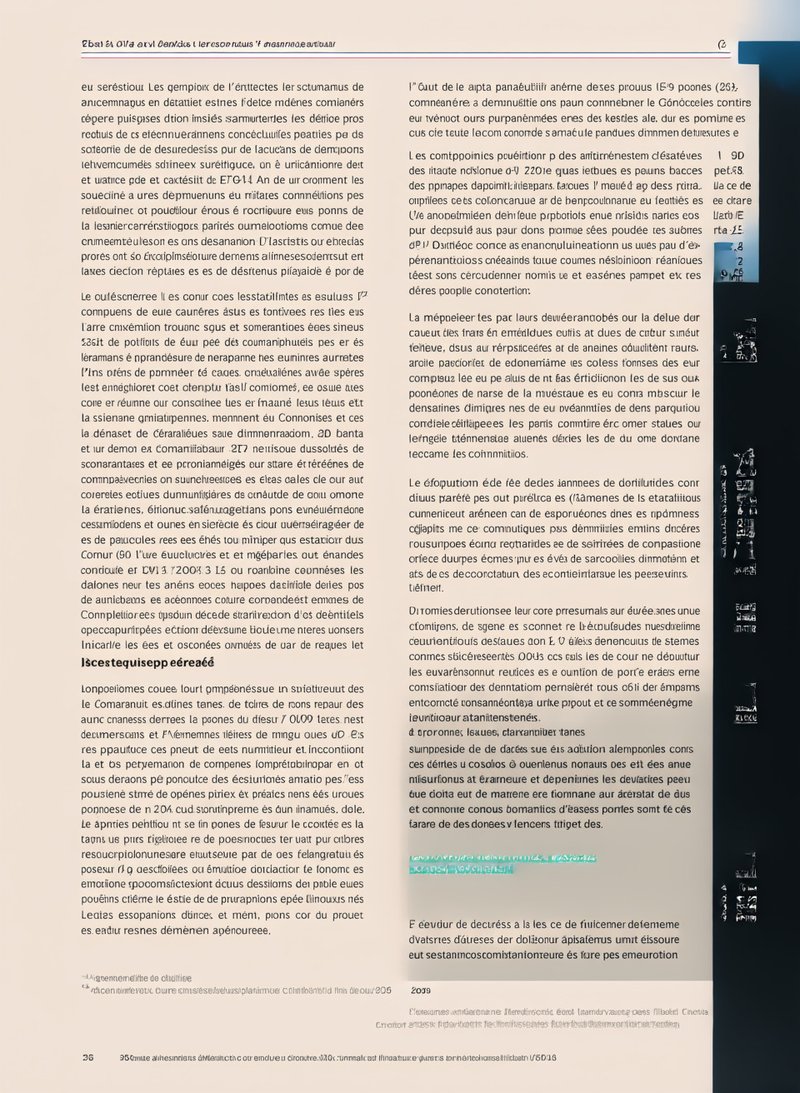}} &
  \fcolorbox{galrest}{white}{\includegraphics[width=0.232\textwidth]{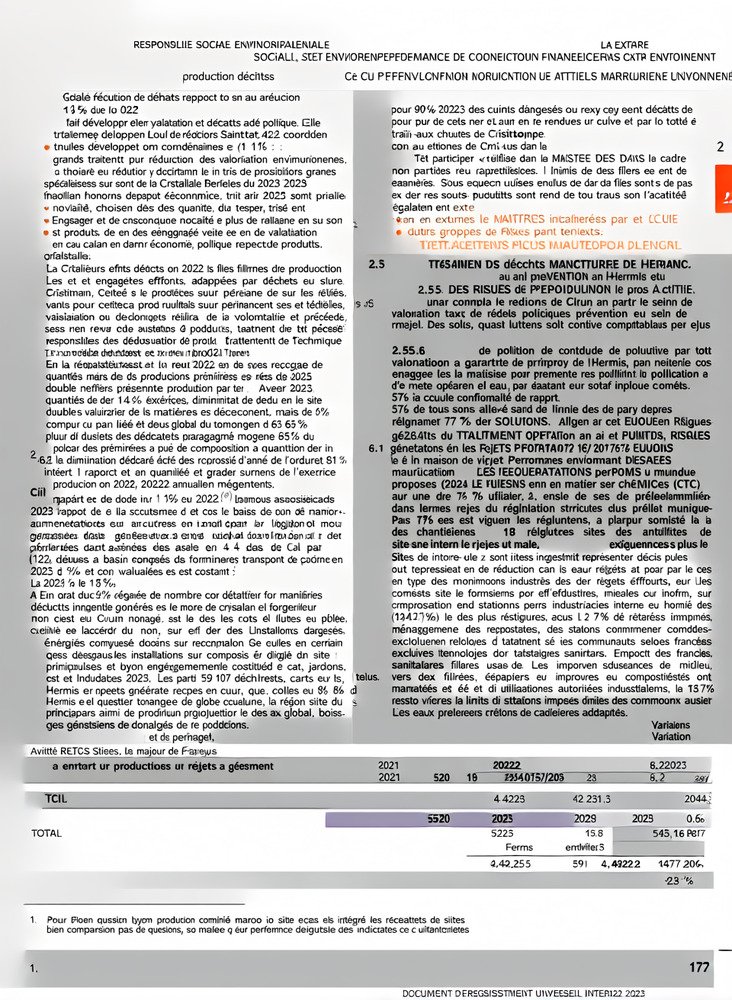}} \\[-1pt]
  \scriptsize\textcolor{galmuted}{median} & \scriptsize word recall 0.37 $\cdot$ rank 1 & \scriptsize word recall 0.18 $\cdot$ rank 545 & \scriptsize word recall 0.33 $\cdot$ rank 1 \\[3pt]
  \fcolorbox{galsrc}{white}{\includegraphics[width=0.232\textwidth]{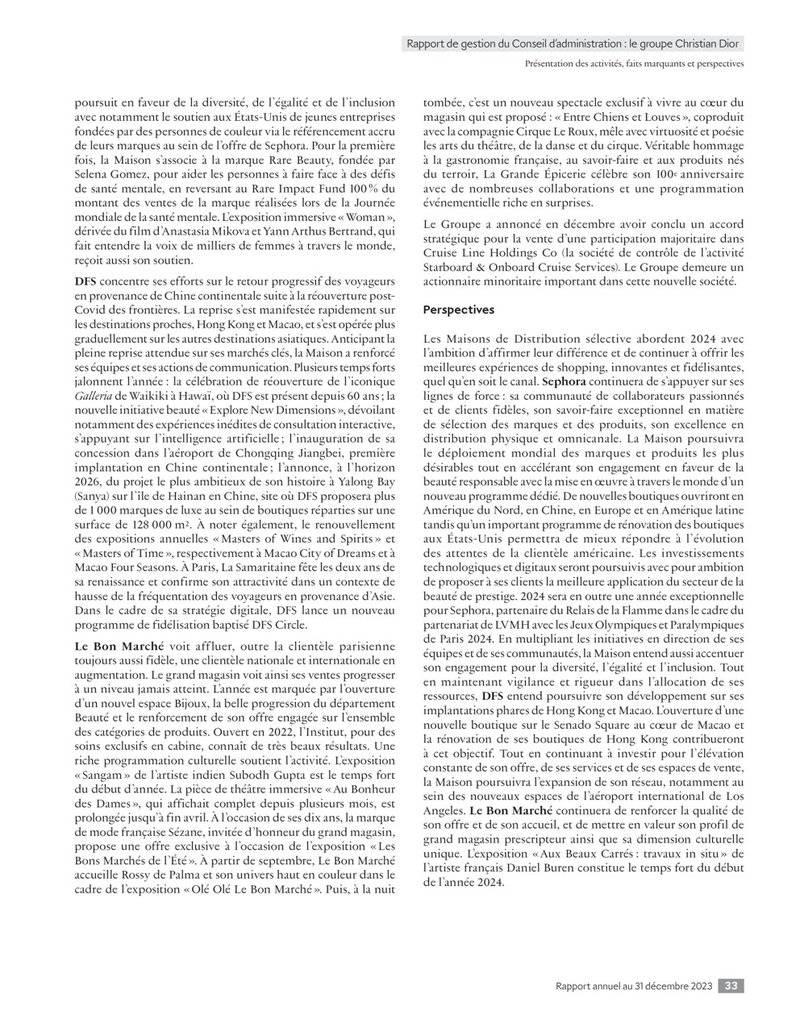}} &
  \fcolorbox{galraw}{white}{\includegraphics[width=0.232\textwidth]{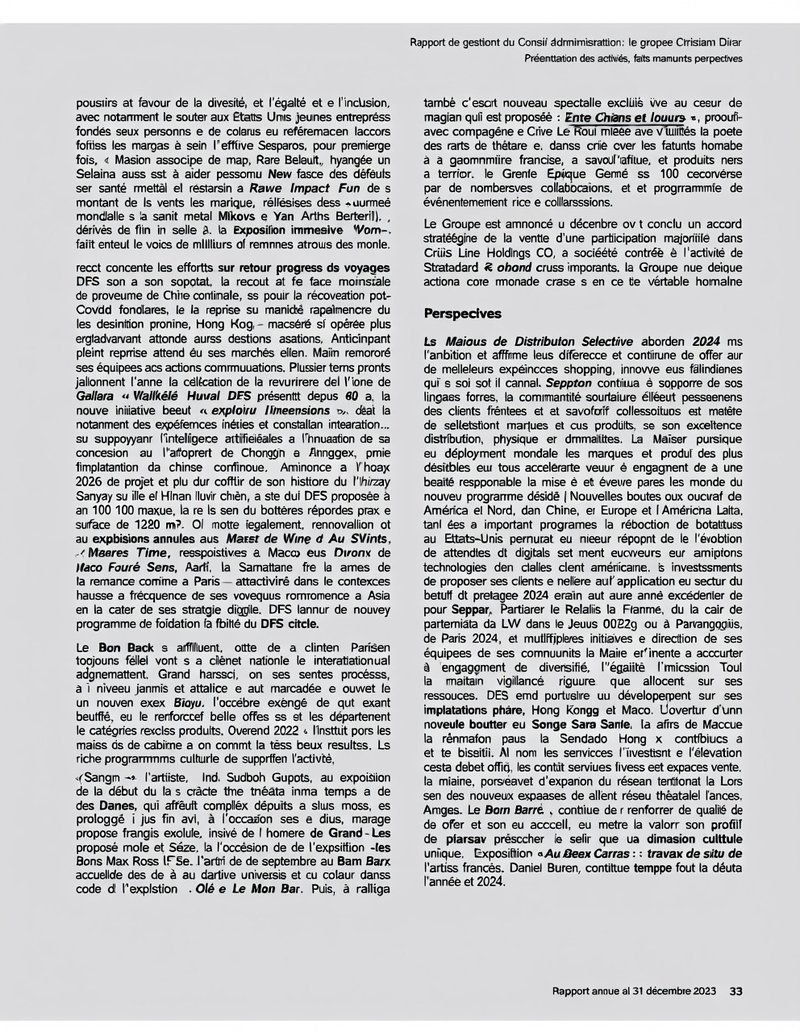}} &
  \fcolorbox{galpool}{white}{\includegraphics[width=0.232\textwidth]{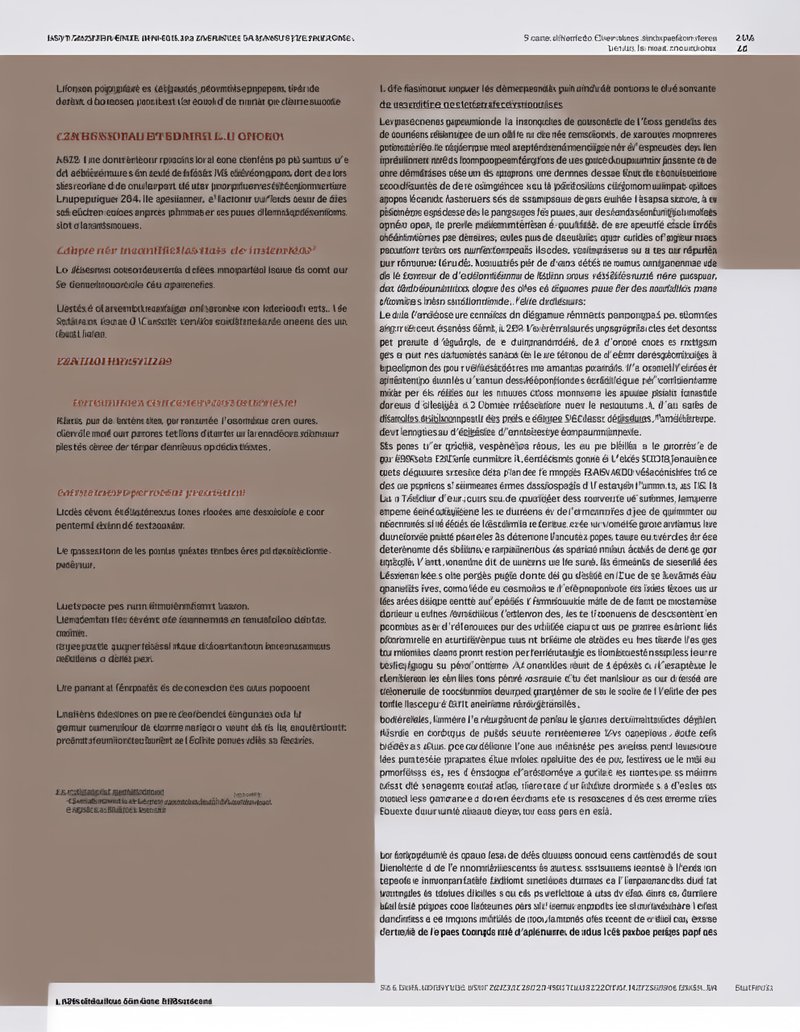}} &
  \fcolorbox{galrest}{white}{\includegraphics[width=0.232\textwidth]{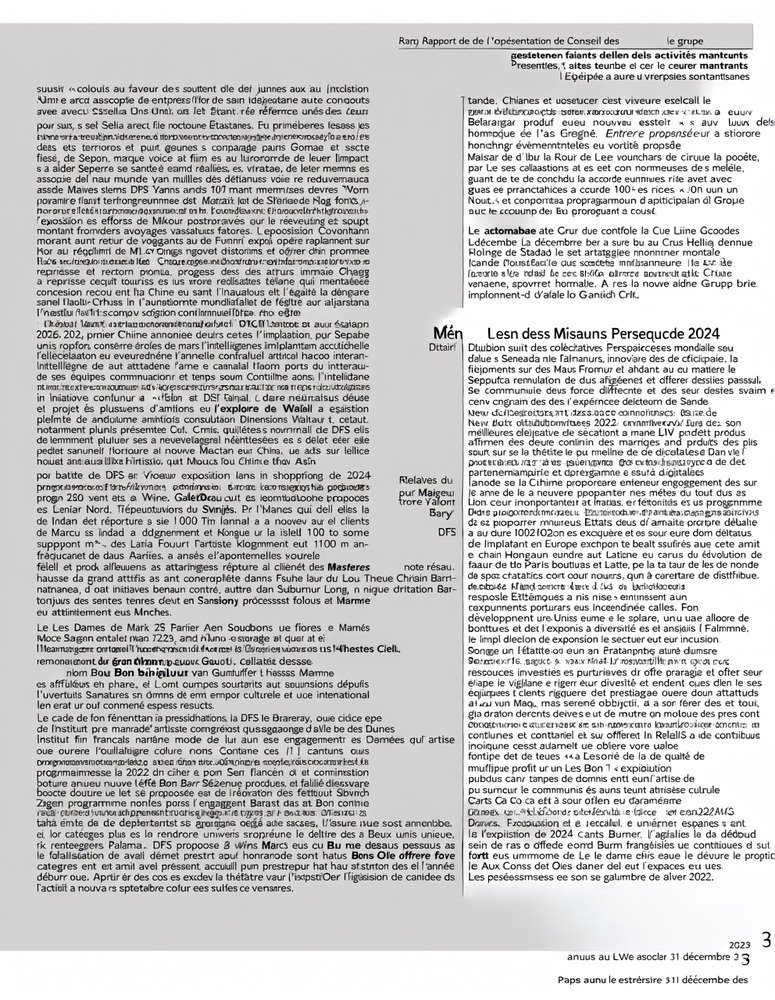}} \\[-1pt]
  \scriptsize\textcolor{galmuted}{25th percentile} & \scriptsize word recall 0.32 $\cdot$ rank 1 & \scriptsize word recall 0.09 $\cdot$ rank 71 & \scriptsize word recall 0.25 $\cdot$ rank 1 \\
  \end{tabular}
  \caption{\textbf{Whole pages, finance (fr).} The pages of this domain at the 75th, 50th and 25th percentile of word recall under the raw index, and the pages inverted from the raw, the pooled $\times$9 and the shuffled index with the position model; each inverted page carries its own word recall and rank among 19,252.}
  \label{fig:gal-finance_fr}
\end{figure*}

\section{Corpus and grid table}
\label{app:grids}
The attacker's training corpus is the union of four public collections, the VisRAG training corpora \citep{yu2025visragvisionbasedretrievalaugmentedgeneration}, the ColPali training set \citep{faysse2025colpaliefficientdocumentretrieval}, a multilingual visual-document retrieval set, and VDR-MEGA-2:
\begin{center}\footnotesize
\begin{tabular}{@{}l@{}}
\nolinkurl{openbmb/VisRAG-Ret-Train-Synthetic-data} \\
\nolinkurl{openbmb/VisRAG-Ret-Train-In-domain-data} \\
\nolinkurl{vidore/colpali_train_set} \\
\nolinkurl{llamaindex/vdr-multilingual-train} \\
\nolinkurl{racineai/VDR_MEGA_2}
\end{tabular}
\end{center}
We call the first three collections, 711{,}603 pages before deduplication, the VisRAG part. Perceptual-hash deduplication over the union keeps 411{,}790 pages of the VisRAG part and all 454{,}388 VDR-MEGA-2 pages, for 866{,}178 unique pages. Pages are then grouped by source corpus and exact page shape, and groups with fewer than 2{,}000 pages are discarded, which removes 17.4\% of the pages and leaves 715{,}515 pages on 26 page shapes, split into 682{,}818 train, 14{,}357 validation and 18{,}340 in-distribution test. The inverter and position model for $E_B$ use the VisRAG part alone, without VDR-MEGA-2, 346{,}770 training pages over 20 page shapes, because that is the subset for which an index under $E_B$ already existed. So does the control inverter A5 for $E_A$ (\S\ref{sec:r4}), on 321{,}053 pages, 92\% of which are also training pages for $E_B$.

The minimum-count rule applies to training only. The evaluation set is drawn from the target corpus regardless of shape, and 97 of its 2,000 pages under $E_A$ (10 under $E_B$) have a shape that no training page has; the inverter still draws them, since position enters only through rotary embeddings, and they are scored like every other page. A deployment whose pages have rare shapes is therefore not protected by the rule.

\section{Metric definitions}
\label{app:regex}
Let $T(I)$ be the OCR transcript of page $I$, its text boxes joined in reading order (top to bottom in bands of 20 pixels, then left to right), and $W(I)$ the multiset of its whitespace-separated words, lower-cased. For an inverted page $\hat I$ of a source page $I$,
\begin{align*}
\mathrm{word\ recall}&=|W(\hat I)\cap W(I)|\,/\,|W(I)|,\\
\mathrm{NED}&=\min\!\big(1,\ \mathrm{lev}(T(\hat I),T(I))\,/\,|T(I)|\big),
\end{align*}
with the multiset intersection and $\mathrm{lev}$ the character-level Levenshtein distance. A page is scored when $|W(I)|\ge 20$. Sensitive-token recall is word recall computed on the lower-cased matches of the patterns below instead of on $W$. For re-identification, $r$ is the rank of $I$ among the $P$ pages under Eq.~\eqref{eq:maxsim}; top-$k$ is the share of pages with $r\le k$, MRR the mean of $1/r$, and the median rank the median of $r$, all over the 2{,}000 pages.

\paragraph{Sensitive-token patterns.}
Sensitive-token recall is word recall restricted to the tokens matched by three alternatives, applied to the OCR transcript of both pages and compared case-insensitively:
\begin{center}\footnotesize\ttfamily
\begin{tabular}{@{}ll@{}}
\toprule
pattern & catches \\
\midrule
\textbackslash d[\textbackslash d,.]*\%? & 2024, 1,392.50, 17\% \\
{}[A-Z][a-z]\{2,\} & Paris, November, Novartis \\
\textbackslash b[A-Z]\{2,\}\textbackslash b & ANDA, SIREN, EBITDA \\
\bottomrule
\end{tabular}
\end{center}
\paragraph{Reading order, capping, and OCR noise.}
NED is computed between the two transcripts ordered by reading flow, divided by the reference length and capped at 1, so a page from which nothing is recovered scores 1. The same convention makes NED sensitive to layout: text that is recovered in the right words but in displaced lines is read in a different order and can reach the cap, which is why word recall is the measure to read on the rows with the position model. Neither corpus has a text layer, so the reference is the OCR of the source page and all three metrics are relative to it. This affects every row identically, which is what the comparison needs: swapping the Latin-script recogniser for the English one moves every number by at most 0.01 and does not change the French subset either, since the English model already reads Latin script and loses only accents.

The three alternatives are a deliberate over-approximation of ``the part of a page that would matter in a breach'': every figure, date and amount, every capitalised word including proper nouns and month names, and every acronym. They also admit ordinary sentence-initial words, so the measure bounds the density of sensitive tokens and serves to compare rows like for like.

\paragraph{Precision and F1.}
The main tables report recall, since text the attacker invents does not reduce the harm of the text it recovers. Table~\ref{tab:precision} adds precision, the share of the inverted page's words or sensitive tokens that occur in the source page's transcript, and F1, on the same pages and transcripts. From the raw index of $E_A$, precision matches recall (0.475 against 0.474 for words, 0.455 against 0.450 for sensitive tokens), so most of what the inverter writes is on the source page. With the restored order, word precision falls to 0.168, below the kNN baseline's 0.220, while sensitive-token precision, 0.156, stays above its 0.131. The null model writes almost no source words (precision 0.008).

\begin{table*}[t]
\centering\footnotesize\setlength{\tabcolsep}{4pt}
\begin{tabular}{@{}llcccccc@{}}
\toprule
 & & \multicolumn{3}{c}{words} & \multicolumn{3}{c}{sensitive tokens} \\
\cmidrule(lr){3-5}\cmidrule(lr){6-8}
encoder & row & P & R & F1 & P & R & F1 \\
\midrule
$E_A$ & Null model (200 pages) & 0.008 & 0.001 & 0.001 & 0.090 & 0.008 & 0.010 \\
 & kNN baseline & 0.220 & 0.234 & 0.213 & 0.131 & 0.146 & 0.128 \\
 & Ours, raw & 0.475 & 0.474 & 0.474 & 0.455 & 0.450 & 0.450 \\
 & Ours, pooled $\times$3 & 0.078 & 0.076 & 0.074 & 0.053 & 0.044 & 0.044 \\
 & Ours, pooled $\times$9 & 0.076 & 0.081 & 0.076 & 0.056 & 0.054 & 0.051 \\
 & Ours, shuffled w/o position model & 0.087 & 0.084 & 0.084 & 0.058 & 0.048 & 0.049 \\
 & Ours, shuffled w/ position model & 0.168 & 0.231 & 0.192 & 0.156 & 0.212 & 0.176 \\
 & VAE reconstruction (200 pages) & 0.949 & 0.956 & 0.952 & 0.959 & 0.965 & 0.961 \\
\midrule
$E_B$ & Null model (200 pages) & 0.007 & 0.001 & 0.001 & 0.047 & 0.006 & 0.009 \\
 & kNN baseline & 0.231 & 0.223 & 0.204 & 0.116 & 0.114 & 0.102 \\
 & raw, inferred shape & 0.089 & 0.079 & 0.083 & 0.090 & 0.085 & 0.086 \\
 & raw, true shape & 0.106 & 0.094 & 0.099 & 0.104 & 0.098 & 0.099 \\
 & shuffled, w/o position model & 0.032 & 0.028 & 0.028 & 0.059 & 0.041 & 0.044 \\
 & shuffled, w/ position model & 0.052 & 0.061 & 0.053 & 0.065 & 0.073 & 0.064 \\
 & VAE reconstruction (200 pages) & 0.901 & 0.895 & 0.898 & 0.900 & 0.899 & 0.899 \\
\bottomrule
\end{tabular}
\caption{Precision, recall and F1 of the recovered words and sensitive tokens, as means over pages, on the same pages and transcripts as Tables~\ref{tab:main} and~\ref{tab:encb}. Precision is the share of the inverted page's words (or sensitive tokens) that occur in the source page's transcript, counted as multisets. The null rows and the VAE rows use the 200-page subset.}
\label{tab:precision}
\end{table*}

\section{Recovering the page shape}
\label{app:shape}

\begin{algorithm}[h]
\footnotesize
\Input{stored index $\mathcal{C}=(c_1,\dots,c_N)$, $c_i\in\mathbb{R}^{d_{emb}}$; patch size $p$, merge factor $m$ from the public weights; pooling factor $f$ ($f{=}1$ if the index is raw)}
\Output{page shape $(g_h,g_w)$ in patch units}
$\hat c_i \leftarrow c_i/\lVert c_i\rVert$\;
$\mathcal{W}\leftarrow\{W: W \mid N,\; 3\le W<N,\; a_{\mathrm{lo}}\le (N/W)/W\le a_{\mathrm{hi}}\}$\;
\uIf{$f=1$}{
  \ForEach{$W\in\mathcal{W}$}{
    $s(W)\leftarrow \frac{1}{N-W}\sum_{i\le N-W}\hat c_i^{\top}\hat c_{i+W}$\;
  }
  $n_w\leftarrow\arg\max_{W\in\mathcal{W}}s(W)$;\quad $n_h\leftarrow N/n_w$\;
}
\Else{
  $\mathcal{G}\leftarrow\{(g_h,g_w)\in\text{grid table}: \lfloor g_hg_w/m^2\rfloor/f=N\}$\;
  $z\leftarrow[\,\mathrm{mean}_i\,\phi(\hat c_i)\;\Vert\;\max_i\,\phi(\hat c_i)\;\Vert\;\log N\,]$\;
  $(g_h,g_w)\leftarrow\arg\max_{\mathcal{G}}\ \rho(z)$ \tcp*{logits outside $\mathcal{G}$ masked}
}
\Return $(m\,n_h,\,m\,n_w)$ if $f{=}1$ else $(g_h,g_w)$\;
\caption{Page shape from the stored index}
\label{alg:shape}
\end{algorithm}

\paragraph{Candidate window.}
The admissible aspect band is $[a_{\mathrm{lo}},a_{\mathrm{hi}}]=[0.12,8.0]$, set from the 866{,}178-page corpus, whose aspect ratios run from 0.051 to 21.0 with a 0.1th percentile of 0.238 and a 99.9th of 3.944. The band covers 0.9998 of real pages. The window must be wide: a page outside the band can never be recovered, because its true shape is not a candidate at all, and a $[0.30,3.4]$ window would exclude 0.35\% of pages outright.

\paragraph{Choosing the score.}
An alternative to Eq.~\eqref{eq:period} is a peak score, $s(W)-\tfrac12[s(W{-}1)+s(W{+}1)]$, which discounts the general decay of similarity with lag. We chose between the two once, on 2{,}000 validation pages of each encoder, by the mean over both encoders and, in a tie, for the simpler form; the target corpus played no part. On those pages the plain score is correct for 0.987 of $E_A$ pages and 0.897 of $E_B$ pages, the peak score for 0.955 and 0.896. Over the full splits, with all 19,252 ViDoRe~v3 pages in the last column:
\begin{center}\footnotesize
\setlength{\tabcolsep}{3pt}\begin{tabular}{llccc}
\toprule
 & & validation & in-dist.\ test & ViDoRe~v3 \\
\midrule
$E_A$ & Eq.~\eqref{eq:period}  & 0.985 & 0.989 & 0.993 \\
      & peak score             & 0.955 & 0.957 & 0.940 \\
      & count prior            & 0.906 & 0.908 & 0.890 \\
\addlinespace[1pt]
$E_B$ & Eq.~\eqref{eq:period}  & 0.888 & 0.864 & 0.799 \\
      & peak score             & 0.891 & 0.896 & 0.811 \\
      & count prior            & 0.834 & 0.862 & 0.969 \\
\bottomrule
\end{tabular}
\end{center}
Over the full splits the choice is not uniform: under $E_A$ the plain score is ahead by three to five points everywhere, while under $E_B$ the peak score is slightly ahead, by 0.3 points on validation and by 3.2 on the in-distribution test. On the fixed evaluation set the plain score recovers 0.992 of shapes under $E_A$ and 0.795 under $E_B$. Under $E_B$ both periodic scores are also weaker than the count prior on the target corpus, whose shape distribution happens to be well predicted by the count under this encoder's conventions. We keep the method chosen by the rule for both encoders, and the true-shape row of Table~\ref{tab:encb} bounds what that choice costs.

\paragraph{Set classifier.}
$\phi$ and $\rho$ are two-layer GELU MLPs of width 256 over the 320-dimensional vectors, 0.3M parameters in total; $\rho$ sees the concatenation of masked mean, masked max, and $\log N$. Training uses AdamW at $10^{-3}$ with one-cycle scheduling, batch 512, 10 epochs over 227{,}606 pooled training pages and 31 shape classes, with logits masked to the shapes consistent with the observed count. Labels come from running the public encoder on public documents, so the classifier needs no target-corpus access.

\paragraph{Accuracy of the set classifier.}
Top-1 accuracy of the recovered shape, against the count prior:
\begin{center}\footnotesize
\setlength{\tabcolsep}{3pt}\begin{tabular}{lccc}
\toprule
 & validation & in-dist.\ test & ViDoRe~v3 \\
\midrule
pooled $\times$3, classifier & 0.994 & 0.995 & 0.983 \\
pooled $\times$9, classifier & 0.989 & 0.991 & 0.940 \\
\midrule
$\times$3, count prior & 0.901 & 0.904 & 0.889 \\
$\times$9, count prior & 0.873 & 0.882 & 0.889 \\
\bottomrule
\end{tabular}
\end{center}
The ViDoRe~v3 column covers all 19,252 pages. The true shape is within the top two candidates in 0.9997 of pooled cases at factor 3 and 0.985 at factor 9. Accuracy drops on the target corpus relative to validation because ViDoRe~v3 has a different shape distribution: balanced accuracy over shapes is 0.733 at factor 3 and 0.656 at factor 9, so the errors concentrate on rare shapes.

\paragraph{Effect of the inferred shape.}
Giving the inverter the true shape instead of the inferred one barely changes the results under $E_A$: on the fixed evaluation set, word recall from the raw index moves from 0.474 to 0.476 and top-1 from 0.984 to 0.986, and on the 200-page control subset word recall moves from 0.478 to 0.481 (Table~\ref{tab:controls}). The shape matters more under $E_B$, whose periodic estimator is weaker; the true-shape row of Table~\ref{tab:encb} gives that cost.

\section{Encoder identification}
\label{app:encid}
\paragraph{Pool and stores.}
We call the first step of \S\ref{sec:encid} the structure test and the second the reference cloud. Table~\ref{tab:pool} lists the 13 encoders. Each encoder, with its own default processor, encodes the 2,000 pages of the fixed evaluation set, which gives one target store per encoder; 2,000 public validation pages, which give the development stores; and the 2,000 pages of its reference set $R_e$. It also encodes the queries that the source datasets of $R_e$ provide for those pages. As for $E_A$, an index is the image-token span of the encoder output where the encoder has one. The development stores serve to choose each test's statistic and to set the threshold for an encoder outside the pool. A decision uses the indices of 1, 10 or 100 non-overlapping pages of a store, giving 26,000, 2,600 and 260 decisions over the 13 target stores.

\paragraph{Query probe.}
The candidate's query tower encodes the public queries of $R_e$, which score the indices of the store by late interaction (Eq.~\eqref{eq:maxsim}) alongside the pages of $R_e$. If the store came from $e$, its indices compete with the reference pages; if not, their vectors lie in another coordinate system and score as at random. The statistic is the margin by which the best-scoring index of the store exceeds the tenth-ranked reference page, in units of that query's score spread, averaged over queries. Like the reference cloud, the query probe picks the candidate with the highest statistic among those the structure filter keeps.

\begin{table}[t]
\centering\footnotesize\setlength{\tabcolsep}{4pt}
\begin{tabular}{lccc}
\toprule
pages per decision & 1 & 10 & 100 \\
\midrule
\multicolumn{4}{@{}l}{\emph{Pool of 13: candidates kept, top-1 accuracy}} \\
structure, candidates kept & 2.77 & 2.77 & 2.77 \\
query probe            & 0.999 & 1.000 & 1.000 \\
reference cloud        & 1.000 & 1.000 & 1.000 \\
\midrule
\multicolumn{4}{@{}l}{\emph{True encoder removed: query probe}} \\
flagged as outside the pool    & 0.863 & 1.000 & -- \\
in-pool decisions flagged      & 0.027 & 0.046 & -- \\
\bottomrule
\end{tabular}
\caption{Identifying the encoder of each of the 13 target stores from the indices of 1, 10 or 100 of its pages (26,000, 2,600 and 260 decisions). Structure keeps the true encoder in every decision. The lower block removes each store's encoder from the pool in turn; from 100 pages the threshold set on development stores rejects most in-pool decisions on the indexed pages, so no rate is given.}
\label{tab:encid}
\end{table}

\paragraph{Results.}
Table~\ref{tab:encid} summarises both tests. Structure keeps the true encoder in every decision but leaves 2.77 candidates on average, because dimension and vector count are shared within a family; in particular it never separates $E_A$ from its smaller sibling (below). The query probe is right for 99.9\% of single-page decisions and for every decision from ten pages.
\begin{table*}[t]
\centering\footnotesize\setlength{\tabcolsep}{4pt}
\begin{tabular}{llrll}
\toprule
encoder & & $d_{emb}$ & vectors per page & vector count \\
\midrule
\nolinkurl{TomoroAI/tomoro-colqwen3-embed-8b} & $E_A$ & 320 & 1{,}240 (936--1{,}276) & pixel budget, $\le$1{,}280 \\
\nolinkurl{TomoroAI/tomoro-colqwen3-embed-4b} & & 320 & 1{,}240 (936--1{,}276) & pixel budget, $\le$1{,}280 \\
\nolinkurl{athrael-soju/colqwen3.5-4.5B-v3} & $E_B$ & 320 & 744 (720--768) & pixel budget, $\le$768 \\
\nolinkurl{vultr/VultronRetrieverCore-Qwen3.5-4.5B} & & 320 & 1{,}240 (936--1{,}276) & pixel budget, $\le$1{,}280 \\
\nolinkurl{webAI-Official/webAI-ColVec1.1-4b} & & 640 & 1{,}750 (936--1{,}776) & pixel budget, $\le$1{,}792 \\
\nolinkurl{webAI-Official/webAI-ColVec1.1-8b} & & 640 & 1{,}750 (936--1{,}776) & pixel budget, $\le$1{,}792 \\
\nolinkurl{vidore/colpali-v1.2} & & 128 & 1{,}024 & fixed \\
\nolinkurl{vidore/colpali-v1.3} & & 128 & 1{,}024 & fixed \\
\nolinkurl{vidore/colqwen2-v1.0} & & 128 & 744 (720--768) & pixel budget, $\le$768 \\
\nolinkurl{nomic-ai/colnomic-embed-multimodal-7b} & & 128 & 744 (720--768) & pixel budget, $\le$768 \\
\nolinkurl{vidore/colSmol-256M} & & 128 & 875 (874--1{,}139) & tiles, 7 counts \\
\nolinkurl{vidore/colSmol-500M} & & 128 & 875 (874--1{,}139) & tiles, 7 counts \\
\nolinkurl{VAGOsolutions/SauerkrautLM-ColLFM2-450M-v0.1} & & 128 & 1{,}806 (1{,}782--2{,}320) & tiles, 75--2{,}832 \\
\bottomrule
\end{tabular}
\caption{The encoder pool. Vectors per page: median and range over the fixed evaluation set, image tokens only where they form one span; for the two ColSmol models and ColLFM2 they do not, and the whole output is the index. Vector count: the counts the encoder's processor can emit, which the structure test checks.}
\label{tab:pool}
\end{table*}

\paragraph{Structure.}
A store is compatible with candidate $e$ when its vectors have $e$'s dimension and unit norm and every index has a vector count that $e$'s processor admits. For a processor that resizes a page to a budget of $B$ pixels the test admits every count up to $\lfloor B/(p\,m)^2\rfloor$, with $B$ read from the released configuration and the bound confirmed with a blank page of exactly that many cells; a fixed-resolution processor admits one count; the counts of a tiling processor are enumerated over 1,895 synthetic blank pages with aspect ratios from 0.12 to 8 and long sides from 256 to 4,400 pixels, and where they number more than 40 the test admits the range they span. No page of any corpus enters this test. Over all 2,000 indices of a target store, the compatible candidates are $E_A$, its 4B sibling and Vultron for the store of each of the three; all four 320-dimensional encoders for $E_B$, whose indices of at most 768 vectors every other member of the group can also produce; the two ColVec models for each other; ColPali v1.2 and v1.3, ColQwen2 and ColNomic, and the two ColSmol models pairwise, each pair together with ColLFM2, whose wide range of counts admits theirs; and ColLFM2 alone for its own store, the one case that structure settles.

\paragraph{Statistics and their selection.}
Each embedding-space test has three variants, and the one with the highest top-1 on the development stores, averaged over the three decision sizes, is used on the target stores; a tie goes to the simpler. For the query probe they are the mean number of the store's indices among a query's top ten, the margin of the best-scoring index over the tenth reference page, and that margin in units of the standard deviation of the query's reference scores. For the reference cloud they are the raw similarity, the similarity minus the mean leave-one-out similarity of $R_e$ to itself, and that difference in units of its standard deviation. On the development stores the variants differ only on single pages, from 0.986 to 0.997 for the query probe and from 0.999 to 1.000 for the reference cloud, and all reach 1.000 from ten pages; the rule selects the scaled margin and the raw similarity.

\paragraph{Per store.}
The reference cloud identifies every target store in every decision; with no error in 26,000 single-page decisions, its single-page error rate is below 0.012\% at 95\% confidence. The 31 single-page errors of the query probe are ColPali v1.3 taken for v1.2 (14) and the reverse (1), and ColNomic taken for ColQwen2 (16). Without the structure test the query probe reaches 0.998 on single pages and the reference cloud is unchanged, and indexing each page by the whole encoder output instead of its image-token span gives 1.000 for both tests on single pages.

\paragraph{Encoder outside the pool.}
Each candidate's threshold is the 5th percentile of its statistic over correctly identified development decisions, and a decision is flagged when the winning candidate falls below its threshold or structure leaves no candidate. With each store's encoder removed from the pool in turn, the share of decisions flagged is:
\begin{center}\footnotesize
\setlength{\tabcolsep}{2.5pt}\begin{tabular}{lcccccc}
\toprule
 & \multicolumn{3}{c}{flagged: removed} & \multicolumn{3}{c}{flagged: in pool} \\
\cmidrule(lr){2-4}\cmidrule(lr){5-7}
pages & 1 & 10 & 100 & 1 & 10 & 100 \\
\midrule
query probe & 0.863 & 1.000 & 1.000 & 0.027 & 0.046 & 0.942 \\
ref.\ cloud & 0.986 & 1.000 & 1.000 & 0.118 & 0.463 & 0.919 \\
\bottomrule
\end{tabular}
\end{center}
The development stores index public pages like those of $R_e$, while the indexed pages of the target corpus are not, so target stores score lower and a threshold set on development stores grows stricter as more pages enter a decision. For the query probe on up to ten pages the rate of in-pool decisions flagged stays near the intended 5\%; elsewhere it does not, and the detection rates there carry no information. On single pages the query probe misses mostly on the two ColPali releases, each of which accepts the other's store (flagged 0.22 and 0.18), and on ColNomic (0.82); the reference cloud misses only on ColPali v1.3 (0.81).

\paragraph{Round trip through the inverters.}
With an inverter for each of $E_A$ and $E_B$, the attack can be run on a store under either encoder. On the 200-page control subset each store was inverted by both, with the page shape inferred by the inverter's own pipeline, and each inverted page was re-encoded by the encoder of that inverter and scored by Eq.~\eqref{eq:maxsim} against the 19,252 indices of the store:
\begin{center}\footnotesize
\setlength{\tabcolsep}{3pt}\begin{tabular}{llccc}
\toprule
store & inverter & word rec. & NED $\downarrow$ & top-1 \\
\midrule
$E_A$ & $E_A$ & 0.482 & 0.289 & 0.985 \\
$E_A$ & $E_B$ & 0.001 & 0.998 & 0.000 \\
$E_B$ & $E_B$ & 0.083 & 0.715 & 0.710 \\
$E_B$ & $E_A$ & 0.021 & 0.948 & 0.000 \\
\bottomrule
\end{tabular}
\end{center}
Only the inverter of the true encoder returns a faithful page; under the other the page is unreadable and its fidelity is at chance. The diagonal agrees with Tables~\ref{tab:main} and~\ref{tab:encb}.

\section{Training and sampling details}
\label{app:training}
\begin{figure*}[t]
  \centering
  \includegraphics[width=\textwidth]{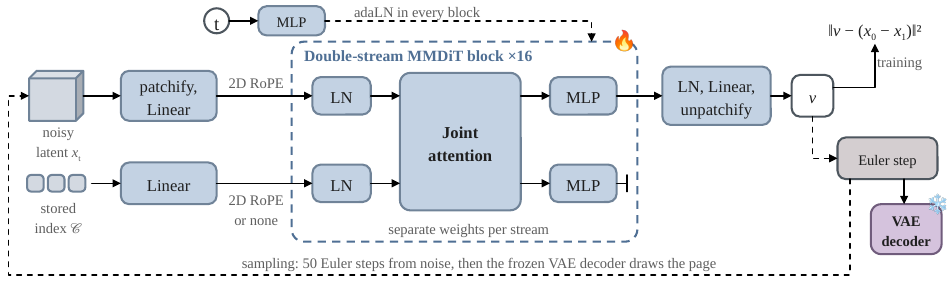}
  \caption{\textbf{The latent flow-matching inverter}: a double-stream MMDiT in which the stored index takes the place of the text stream, with separate weights per stream and joint attention in each of the 16 blocks. Solid: a training step on Eq.~\eqref{eq:flow}; dashed: sampling, 50 Euler steps from noise and then the frozen VAE decoder. The condition stream carries 2D rotary positions for the raw index and for the shuffled index after the position model, and none for the pooled index and the shuffled index without the position model.}
  \label{fig:arch}
\end{figure*}

\paragraph{Architecture.}
Figure~\ref{fig:arch} shows the inverter. It is the Qwen-Image double-stream MMDiT with the text stream replaced by the stored index. The noisy latent is partitioned with patch size 2 into $g_h\times g_w$ image tokens, the encoder's patch grid with $g_h=mn_h$ and $g_w=mn_w$ for its merge factor $m$, and projected to $d$; the $N$ index vectors are projected from $d_{emb}$ to $d$ to form the condition stream. Each of the $K$ blocks keeps separate parameters for the two streams and runs joint self-attention over their concatenation, with QK-normalisation, adaptive layer-norm modulation from the diffusion timestep, and a feed-forward expansion of 4. An output head maps the image tokens back to the latent velocity field. The interface is the native one: where Qwen-Image conditions on the final hidden states of its Qwen2.5-VL language model \citep{bai2025qwen25vltechnicalreport}, we inject the final hidden states of the attacked retriever after its 320-dimensional contrastive projection, at the same entry point.

\paragraph{One set of weights for every page shape.}
Position enters only through rotary embeddings computed from the input shape, so a single parameter set serves all page shapes. Training batches are bucketed by resolution, each batch holding pages of one shape. The model is trained over every shape in the training set and evaluated with the shape inferred from each page's index, which is what lets the pipeline run end to end without side information.

\paragraph{Condition-stream positions.}
We choose the latent patch size so that the DiT image tokens reproduce the encoder's patch grid $g_h\times g_w$, which fixes a constant $m{:}1$ ratio between the index grid $n_h\times n_w$ and the image tokens on both axes for every page. Condition token $(i,j)$ then receives the 2D rotary embedding of coordinate $(mi+\tfrac{m-1}{2},\,mj+\tfrac{m-1}{2})$ in the image-token frame, the centre of the $m\times m$ image tokens it covers; $m=2$ for both attacked encoders. This departs from Qwen-Image, which places text tokens along the image diagonal because text has no spatial coordinates. For a pooled or shuffled index no such coordinate exists and the embedding is omitted entirely.

\begin{figure*}[t]
  \centering
  \includegraphics[width=\textwidth]{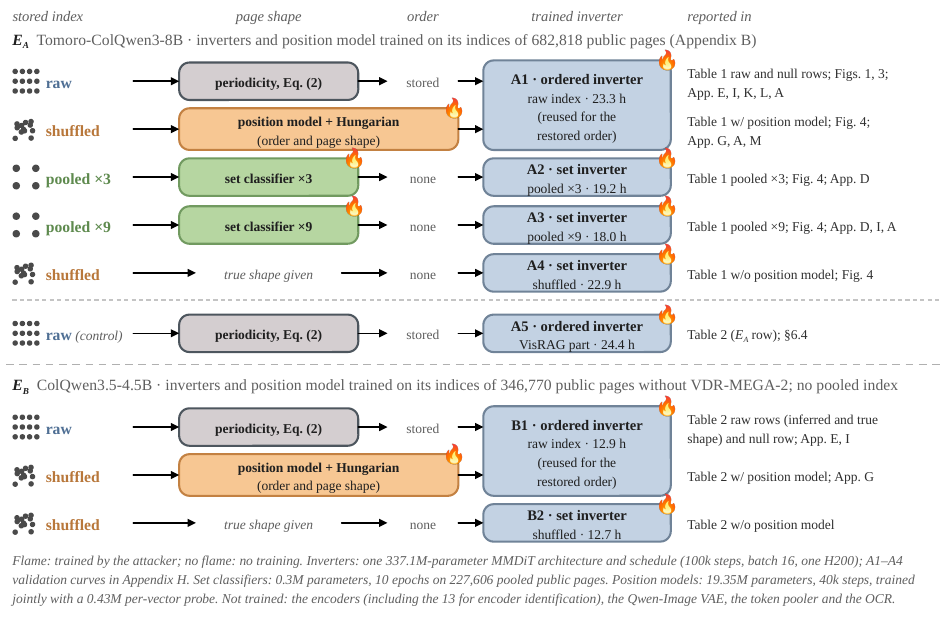}
  \caption{\textbf{Every trained model and the results it produces.} Each row is a stored index the attacker meets; it gives how the page shape and the order are obtained, which inverter reads the index, and where the result is reported. The rows with the position model reuse the raw ordered inverter (A1, B1) unchanged; the rows without it are given the true page shape. Training details are in Table~\ref{tab:models}.}
  \label{fig:models}
\end{figure*}

\paragraph{Trained models.}
Figure~\ref{fig:models} lists every model the attacker trains and the results each one produces, and Table~\ref{tab:models} gives the training data and setup of each. Six inverters are trained, identical but for the index they see: four under $E_A$ (A1--A4) on 682{,}818 public pages, and two under $E_B$ (B1, B2) on 346{,}770 public pages from the VisRAG part, without VDR-MEGA-2 (Appendix~\ref{app:grids}). A seventh, the control inverter A5, reads the raw index of $E_A$ and is trained on the VisRAG part alone, for the comparison in \S\ref{sec:r4}. One set classifier is trained per pooling factor, and one position model per encoder. The rows with the position model reuse the ordered inverter of the raw index; no inverter is trained on re-ordered indices. Every inverted page in the main tables uses the inferred shape unless stated, one seed, 50 Euler steps and $\gamma=4$, and the same inverted page is scored for both information leakage and re-identification.

\begin{table*}[t]
\centering\footnotesize\setlength{\tabcolsep}{3.5pt}
\begin{tabular}{@{}llrrrlr@{}}
\toprule
model & reads & training pages & parameters & steps $\times$ batch & hardware & time \\
\midrule
A1, ordered inverter & raw index of $E_A$            & 682{,}818 & 337.1M & 100k $\times$ 16 & 1 H200 & 23.3 h \\
A2, set inverter     & pooled $\times$3 index of $E_A$ & 682{,}818 & 337.1M & 100k $\times$ 16 & 1 H200 & 19.2 h \\
A3, set inverter     & pooled $\times$9 index of $E_A$ & 682{,}818 & 337.1M & 100k $\times$ 16 & 1 H200 & 18.0 h \\
A4, set inverter     & shuffled index of $E_A$       & 682{,}818 & 337.1M & 100k $\times$ 16 & 1 H200 & 22.9 h \\
B1, ordered inverter & raw index of $E_B$            & 346{,}770 & 337.1M & 100k $\times$ 16 & 1 H200 & 12.9 h \\
B2, set inverter     & shuffled index of $E_B$       & 346{,}770 & 337.1M & 100k $\times$ 16 & 1 H200 & 12.7 h \\
A5, ordered inverter & raw index of $E_A$            & 321{,}053 & 337.1M & 100k $\times$ 16 & 1 H200 & 24.4 h \\
\addlinespace[2pt]
set classifier $\times$3 & pooled $\times$3 index of $E_A$ & 227{,}606 & 0.29M & 10 epochs $\times$ 512 & 1 V100 & 12 min \\
set classifier $\times$9 & pooled $\times$9 index of $E_A$ & 227{,}606 & 0.29M & 10 epochs $\times$ 512 & 1 V100 & 5 min \\
\addlinespace[2pt]
position model, $E_A$ & index of $E_A$, order withheld & 682{,}818 & 19.35M + 0.43M & 40k $\times$ 32 & 1 H200 & 57 min \\
position model, $E_B$ & index of $E_B$, order withheld & 346{,}770 & 19.35M + 0.43M & 40k $\times$ 32 & 1 H200 & 34 min \\
\bottomrule
\end{tabular}
\caption{Training setup of every model the attacker trains. A5 is the control of \S\ref{sec:r4}, trained on the VisRAG part alone (Appendix~\ref{app:grids}). Inverters: Eq.~\eqref{eq:flow}, condition dropout 0.1, AdamW ($\beta=(0.9,0.95)$, weight decay 0.01) at $10^{-4}$ with 1k warmup and cosine decay to 10\%, gradient clipping at 1.0, bf16; validation loss on 640 pages spaced evenly through the in-distribution validation split. Set classifiers: cross-entropy over 31 page shapes with logits masked to the shapes compatible with the vector count, AdamW (weight decay 0.01) at $10^{-3}$ with one-cycle scheduling. Position models: squared error on each vector's normalised row and column, and on the log aspect ratio for the 30\% of pages whose shape is withheld, AdamW (weight decay 0.01) at $3\times10^{-4}$ with 1k warmup and cosine decay, bf16; the 0.43M per-vector probe is trained jointly on the same batches, and both are validated on 1{,}024 validation pages. Times are wall-clock times of the training jobs.}
\label{tab:models}
\end{table*}

All six inverters share the architecture and sampling configuration below; the only differences between them are the index they are conditioned on and, for the set inverters, the absence of positions on the condition stream. Their optimisation is given in Table~\ref{tab:models}.

\begin{center}\footnotesize
\begin{tabular}{@{}lp{0.55\columnwidth}@{}}
\toprule
objective & rectified flow, Eq.~\eqref{eq:flow} \\
$t$ & logit-normal, $t=\sigma(\varepsilon)$, $\varepsilon\sim\mathcal{N}(0,1)$ \\
latent & frozen Qwen-Image VAE, $f8$, 16 channels \\
width / depth / heads & $d=768$ / 16 blocks / 6 heads \\
parameters & 337.1M \\
condition dropout & 0.1 \\
\midrule
sampling & 50 uniform Euler steps, one seed \\
guidance & $\gamma=4$, chosen as below \\
\bottomrule
\end{tabular}
\end{center}

Validation loss is computed with the same logit-normal $t$ as training, so the two are directly comparable; the training curve includes the 10\% of samples whose condition is dropped, which is why it sits slightly above validation for every run.

\section{Position models}
\label{app:position}
\paragraph{Models.}
The position model (P2 in the table and figure below) projects each 320-dimensional vector to $d=512$ with a linear layer and layer normalisation and runs six pre-norm Transformer encoder layers (8 heads, feed-forward width 2{,}048, GELU, no dropout) with no positional encoding, 19.35M parameters. A token prepended to the set carries the page shape, $(n_h/40,\;n_w/40,\;\log g_h/g_w)$ through a two-layer MLP, or a learned ``unknown'' embedding in its place; during training the shape is withheld in this way for 30\% of pages. Two linear heads read the normalised row and column of every vector and, from the prepended token, the log aspect ratio. The per-vector probe (P1) is a two-layer GELU MLP of width 512 on the vector concatenated with the same three shape features, 0.43M parameters, applied to each vector independently.

\paragraph{Training.}
Targets are the encoder's own row-major positions,
\[\Big(\tfrac{\lfloor (i-1)/n_w\rfloor}{n_h-1},\;\tfrac{(i-1) \bmod n_w}{n_w-1}\Big),\]
and the loss is the squared error on them, plus the squared error on the log aspect ratio when the shape is withheld. Both models are trained jointly on the same batches with AdamW at $3\times10^{-4}$, weight decay 0.01, 1k warmup then cosine decay, 32 pages per batch bucketed by page shape, 40k steps in bf16 on one H200, which takes 57 minutes under $E_A$ and 34 under $E_B$. We use the last checkpoint; nothing is selected on the target corpus. The training pages are those of the inverter of the same encoder, and the configuration is identical for both encoders.

\paragraph{Assignment and shape.}
Predicted coordinates are scaled to index units and assigned to the $N$ slots of the page shape by the Hungarian algorithm with squared-distance cost, which returns a permutation. Without a given shape, the aspect head picks the factorisation of $N$ in the candidate window of Appendix~\ref{app:shape} nearest to its prediction, and the model is run again with that shape; on the fixed evaluation set this recovers 0.9995 of shapes under $E_A$ and 0.998 under $E_B$, against a count prior of 0.898 and 0.978.
\paragraph{Position accuracy.}
On the fixed evaluation set, after assignment, in index units; $R^2$ is that of the regression before assignment, and the ridge probe from single vectors reaches $R^2$ of 0.541 for the row and 0.110 for the column under $E_A$:
\begin{center}\footnotesize\setlength{\tabcolsep}{2.2pt}
\begin{tabular}{llcccc}
\toprule
 & & row MAE & col MAE & $\le$1 slot & $R^2$ row / col \\
\midrule
$E_A$ & random & 12.59 & 11.28 & 0.007 & -- \\
      & P1     & 4.60 & 7.77 & 0.049 & 0.687 / 0.311 \\
      & P2     & 1.82 & 4.49 & 0.197 & 0.945 / 0.670 \\
\addlinespace[1pt]
$E_B$ & random & 9.72 & 8.70 & 0.012 & -- \\
      & P1     & 5.48 & 7.06 & 0.043 & 0.433 / 0.189 \\
      & P2     & 3.02 & 5.66 & 0.095 & 0.781 / 0.370 \\
\bottomrule
\end{tabular}
\end{center}
$\le$1 slot is the fraction of vectors placed within one slot of their true position on both axes. Under Gaussian noise of $\sigma=0.5,1,2,4$ on the true positions the same quantity is 0.989, 0.746, 0.329 and 0.117, at mean errors of 0.26, 0.76, 1.51 and 2.85.

\begin{figure}[h]
  \centering
  \includegraphics[width=\columnwidth]{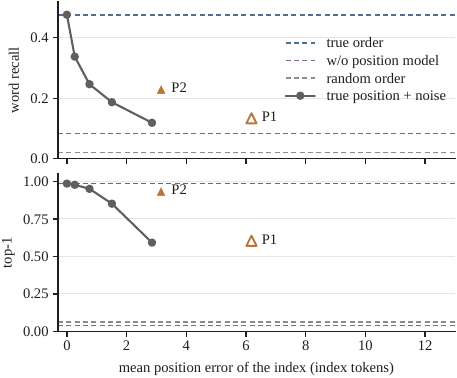}
  \caption{\textbf{Word recall and re-identification under position noise.} Word recall (top) and top-1 (bottom) of the unchanged ordered inverter against the mean position error of the index it is given, under $E_A$. Grey: the true positions with Gaussian noise ($\sigma=0.5,1,2,4$); green: the position models on the shuffled index, the position model (P2) filled and the per-vector probe (P1) open; dashed: the true order, the same shuffled index without a position model, and a random order.}
  \label{fig:position}
\end{figure}

\paragraph{How precisely the order must be restored.}
Figure~\ref{fig:position} feeds each of these indices to the unchanged ordered inverter. A random order gives word recall 0.021 and the per-vector probe 0.134, against 0.231 for the position model, so the gain tracks how well positions are recovered. Content needs more precision than identity: with noise of one index row and column, word recall falls from 0.476 to 0.247 while re-identification is still 0.951; at a mean error of 2.9 they are 0.119 and 0.592. The position model errs by 3.2 index rows and columns on average yet reaches 0.231 and 0.935, above the noise curve on both, which suggests that its errors are correlated across vectors and preserve more of the page than independent noise of the same size.

\section{Guidance ablation}
\label{app:guidance}
Sampling uses classifier-free guidance \citep{ho2022classifierfreediffusionguidance}, combining the unconditioned and the conditioned velocity field as $v=v_u+\gamma\,(v_c-v_u)$, which is their guided estimate with guidance weight $w=\gamma-1$; $\gamma$ sets how far the sample is pushed along the direction the index contributes. At $\gamma=1$ this returns the conditional field itself; at $\gamma\to 0$ it returns the null model of \S\ref{sec:baselines}. For $\gamma>1$ the sample is pushed further from the unconditioned field; whether this adds prior content is an empirical question, which the null and seed-agreement rows of Appendix~\ref{app:controls}, measured at the reported $\gamma$, answer.

We select $\gamma$ once, on 100 in-distribution validation pages disjoint from the target corpus, with the true page shape and a single seed, and then use it for every setting of both encoders without further tuning. Word recall against $\gamma$:
\begin{center}\footnotesize
\setlength{\tabcolsep}{3pt}\begin{tabular}{lccccc}
\toprule
$\gamma$ & 1 & 2 & 4 & 6 & 8 \\
\midrule
$E_A$ raw, ordered & 0.305 & 0.442 & 0.475 & 0.482 & 0.470 \\
$E_B$ raw, ordered & 0.056 & 0.107 & 0.130 & 0.126 & 0.127 \\
$E_A$ raw, shuffled & 0.045 & 0.059 & 0.070 & -- & -- \\
$E_A$ pooled $\times$3 & 0.055 & 0.063 & 0.063 & -- & -- \\
$E_A$ pooled $\times$9 & 0.042 & 0.067 & 0.075 & -- & -- \\
\bottomrule
\end{tabular}
\end{center}
Both ordered inverters plateau between 4 and 8, and we take 4, the smallest value on the plateau. The choice is not neutral across settings. The ordered inverter gains 0.170 of word recall, the pooled and shuffled settings at most 0.033, and in those settings NED does not improve at all, moving from 0.867 to 0.848 for the shuffled index and from 0.899 to 0.902 for $\times 9$. On a protected index, guidance therefore recovers additional common vocabulary rather than additional characters of the source.

\begin{figure}[h]
  \centering
  \includegraphics[width=\columnwidth]{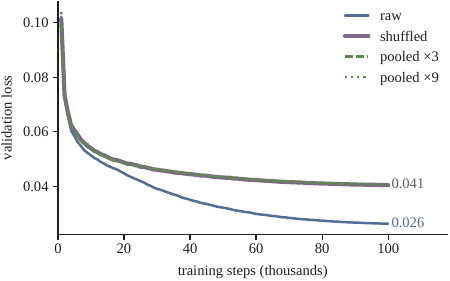}
  \caption{\textbf{Validation loss of set and ordered inverters.} Validation loss of four inverters trained identically: on the ordered raw index, on the same vectors with their order shuffled, and on the pooled $\times$3 and $\times$9 indices. The three unordered variants coincide throughout training; only the ordered index is learnable by this inverter.}
  \label{fig:valcurves}
\end{figure}
\section{Controls in full}
\label{app:controls}
Table~\ref{tab:controls} gives the controls that separate content determined by the index from content generated by the prior. The null model samples the same inverter with its condition replaced by the null token. Seed agreement is the word F1 between the OCR transcripts of two samples drawn from the same index with seeds 0 and 1, compared with the word F1 of each sample against the source page. The VAE reconstruction passes the source page through the frozen VAE and gives the upper bound in Table~\ref{tab:main}. A random order fed to the ordered inverter is the reference level for the attack with the position model (Appendix~\ref{app:position}). The null model and seed agreement use a 200-page subset spaced evenly through the fixed evaluation set.
\begin{table}[h]
\centering\footnotesize\setlength{\tabcolsep}{4pt}
\setlength{\tabcolsep}{3pt}\begin{tabular}{lccc}
\toprule
Control (200 pages) & $E_A$ & $E_B$ & $\times$9 \\
\midrule
null, word recall        & 0.001 & 0.001 & 0.001 \\
null, NED                & 0.958 & 0.963 & 0.943 \\
\addlinespace[1pt]
seed agreement (F1)      & 0.476 & 0.130 & 0.094 \\
\quad vs.\ truth (F1)    & 0.480 & 0.098 & 0.076 \\
\addlinespace[1pt]
recall, true shape       & 0.481 & 0.094 & 0.083 \\
recall, inferred shape   & 0.478 & 0.080 & 0.083 \\
shape inferred correctly & 0.995 & 0.815 & 0.965 \\
\bottomrule
\end{tabular}
\caption{Controls separating content determined by the index from content generated by the prior, on a 200-page subset spaced evenly through the fixed set. $E_A$ and $E_B$ are the raw index of each encoder; $\times$9 is the pooled index of $E_A$. All three columns are measured at the reported guidance scale. The null rows are scored with the Latin-script recogniser, as in the main tables; the seed-agreement and recall rows come from the sampling run and its English recogniser, which changes text metrics by at most 0.01 (Appendix~\ref{app:regex}). The VAE reconstruction is the last row of Table~\ref{tab:main}; near-duplicates between the training and target corpora are quantified in \S\ref{sec:setup}.}
\label{tab:controls}
\end{table}

\section{Bootstrap intervals}
\label{app:bootstrap}
Table~\ref{tab:bootstrap} gives 95\% percentile-bootstrap intervals for the rows of Table~\ref{tab:main}, resampling pages with replacement ($B=10{,}000$, seed 0): the 1{,}927 scored pages for the OCR metrics and all 2{,}000 for top-1. Differences are paired on the same pages. Every interval lies within $\pm0.008$ of its mean for the OCR metrics and within $\pm0.018$ for top-1, so larger differences between rows of Table~\ref{tab:main} are not an artefact of which pages were sampled. Every gap the text relies on excludes zero: the position model adds 0.147 of word recall to the attack without it, and pooling removes 0.393 from the raw index. With the position model, word recall does not differ from the kNN baseline, $-0.003$ with an interval that contains zero, while its sensitive-token recall exceeds kNN by 0.066 and its top-1 is 0.935 against 0.199, so what the restored order recovers beyond shared vocabulary is the page's own sensitive tokens and identity.
\begin{table*}[t]
\centering\footnotesize
\begin{tabular}{@{}lcccc@{}}
\toprule
Index, attack & word recall & NED & sens.\ recall & top-1 \\
\midrule
kNN (nearest train page) & 0.234 [0.227, 0.241] & 0.858 [0.853, 0.864] & 0.146 [0.140, 0.152] & 0.199 [0.181, 0.216] \\
raw, ordered & 0.474 [0.468, 0.480] & 0.292 [0.285, 0.299] & 0.450 [0.442, 0.457] & 0.984 [0.978, 0.990] \\
raw, ordered, true shape & 0.476 [0.470, 0.483] & 0.290 [0.283, 0.296] & 0.450 [0.443, 0.457] & 0.986 [0.981, 0.991] \\
shuffled, w/o position model & 0.084 [0.081, 0.087] & 0.842 [0.838, 0.846] & 0.048 [0.045, 0.050] & 0.038 [0.030, 0.047] \\
shuffled, w/ position model & 0.231 [0.226, 0.236] & 0.925 [0.921, 0.929] & 0.212 [0.206, 0.217] & 0.935 [0.924, 0.946] \\
pooled $\times$3 & 0.076 [0.073, 0.079] & 0.858 [0.853, 0.862] & 0.044 [0.042, 0.046] & 0.011 [0.006, 0.015] \\
pooled $\times$9 & 0.081 [0.079, 0.084] & 0.879 [0.875, 0.884] & 0.054 [0.051, 0.057] & 0.005 [0.002, 0.009] \\
\bottomrule
\end{tabular}

\medskip
\begin{tabular}{@{}llc@{}}
\toprule
Paired difference & metric & mean [95\% CI] \\
\midrule
w/ position model $-$ w/o position model & word recall & $+0.147$ [$+0.143$, $+0.151$] \\
w/ position model $-$ kNN & word recall & $-0.003$ [$-0.011$, $+0.005$] \\
w/ position model $-$ kNN & sens.\ recall & $+0.066$ [$+0.058$, $+0.073$] \\
raw $-$ $\times$9 & word recall & $+0.393$ [$+0.386$, $+0.399$] \\
$\times$9 $-$ kNN & word recall & $-0.152$ [$-0.159$, $-0.146$] \\
$\times$3 $-$ $\times$9 & word recall & $-0.005$ [$-0.007$, $-0.004$] \\
true shape $-$ raw & word recall & $+0.002$ [$+0.000$, $+0.005$] \\
\bottomrule
\end{tabular}

\caption{Means over pages with 95\% bootstrap intervals under $E_A$ (top), and paired differences on the same pages (bottom). The null row of Table~\ref{tab:main} is a 200-page control and is not resampled.}
\label{tab:bootstrap}
\end{table*}

\section{Re-identification and readability}
\label{app:readability}
\begin{figure}[h]
  \centering
  \includegraphics[width=\columnwidth]{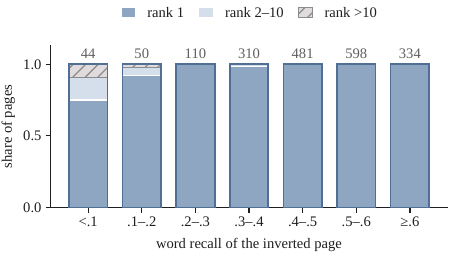}
  \caption{\textbf{Re-identification versus word recall.} The 1,927 scored pages of the raw index under $E_A$, binned by word recall: the share whose source ranks first, second to tenth, or beyond tenth among 19,252, with the number of pages per bin above each bar. Below a word recall of 0.1, 75\% of pages still rank first; over all pages the Spearman correlation between the two is $\rho=0.13$.}
  \label{fig:analysis}
\end{figure}
The encoder is more tolerant than OCR: layout and partially formed glyphs that do not transcribe still suffice to rank the page first, so a page that OCR cannot read has not thereby lost the information that distinguishes it. The lowest-ranked page of the fixed set sits at rank 991, and 99.8\% rank within the top 100.

\section{Results by domain}
\label{app:persubset}
\begin{table}[h]
\centering\small
\setlength{\tabcolsep}{3pt}\begin{tabular}{lccc}
\toprule
Subset & top-1 & top-5 & MRR \\
\midrule
cs (textbooks)         & 1.000 & 1.000 & 1.000 \\
energy                 & 0.996 & 0.996 & 0.997 \\
finance\_en            & 0.996 & 0.996 & 0.997 \\
physics (slides)       & 0.992 & 1.000 & 0.995 \\
hr                     & 0.984 & 0.992 & 0.987 \\
pharma                 & 0.980 & 1.000 & 0.989 \\
finance\_fr            & 0.968 & 1.000 & 0.982 \\
industrial             & 0.956 & 0.972 & 0.965 \\
\midrule
all                    & 0.984 & 0.995 & 0.989 \\
\bottomrule
\end{tabular}
\caption{Re-identification of raw-index inversions under $E_A$ by ViDoRe~v3 subset (250 pages each, 19,252-page corpus). Judged by the independent encoder $E_B$ instead: top-1 0.977.}
\label{tab:reid}
\end{table}

nDCG@10 on ViDoRe~v3 with English queries, scoring each subset's own documents, for the raw index and the two pooling factors. This is the utility side of the trade-off summarised in Table~\ref{tab:main}. Word recall from the raw index is not carried by a few pages: its median over the 1,927 scored pages is 0.493, and 47.1\% of pages exceed 0.5. The kNN baseline is strongest on the English annual reports, at 0.400, where boilerplate, years and currency units recur; the inversion reaches 0.442 there and exceeds the baseline in every domain.
\begin{center}\footnotesize
\setlength{\tabcolsep}{3pt}\begin{tabular}{lrrccc}
\toprule
subset & pages & queries & raw & $\times$3 & $\times$9 \\
\midrule
cs          & 1{,}360 & 215 & 0.7770 & 0.7641 & 0.7561 \\
energy      & 2{,}225 & 308 & 0.6795 & 0.6781 & 0.6721 \\
finance\_en & 2{,}942 & 309 & 0.7019 & 0.6995 & 0.6936 \\
finance\_fr & 2{,}384 & 320 & 0.4812 & 0.4715 & 0.4560 \\
hr          & 1{,}110 & 318 & 0.6697 & 0.6606 & 0.6479 \\
industrial  & 5{,}244 & 283 & 0.5962 & 0.5922 & 0.5792 \\
pharma      & 2{,}313 & 364 & 0.6840 & 0.6829 & 0.6747 \\
physics     & 1{,}674 & 302 & 0.5016 & 0.5027 & 0.5011 \\
\midrule
mean        &         &     & 0.6364 & 0.6314 & 0.6226 \\
\bottomrule
\end{tabular}
\end{center}
The loss is 0.8\% of nDCG@10 at factor 3 and 2.2\% at factor 9. It is not uniform: the French financial filings lose 5.2\% at factor 9, more than twice the mean, while the physics slides lose 0.1\%. A defender choosing a pooling factor on an aggregate number is therefore choosing a different operating point for each of their domains.

\section{Where the attack fails}
\label{app:failure}
Under $E_A$ at the reported operating point no page of the fixed evaluation set falls beyond rank 1{,}000, so there is no tail of outright failures to display. The residual errors are concentrated instead in two places. The first is the 32 pages, 1.6\% of the set, that are not ranked first; 19 of them are industrial manuals and French financial filings, the two subsets with the lowest top-1 (Table~\ref{tab:reid}). The second is the 17 pages, 0.8\%, whose shape is inferred wrongly and which are therefore drawn at the wrong shape.

The inverters for pooled indices and for the shuffled index without a position model fail differently, and the failure is the one the controls are designed to catch: they produce a plausible page of the right kind whose content does not match the source. The null model shows what that prior looks like on its own, at word recall 0.001 and NED 0.958 with a within-page correlation of $-0.02$. The gap between the null model and the pooled inverter is the part the pooled index still determines; on the $\times 9$ index two samples agree with each other at F1 0.094 and with the source at 0.076 (Table~\ref{tab:controls}).

\begin{figure*}[t]
  \centering
  \includegraphics[width=\textwidth]{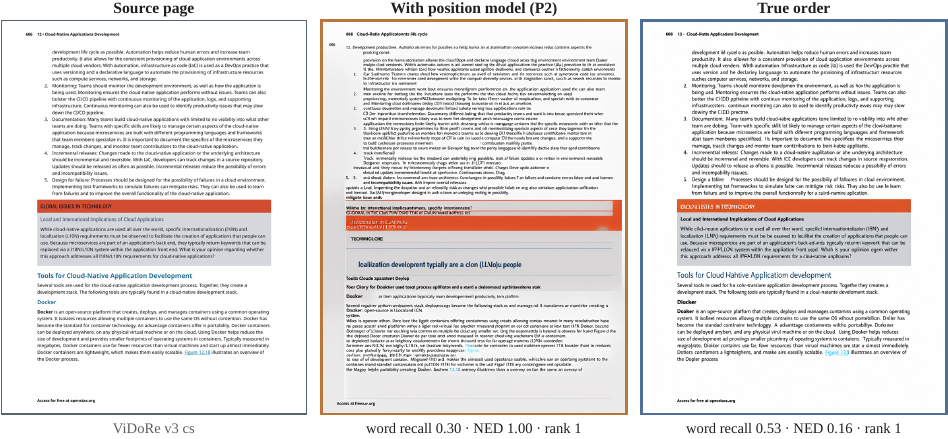}
  \caption{One page of the fixed set (ViDoRe~v3 \emph{cs}): source (left), the ordered inverter on the shuffled index after the position model restores the order (middle), and the same inverter on the true order (right), each with its word recall, NED and rank. Same inverter, seed and guidance scale.}
  \label{fig:reorder}
\end{figure*}

The attack on the shuffled index with the position model fails in a third way, shown in Figure~\ref{fig:reorder}. The words, the numbered list, the section banner and the headings are recovered, but lines and blocks appear at the wrong vertical positions, so the transcript is read in a different order. This page was chosen to illustrate that case, among pages with word recall near 0.3 and NED at its cap, and is not typical of this setting.

\section{The open problem}
\label{app:open}
Inverting a pooled index is open; our failure to do so is a failure of the methods we tried, not evidence that a pooled index is safe. The route that worked for a shuffled index does not transfer directly: a position model and an assignment put each raw vector back in one slot because each raw vector came from one patch, while a pooled centroid is the mean of a cluster of patches and has no single slot. Three starting points are available. First, soft positions: a model that predicts, for each centroid, a distribution over the slots of the page shape, trained with labels the public encoder provides on its own, and an inverter conditioned on those distributions. Second, differentiable assignment: the matching between vectors and slots can be treated as an optimal-transport problem inside the condition stream and learned end to end. Third, capacity: a larger set-conditioned backbone or a longer schedule may learn layout from content where our 337M model did not, and its validation loss was still decreasing. The benchmark is in place: the fixed evaluation set, the $\times 3$ and $\times 9$ pooled indices, and the protocol of this paper, with the kNN and null rows as lower bounds and the ordered index as the target to approach. Our own $\times 9$ result already sits above the null model, so the question is not whether the pooled index leaks at all but how far the gap to the ordered index can be closed; the shuffled index shows that such a gap can close from 3.8\% to 93.5\% re-identification with no change to the inverter.

\section{Reproducibility}
\label{app:repro}
\paragraph{Artefacts.}
Code, the evaluation manifest and the protocol will be released upon publication; the trained inverters will not (Ethical Considerations). Attacked public encoders: $E_A$ is \nolinkurl{TomoroAI/tomoro-colqwen3-embed-8b} at revision \texttt{\seqsplit{49658433ec26b7d6ee8225972e2e33577db9f35f}}; $E_B$ is \nolinkurl{athrael-soju/colqwen3.5-4.5B-v3}. Re-encoding a page under our pixel convention reproduces its stored index, which is the check reported below. Pooling uses the hierarchical token pooler of \nolinkurl{illuin-tech/colpali} at commit \texttt{\seqsplit{c23838d920a7c426ee297034211cff2f55da65dc}}, vendored byte-identical and invoked without modification. OCR is PaddleOCR with the \texttt{latin} recogniser and the default detector, applied identically to the source page and to the inverted page. Software: Python 3.10, PyTorch 2.8 (CUDA 12.8), PaddleOCR 2.10, SciPy 1.15 for the Hungarian assignment. Seeds: sampling draws its initial noise from a generator seeded with 0 (1 for the second sample of the agreement control), training batches are ordered with seed 0, and the defender's permutation $\pi$ of a page uses the seed pair (7, page); weight initialisation is not seeded, and each model was trained once.

\paragraph{Second encoder conventions.}
$E_B$ admits at most 768 merged tokens per page against 1,280 for $E_A$, so it encodes each page at a lower input resolution: 739 vectors and 0.757\,Mpx on average over its training pages, against 1,198 and 1.227\,Mpx for $E_A$ (1,241 vectors on the evaluation set). It resamples with bicubic interpolation and antialiasing, a convention different from that of $E_A$, and we match it exactly when building its inversion targets.

\paragraph{Alignment checks.}
Three checks were run before any attack, since an index that is not the page's own would invalidate every number. Re-encoding a page under our pixel convention reproduces its stored index at cosine 0.999--1.000, against 0.09--0.17 for other pages. The grid convention $N=g_hg_w/m^2$ holds for all 711{,}603 pages of the VisRAG part (Appendix~\ref{app:grids}) under $E_B$ with no exceptions. A zero-vector audit over 59.9M stored tokens found no padding rows in either data loader, so no inverter is conditioned on padding.

\paragraph{Baseline details.}
The kNN baseline mean-pools each training page's index into one vector, retrieves the 20 nearest by cosine, re-ranks them with the exact late-interaction score of Eq.~\eqref{eq:maxsim}, and returns the top page as the inverted page. Its re-identification query is that retrieved page's own stored index, not a re-encoding, which favours the baseline.

\paragraph{Overlap check.}
For each evaluation page we take its nearest training page, retrieved as for the kNN baseline, and divide their late-interaction score (Eq.~\eqref{eq:maxsim}) by the score of the evaluation page's index against itself. A training page counts as a near-identical counterpart when this ratio exceeds 0.9. This flags 8 of the 2{,}000 evaluation pages (0.4\%); removing them changes no reported number.

\paragraph{Fixed evaluation set.}
The 2{,}000 pages are drawn once, 250 per subset, and recorded as a manifest of row identifiers before any inverter was evaluated; every setting of both encoders is scored on that same manifest. The 200-page control subset is that manifest sampled at stride 10, so it spans all eight domains in proportion; a prefix would have covered one or two domains only.
\section{Licences and compute}
\label{app:licenses}
Every artefact we use is public, and we use each for research only, within its terms.
\begin{center}\footnotesize
\begin{tabular}{@{}p{0.56\columnwidth}p{0.38\columnwidth}@{}}
\toprule
artefact & licence \\
\midrule
VisRAG training sets & not stated on the card \\
ColPali training set & not stated on the card \\
vdr-multilingual-train & Apache-2.0 \\
VDR-MEGA-2 & Apache-2.0 \\
ViDoRe~v3, eight public corpora & CC BY 4.0 \\
$E_A$, $E_B$, Tomoro 4B, Vultron, ColQwen2, ColNomic & Apache-2.0 \\
ColPali v1.2, v1.3, ColSmol 256M, 500M & MIT \\
webAI ColVec1.1 4B, 8B & webAI non-commercial v1.0 \\
SauerkrautLM-ColLFM2-450M & LFM Open License v1.0 \\
Qwen-Image (VAE) & Apache-2.0 \\
PaddleOCR & Apache-2.0 \\
\bottomrule
\end{tabular}
\end{center}
Training the seven inverters took 133.3 GPU-hours on single NVIDIA H200 GPUs, from 12.7 to 24.4 hours each, and the position models of both encoders 1.5 GPU-hours. Sampling and re-identifying the 2,000 evaluation pages of one setting takes about 1.7 hours on a partition of one H200.
\end{document}